\documentclass[useAMS,usenatbib,twocolumn]{mn2e}
\pdfoutput=1
\usepackage{amssymb}
\usepackage{amsmath}
\usepackage{natbib}
\usepackage[pdftex]{graphicx}
\usepackage{subfigure}
\usepackage{booktabs}
\usepackage{tabularx}
\begin{document}

\title[Vertical-shear instability]
{On the vertical-shear instability in astrophysical discs}
     \author[A. J. Barker \& H. N. Latter]{A. J. Barker\thanks{Email address: ajb268@cam.ac.uk} and H. N. Latter \\
Department of Applied Mathematics and Theoretical Physics, University of Cambridge, Centre for Mathematical Sciences, \\ Wilberforce Road, Cambridge CB3 0WA, UK}
	
\pagerange{\pageref{firstpage}--\pageref{lastpage}} \pubyear{2015}

\maketitle

\label{firstpage}

\begin{abstract}
We explore the linear stability of astrophysical discs exhibiting vertical shear, which arises when there is a radial variation in the temperature or entropy. Such discs are subject to a ``vertical-shear instability'', which recent nonlinear simulations have shown to drive hydrodynamic activity in the MRI-stable regions of protoplanetary discs. We first revisit locally isothermal discs using the quasi-global reduced model derived by Nelson et al. (2013). This analysis is then extended to global axisymmetric perturbations in a cylindrical domain. We also derive and study a reduced model describing discs with power law radial entropy profiles (``locally polytropic discs"), which are somewhat more realistic in that they possess physical (as opposed to numerical) surfaces. In all cases the fastest growing modes have very short wavelengths and are localised at the disc surfaces (if present), where the vertical shear is maximal. An additional class of modestly growing vertically global body modes is excited, corresponding to destabilised classical inertial waves (``r-modes''). 
We discuss the properties of both types of modes, and stress that those that grow fastest occur on the shortest available length scales (determined either by the numerical grid or the physical viscous length). This ill-posedness makes simulations of the instability difficult to interpret. We end with some brief speculation on the nonlinear saturation and resulting angular momentum transport.
\end{abstract}


\section{Introduction}

Accretion through magnetorotational turbulence is only viable
in sufficiently ionised regions of
protoplanetary discs, namely at their inner and outer radii 
\citep{BalbusHawley1998,Armitage2011}. Between 
 1-20 AU (the ``dead zone'')
 non-ideal effects extinguish the MRI,
 and instead accretion may occur via magneto-centrifugally launched
 winds (e.g.~\citealt{Lesur2014,Bai2014}). 
However, identifying additional hydrodynamic mechanisms for
driving turbulence is essential,
 due to its potential impact on the dynamics of solids, and therefore for planet formation.

Though pure Keplerian shear flow is difficult to destabilise, 
several mechanisms have been proposed: subcritical baroclinic
instability \citep{Petersen2007,LesurPap2010}, convective instability
\citep{RudenPap1988,LesurOgilvie2010} and gravitational instability
\citep{Toomre1964,LinPringle1987}, to name but a few. 
 Another mechanism that has recently received attention is the
 ``vertical-shear instability'' (hereafter VSI) which, as its name
 suggests,
 attacks rotating systems that exhibit vertical shear
(\citealt{UrpinBrand1998}; 
\citealt{Urpin2003}). Fundamentally, the VSI is a form of centrifugal
instability and is a 
close cousin of the Goldreich-Schubert-Fricke instability,
originally applied to stellar interiors
\citep{GoldreichSchubert1967,Fricke1968}.
But observations of protostellar discs \citep{AndrewsWilliams2005} and
theoretical models of passively heated discs \citep{ChiangGold1997}
suggest that they too should display destabilising vertical shear,
generated from radial variations in temperature
or entropy. Recent numerical simulations
 indicate that the nonlinear evolution of the
VSI can produce hydrodynamic turbulence and modest levels of angular
momentum transport (\citealt{Nelson2013}; \citealt{StollKley2014}). It
thus could be a potentially key player in the dynamics of
protoplanetary disc dead zones.

The VSI was originally
studied with a local (Boussinesq) approach by
\cite{UrpinBrand1998} and by \cite{Urpin2003}. More recently
\cite{Nelson2013} described it
with a quasi-global model that 
captures the full vertical structure of growing anelastic
modes in radial geostrophic balance (assuming the background is
locally isothermal). Being global in the vertical, but local in the
radial, this model is akin to the commonly used vertically
stratified shearing box. However,
several properties of the linear VSI require further explanation,
especially with respect to its global manifestation in more realistic
disc models. This is a particularly important issue when trying to connect the
linear theory to global simulations, and in
interpreting their nonlinear outcome. Our paper is devoted to
exploring this aspect of the problem.

 We perform linear stability analyses of astrophysical
discs exhibiting global variations in temperature and entropy, and as
a consequence vertical shear. We employ locally
isothermal and polytropic models in both quasi-global and fully global
2D geometries, which revise and extend previous work. 

In agreement with \cite{Nelson2013}, we find that the VSI excites two
classes of modes.
The first class corresponds to classical free inertial waves (r-modes) that are
present in any astrophysical disc \citep{LP1993,KP1995,Kato2001} but which have been destabilised by the
vertical shear. These, referred to as ``body modes'', grow at modest
rates and typically exhibit longer wavelengths (though the radial
wavelength of the waves is still short). 

The second class
corresponds to modes localised to the vertical surfaces of the disc where the
vertical shear is maximal. 
These grow much faster and have very short wavelengths, making them
difficult to resolve numerically. In fact, unless viscosity is
included, the fastest growing modes possess arbitrarily small
wavelengths, making their simulation problematic.
Note that, though
they have been termed
``surface modes'', these are different to the classical surface
gravity waves that appear in polytropic disc models, as they lie in a
different frequency range; they are hence a form of localised low-frequency inertial
wave. Strict
isothermal models do not possess a physical vertical surface and hence do not
support these surface modes. Polytropic disc models do, however, as should
any realistic disc model that possesses a transition between an
optically thick interior and an optically thin ``corona''.

We begin by explaining why a radial variation in entropy or
temperature generally leads to vertical shear in \S~\ref{physics}.
There we also explain why such discs are likely to be unstable. After
defining our basic disc models in \S~\ref{Basiceqns}, we analyse
the resulting VSI in the locally isothermal disc in \S~\ref{VSIiso}
and \ref{2D} and the locally polytropic disc in
\S~\ref{VSIpoly}. Finally, we will discuss the implications of our
results in \S~\ref{Conclusions}, where we also speculate on the
nonlinear evolution of the VSI  and its efficiency at transporting  
angular momentum.

\section{Vertical-shear instability}
\label{physics}

\begin{figure*}
  \begin{center}
     \subfigure[$\Omega$]{\includegraphics[trim=7cm 1cm 6cm 2cm, clip=true,width=0.3\textwidth]{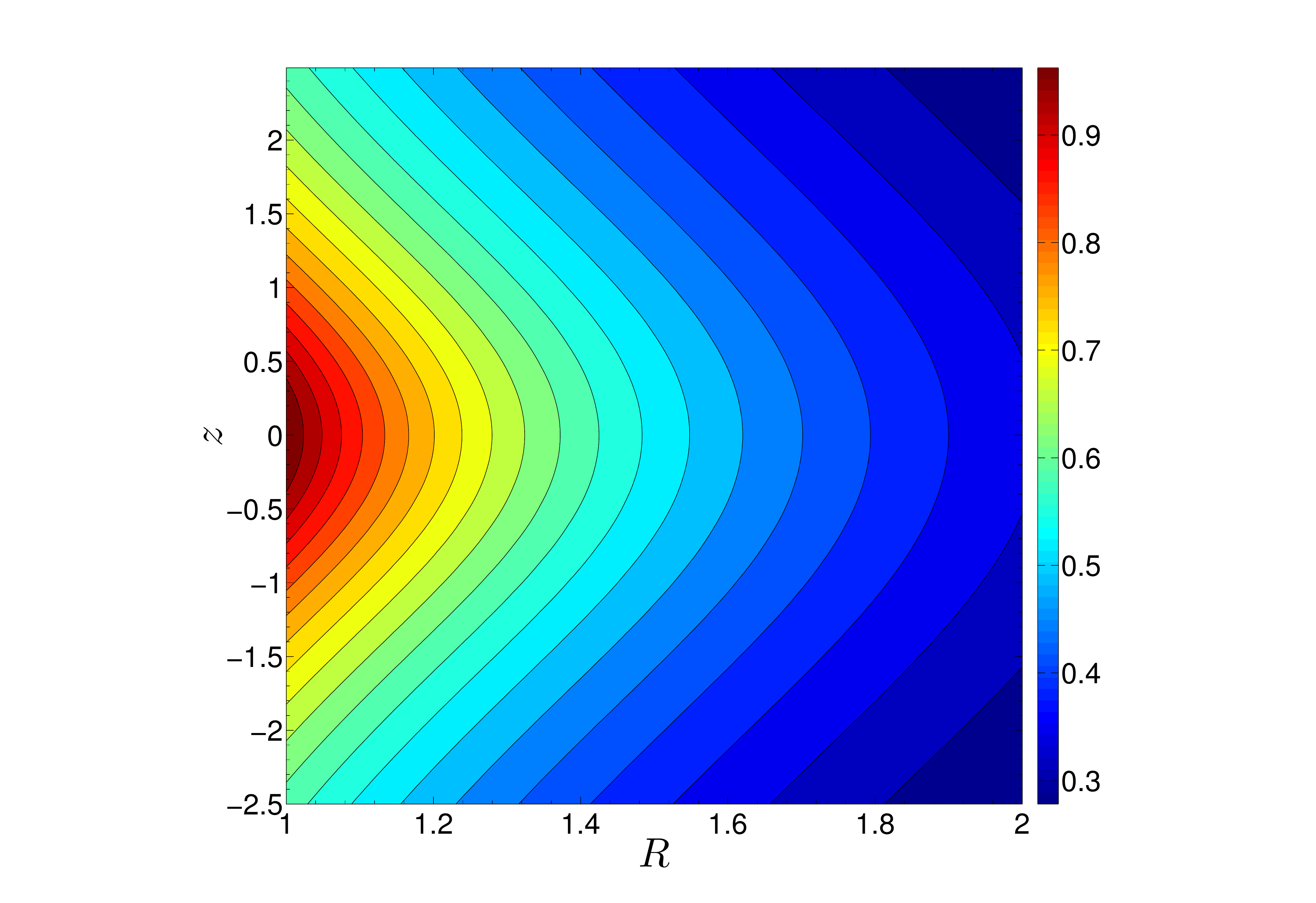} } 
      \subfigure[$\partial_{z}(R\Omega)$]{\includegraphics[trim=7cm 1cm 6cm 2cm, clip=true,width=0.3\textwidth]{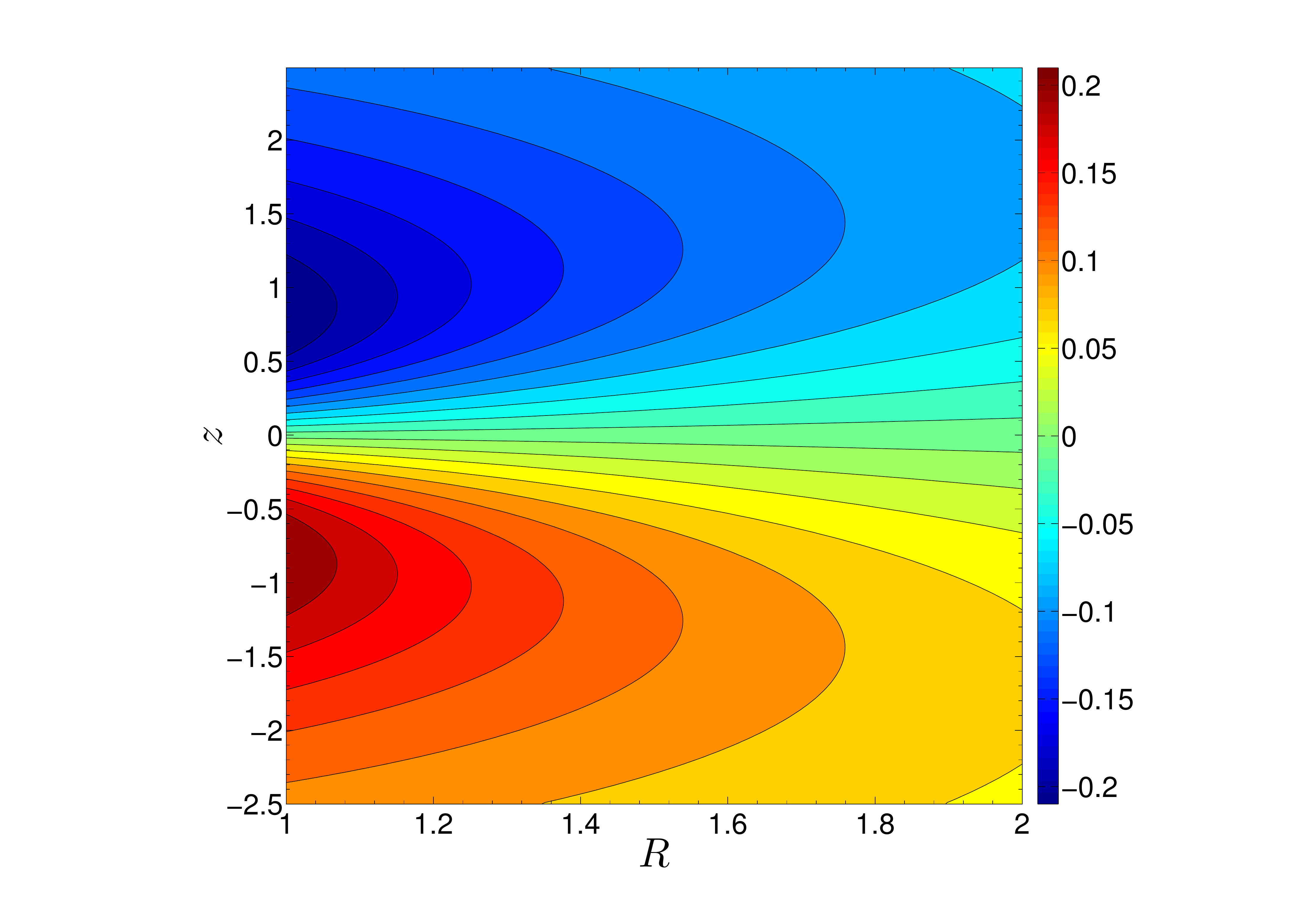} } 
       \subfigure[$\rho$]{\includegraphics[trim=6cm 1cm 6cm 2cm, clip=true,width=0.31\textwidth]{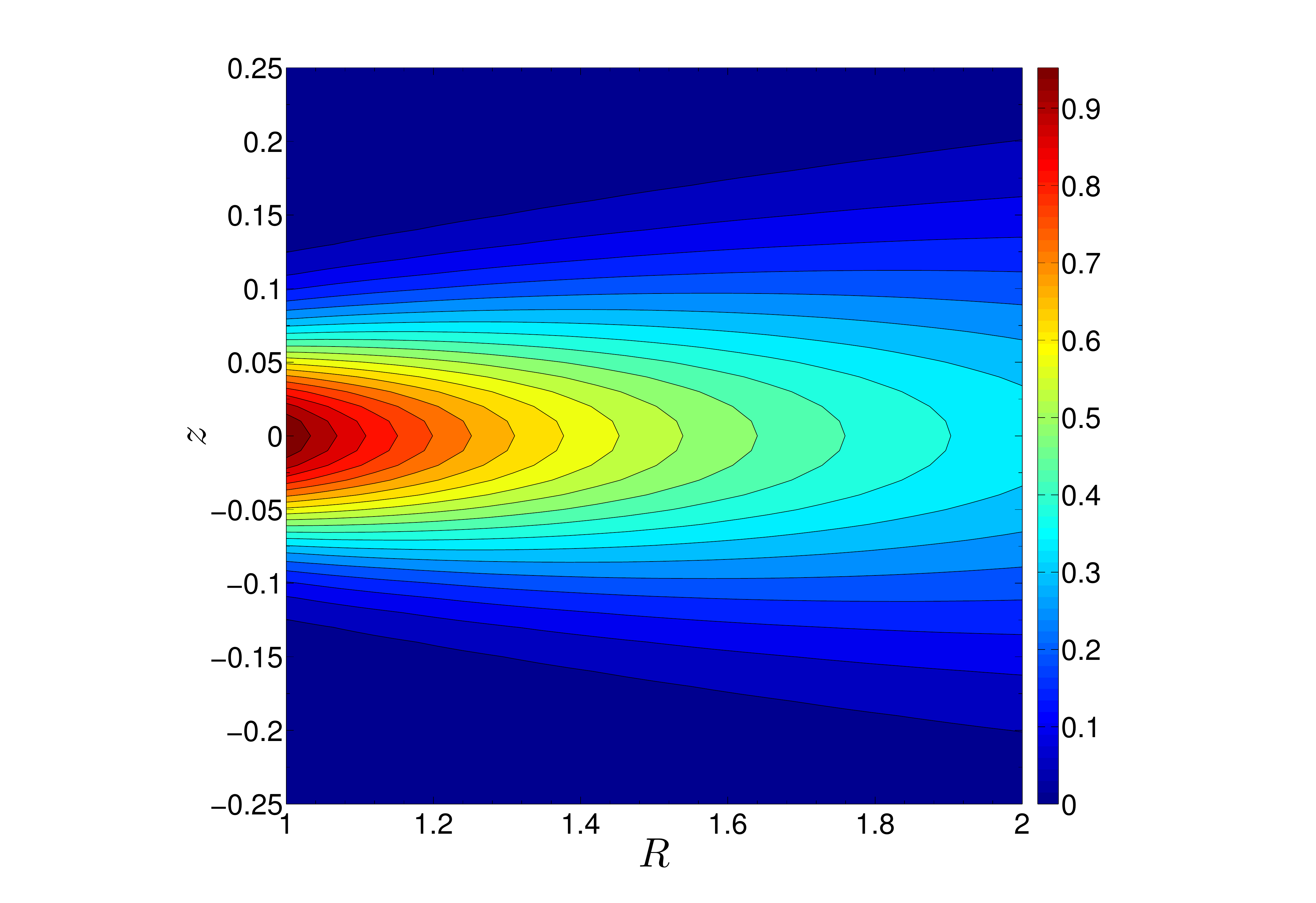} }
    \end{center}
  \caption{Basic state for the locally isothermal disc with $q=-1,
    p=-1.5$ and $c_0=0.05$. The left panel shows a contour plot of
    $\Omega$ on the $(R,z)$ plane. The middle panel is a similar
    contour plot, but this shows the magnitude of the vertical
    shear $\partial_{z}(R\Omega)$, which has a maximum at $|z|\sim 1$ (whereas the scaleheight at the inner radial boundary is 0.05). The right panel shows the density $\rho$.}
  \label{0a}
\end{figure*}

Discs with radial variations in temperature or entropy necessarily possess vertical shear. To see that this must be, consider the ``thermal wind equation" (the azimuthal component of the vorticity equation for the axisymmetric basic state of the disc):
\begin{eqnarray}
\label{TWE}
\partial_{z}(R\Omega^2) &=&-\boldsymbol{e}_{\phi}\cdot \left(\nabla \rho\times\nabla P\right)/\rho^2 \\
&=&\partial_{R}T\partial_{z}S-\partial_{z}T\partial_{R}S.
\end{eqnarray}
Here we have adopted cylindrical polar coordinates centred on the
central object $(R, \phi, z)$ and $\rho, P, S$ and $T$ are the basic
state density, pressure, specific entropy and temperature profiles,
respectively. Eq.~\ref{TWE} tells us that a radial variation in the background
temperature or entropy generates a departure from cylindrical
rotation through the baroclinic terms on the right hand side.
Thus the angular velocity
$\Omega=\Omega(R,z)$, and consequently the disc exhibits 
a weak vertical shear. 
 For
illustration, we show the angular velocity and vertical shear for a disc with a radial variation in temperature in Fig.~\ref{0a}, and the vertical shear for a disc with a radial variation in entropy in Fig.~\ref{0b} (both disc models and the notation adopted are defined in \S~\ref{Basiceqns}).

\subsection{Physical picture}

Vertical shear provides a source of free energy that can drive
hydrodynamic instabilities. How might modes access this free energy?
Consider a ring of fluid at a given location (A) within the disc with
coordinates $(R_{A},z_{A})$, and hence specific angular momentum
$h_{A}=R^2_{A}\Omega(R_{A},z_{A})$. If we vertically displace this ring to a new
position (B) with coordinates $(R_{A},z_{A}+\Delta z)$, then its
specific angular momentum will be conserved as long as viscosity can
be neglected (i.e.\ we assume that $|\Delta z|$ is much larger than the viscous length).
But if the angular momentum of
fluid at the new location $h_B$ is smaller
(larger) than $h_{A}$, then the ring will be pushed 
outwards (inwards) by the centrifugal acceleration
$(h_{A}^2-h_{B}^2)/R_{A}^3$, leading to a dynamical
instability. Given that $h^2_{B}\approx h^2_{A} + \Delta
z \partial_{z}h^2$, instability occurs whenever $\partial_{z}h^2 <0$ (or indeed $>0$), i.e. if
there is any vertical shear. Basically, this interchange of rings of
fluid reduces the total energy of 
the configuration, leading to an instability that transports angular
momentum in order to eliminate the vertical shear.\footnote{Our
  illustrative perturbation is vertical for simplicity; any
  displacement lying within the 
 angle between the rotation axis and a surface of constant angular
 momentum will do (as explained in \citealt{KnoblochSpruit1982}, for
 example).} 
This is a modified form of Rayleigh's argument for
centrifugal instability. Though accretion discs are stable
according to the classical Rayleigh criterion, any vertical
shear permits its circumvention and hence the onset of
instability.

This physical argument works for neutrally stratified discs, but must
be altered when a stable vertical entropy stratification is present,
as exhibited by protoplanetary disc dead zones.
So we next introduce buoyancy, and viscous and
thermal diffusion. Buoyancy forces impede exchanges of the type described,
and thus inhibit any adiabatic (dynamical) instability (cf.\ the
Solberg-H\o iland criterion). Instability is
nevertheless possible
 if the buoyancy forces are eliminated, such as by sufficiently fast
 cooling or thermal diffusion. 
For this to work, displacements $|\Delta z|$ must then be much
 shorter than the thermal diffusion scale. 
 The resulting instability is hence double-diffusive in character, possessing
 lengthscales lying in the range bounded from below by the
 viscous length and above by the thermal diffusion length.
 Originally identified in the 1960's and applied to stellar interiors
 \citep{GoldreichSchubert1967,Fricke1968}, 
 only much later was it recognised that such 
an instability could emerge in astrophysical discs
 \citep{UrpinBrand1998}.\footnote{\cite{UrpinBrand1998} 
 coined the term ``vertical shear instability''. 
 However, there is a good case for the retention of the
 name ``Goldreich-Schubert-Fricke'' (GSF) instability, 
even if ``VSI'' has the merit of clearly advertising the
underlying physics.}

\subsection{Estimates from a local model}

According to a local Boussinesq analysis, the growth rate of the VSI
is
\begin{eqnarray}
\sigma \approx |\partial_{z}(R\Omega)|\sim \epsilon |q| \Omega,
\end{eqnarray}
where $\epsilon=H/R$ is the disc aspect
 ratio and $q$ is the exponent in the power law for 
temperature (or entropy), so that $T\propto R^q$
\citep{UrpinBrand1998,Urpin2003,Nelson2013}. For $|q|\sim 1$, the VSI
will hence grow, and presumably saturate, relatively quickly on a
timescale not far from the dynamical time for thicker discs.

The VSI afflicts intermediate lengthscales $\ell$ in the range 
\begin{eqnarray}
\ell_{\nu}\,\lesssim\, \ell\, \lesssim\, \ell_{\chi},
\end{eqnarray}
where the viscous and thermal diffusion lengths are defined through
\begin{eqnarray}
\ell_{\nu}= (\nu/\sigma)^{\frac{1}{2}}, \qquad \ell_{\chi}=  (\chi/N_{z})^{\frac{1}{2}}. 
\end{eqnarray}
Here $\nu$ is the kinematic viscosity, $\chi$ is the thermal
diffusivity, and
$N_{z}>0$ is the vertical buoyancy frequency.
On the other hand, 
lengthscales above $\ell_{\chi}$ are stabilised by buoyancy forces (as long as $N_{z}^2>0$); only when these are subdued by sufficiently rapid thermal diffusion is instability possible. 
On lengthscales smaller than $\ell_{\nu}$, viscosity neutralises the excess angular momentum of a fluid element too quickly to allow the mode to grow. The fastest growing modes have vertical ($k_{z}$) and radial wavenumbers ($k_{R}$) that satisfy
\begin{eqnarray} \label{lengths}
\frac{k_{z}}{k_{R}}=\frac{\ell_R}{\ell_z}\lesssim \epsilon\,q,
\end{eqnarray}
so the radial wavelength of the mode $(\ell_R)$ is typically much shorter than
the vertical wavelength $(\ell_z)$ \citep{UrpinBrand1998,Urpin2003}. Hence, being
inertial waves, the group velocity points almost vertically, in
accordance with the direction of transport outlined in the previous subsection.
Note that modes grow at or near the fastest rate on short lengthscales all
the way to the viscous cut-off $\ell_\nu$. 

To give some sense of the numbers, 
the growth rate for the VSI is typically
$\sigma\sim 0.3 \;\mathrm{yr}^{-1}$, 
if $\epsilon\sim 0.05$ at 1 AU and $|q|\sim 1$.
The microscopic kinematic viscosity
at the
 mid-plane of a protoplanetary disc at 1 AU may be estimated as 
$\nu \sim 2.5 \times 10^{5} \mathrm{cm}^2\mathrm{s}^{-1}$, yielding 
$\ell_{\nu}\sim 10^2$ km, much shorter 
than the local disc thickness $H$ (of the order of $10^7$ km).
The thermal diffusivity, on the other hand, 
is significantly larger, $\chi \sim 5 \times
10^{12} \mathrm{cm}^2\mathrm{s}^{-1}$ 
(taking values typical for $H_{2}$ in a Minimum Mass Solar Nebula). 
Note that $\chi$ varies significantly with height in the disc
(e.g.~\citealt{BellLin1994}),
 but we ignore this complication when making simple estimates.
 It is unclear what value $N_{z}$ should take in a dead zone in a
 protoplanetary disc,
 but it is probably much smaller than $\Omega$. Therefore a (very
 crude) lower
 bound on the thermal diffusion length
 is $\l_{\chi}\gtrsim 10^{5}$ km. In all likelihood, however,
 $\ell_\chi\sim H$, not least because thermal diffusion becomes
 stronger as we approach the photosphere. 

These estimates suggest the following ordering: $H
 \sim \ell_\chi \gg \ell_\nu$, with the VSI occurring in the wide gulf
separating the viscous and thermal diffusion lengths. We might
expect the fastest growing modes to localise in regions of greatest
vertical shear;
in any realistic disc model, the magnitude of the vertical shear
 increases with distance from the mid-plane, and takes its maximum value at the disc surface\footnote{This is not the case for the locally isothermal thin disc model discussed in \S \ref{isobasic}, which has no surface and for which the magnitude of the vertical shear takes it maximum value at vertical infinity, where the density is negligible.}. 
So we might expect the fastest growing modes to occur on very short
lengthscales (just above $\ell_{\nu}\sim 100$ km) located near the surface.

\begin{figure}
  \begin{center}
      \subfigure[$\partial_{z}(R\Omega)$]{\includegraphics[trim=6cm 1cm 6cm 2cm, clip=true,width=0.32\textwidth]{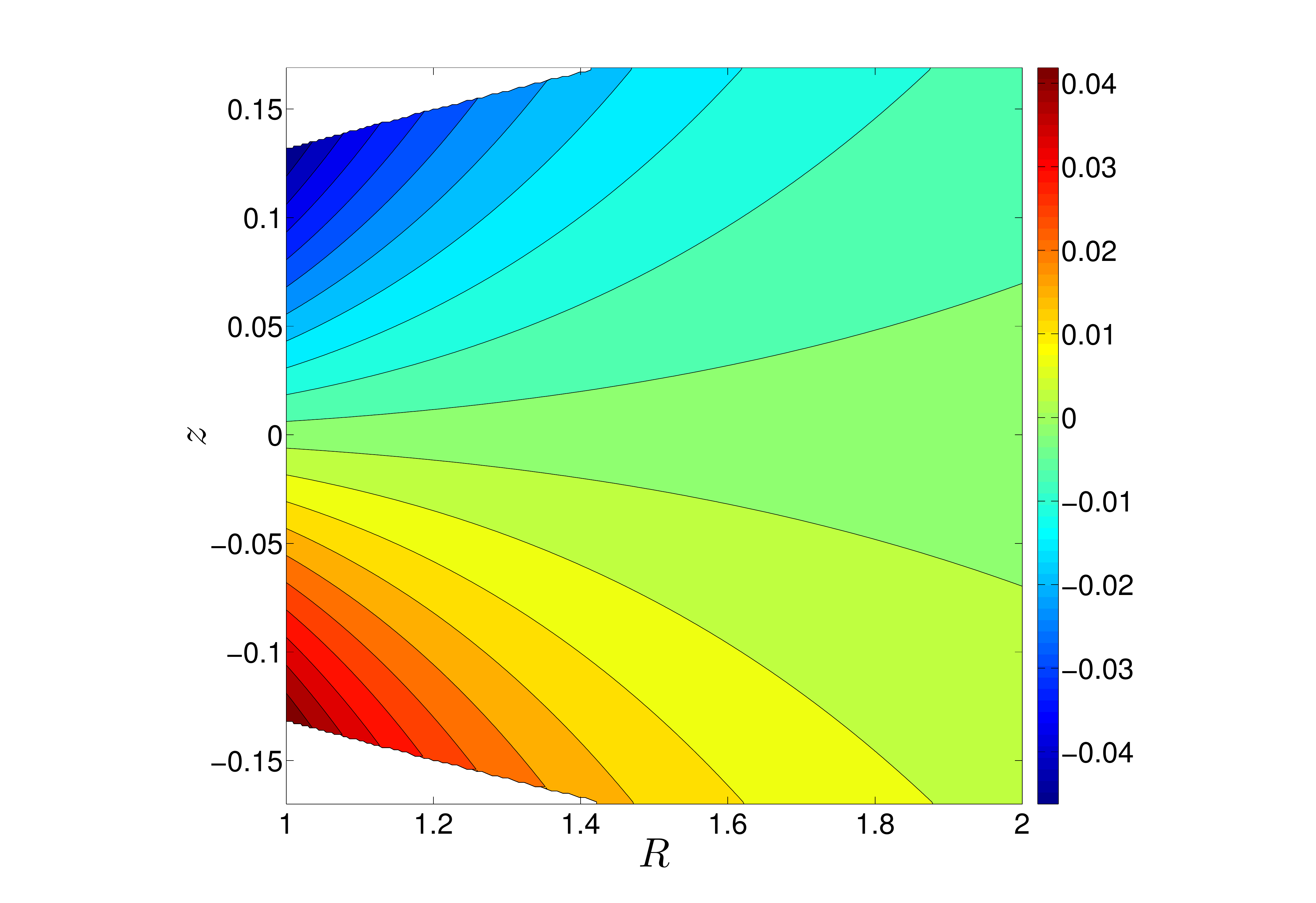} } \\
       \subfigure[$\rho$]{\includegraphics[trim=6cm 1cm 6cm 2cm, clip=true,width=0.32\textwidth]{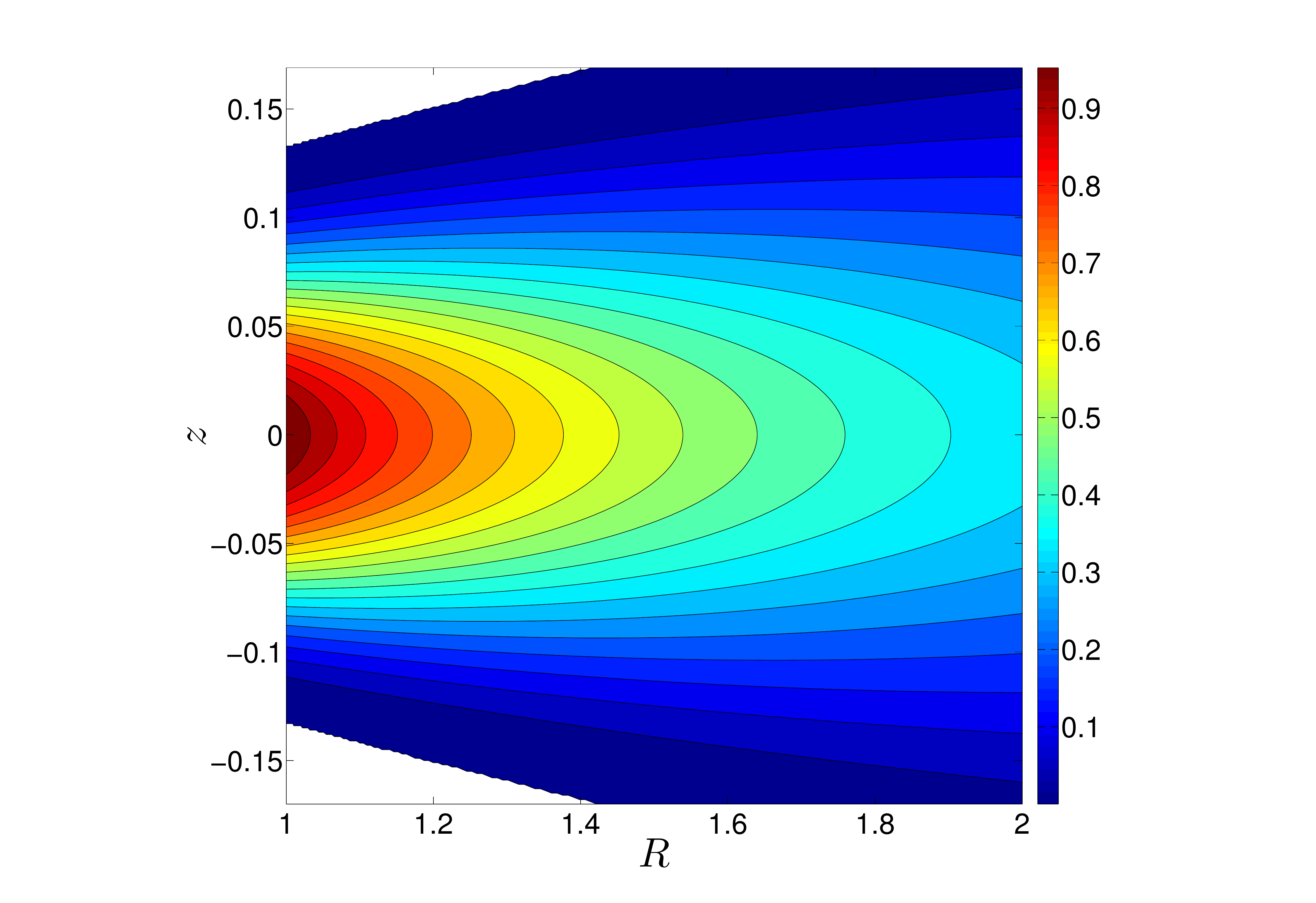} }
    \end{center}
  \caption{Basic state for the locally polytropic disc with $q_s=-1, p=-1.5, a_0=0.05$ and $\gamma=1.4$ (therefore $H(R=1)\approx0.13$). 
  The top panel shows a contour plot of the vertical shear
  $\partial_{z}(R\Omega)$ on the $(R,z)$-plane. For
  given $R$, the magnitude of the vertical shear has a maximum near
  the disc surface. 
The bottom panel shows the density $\rho$. Regions outside the disc surface where $\rho=0$ are coloured white in both panels. (The angular velocity is not shown since the departure of constant $\Omega$ surfaces from the vertical is very small).}
  \label{0b}
\end{figure}

\subsection{Nonlinear evolution}
\label{angmomtransport}

We expect that the VSI will work to eradicate the destabilising
conditions from which it arose (the vertical shear) and ultimately return
the system to a marginally stable, cylindrical rotation
profile. Of course, resisting the VSI will be the driver of the vertical
shear itself: the radiation field of the protostar. In the
struggle between these two opponents, the system will
probably reach a quasi-steady `balance' in which the vertical shear is
diminished (but not entirely removed) and some degree of
hydrodynamical activity remains. The properties of this state will be
determined by the relative efficiency of the VSI in wiping away the shear
versus the shear's thermal driving via the protostar.
Presumably, if the VSI is inefficient then
significant vertical shear persists and, as a consequence, significant
turbulent motions.\footnote{In this case, 
 the fastest growing VSI-driven modes
  probably saturate through secondary shear instabilities --  as is
  the case for fingering convection, another kind of double-diffusive
  instability \citep{Brown2013}. The resulting turbulence in the low
  density regions near the disc surface may best be studied using
  local simulations, since they can access more realistic small
  scales.} 
The locally isothermal simulations of \cite{Nelson2013}
best describe this scenario, because the destabilising gradients are fixed
and cannot be modified by the VSI. On the other hand, if the thermal
driving is weak,
 the VSI should eliminate 
the vertical shear, and subsequently 
its motions will settle down to a much lower
level, because the system is near (if not at)
marginal stability. 
This situation can be observed in simulations where the unstable
equilibria possess long relaxation times \citep{Nelson2013} or in cases where the disc is
not thermally driven \citep{StollKley2014}.

However it is generated, the properties 
of the final quasi-steady state are of key importance
to dead-zones, intervening potentially in the dynamics of solids and
even in angular momentum transport. It certainly should be the focus of future numerical
simulations. Note that this state will evolve slowly,
on the long timescale of the disc's and protostar's
evolutionary track, and also presumably on the shorter timescale
of the protostar's emission variability.

The VSI in protoplanetary discs is challenging to simulate adequately 
because the fastest growing modes have very short lengthscales, many
orders of magnitude shorter than both the disc thickness and
(typically)
the numerical grid
scale. 
Global
simulations are then not only ill-posed but exhibit a nonlinear 
saturation whose characteristic lengthscales are forced to be longer
than is realistic. It may be that
global simulations of the VSI greatly overestimate the amount of power in the largest
lengthscales.
Indeed \cite{StollKley2014} find no convergence in radial angular momentum transport 
($\alpha$) as resolution is
increased, a result that emphasises the prominence of 
the smallest scales.

We can crudely estimate an upper limit for the resulting
  radial 
angular momentum transport
 if we assume that $\ell_R \sim |q|\epsilon\ell_z$ and if 
 nonlinear saturation occurs when the velocity amplitude is of order
$\ell_R \sigma$. Consequently, the turbulent viscosity scales as
$\nu_T \sim \ell_R^2\sigma\sim |q|^3\epsilon^3\,\ell_z^2\Omega.$
To make further progress we must estimate the size of $\ell_z$. 
If the dominant scales at saturation are the largest ones,
then $\ell_z \lesssim H$, which gives an upper bound on the $\alpha$ parameter:
\begin{eqnarray}
\alpha \lesssim |q|^3\epsilon^3,
\end{eqnarray}
For a protoplanetary disc with $\epsilon\approx
0.05$, we get $\alpha\lesssim 10^{-4}$,
which is broadly consistent with the results of current global
simulations \citep{Nelson2013,StollKley2014}.\footnote{Note that many
  of the $\alpha$ calculations in \cite{StollKley2014} are
  two dimensional and questions may be asked of angular
  momentum transport in this case (see arguments in \citealt{Balbus2000}).} If, however, we assume
  instead that the dominant scale at saturation is much shorter
(so that $\ell_z \ll H$), then this is a gross overestimate.

\section{Hydrodynamic equations and basic state profiles}
\label{Basiceqns}

In the previous section, we explained why discs with global radial
variations in temperature or entropy\footnote{We do not set out to
  analyse the stability of local structures in the thermal properties
  of the disc, such as edges or pressure bumps etc.} are likely to be
unstable to the VSI. We now list the equations and describe the
simplified disc models that will be used to analyse the VSI. We begin
with the equations of compressible hydrodynamics for an inviscid
adiabatic fluid:
\begin{eqnarray}
\label{fulleqns1}
&& \left(\partial_{t}+\boldsymbol{u}\cdot\nabla\right)\boldsymbol{u}=-\frac{1}{\rho}\nabla P -\nabla \Phi, \\
&& \left(\partial_{t}+\boldsymbol{u}\cdot\nabla\right) \rho = -\rho \nabla\cdot\boldsymbol{u}, \\
&& \left(\partial_{t}+\boldsymbol{u}\cdot\nabla\right) S =0,
\label{fulleqns2}
\end{eqnarray}
where $\boldsymbol{u}$ is the velocity.
The ideal equation of state is $P=\mathcal{R} \rho T$ (where $\mathcal{R}$
is the gas constant divided by the mean molecular weight).

 If we adopt cylindrical polar coordinates $(R,\phi,z)$, the gravitational potential due to the central object is approximately that of a point-mass
\begin{eqnarray}
\Phi(R,z) &&= -\frac{GM}{\sqrt{R^2+z^2}},\\
 &&\approx -\frac{GM}{R}\left[1 - \frac{z^2}{2 R^2}\right] = \Phi_{0}+\Phi_{2} z^2,
\end{eqnarray}
where in some cases we expand for a thin-disc 
($|z|\ll R$; second line), and we define $\Phi_{0}=-GM/R$ and $\Phi_{2}=GM/(2R^3)$. 

\subsection{Basic state}

The axisymmetric basic state of the differentially rotating disc has $\boldsymbol{u}=R\Omega(R,z) \boldsymbol{e}_{\phi}$, and satisfies the equations of radial and vertical force balance:
\begin{eqnarray}
\label{forcebalance}
-R\Omega^2\boldsymbol{e}_{R} &=& -\frac{1}{\rho}\nabla P+\nabla \Phi.
\end{eqnarray}
Taking the curl then provides the thermal wind equation (Eq.~\ref{TWE}).
For simplicity, and to allow some analytical reduction and exploration, we restrict ourselves to studying two simple models (as in \citealt{Nelson2013}): the locally isothermal disc with a radial power law in temperature, and the locally polytropic disc with a power law entropy function. In both cases, we want to consider a disc with mid-plane density
\begin{eqnarray}
\rho_{m}(R) &=& \rho_0 \left(\frac{R}{R_{0}}\right)^{p},
\end{eqnarray}
where $\rho_{0}$ is the mid-plane density at a radius $R_{0}$. It turns out that the value of $p$ is unimportant for the VSI.

\subsubsection{Locally isothermal disc with a radial power law in temperature}
\label{isobasic}

The first model that we will discuss is a locally isothermal disc with $P=c_{s}^2(R)\rho$ (where $c_{s}$ is the isothermal sound speed), in which the temperature depends only on cylindrical radius as the power law
\begin{eqnarray}
T(R)=T_{0}\left(\frac{R}{R_{0}}\right)^{q},
\end{eqnarray}
and similarly $c^2_{s}(R)=c^2_{0}\left(R/R_{0}\right)^{q}$, where $c^2_{0}=\mathcal{R}T_{0}$ is the square of the isothermal sound speed at a radius $R_{0}$. The corresponding density profile that satisfies Eq.~\ref{forcebalance} is
\begin{eqnarray}
\rho(R,z)=\rho_m (R) \exp \left(\frac{1}{c^2_{s}(R)}\left[\Phi_{0}(R)-\Phi(R,z)\right]\right) .
\end{eqnarray}
Note that the disc has no surface and formally extends to infinity.
The angular velocity that satisfies Eq.~\ref{forcebalance} is
\begin{eqnarray}
\Omega(R,z)=\Omega_{0}(R)\left(1+q+(p+q)\frac{H^2}{R^2}-\frac{q R}{\sqrt{R^2+z^2}}\right)^{\frac{1}{2}},
\end{eqnarray}
where $\Omega_{0}(R)=\sqrt{GM/R^3}$ and $H(R)=c_{s}(R)/\Omega_{0}(R)\propto (R/R_{0})^{(q+3)/2}$ is the local disc scaleheight. The angular velocity therefore depends on $z$ whenever $q\ne 0$. We set out to consider $p$ and $q$ so that these discs are stable according to the Solberg-H\o iland criteria governing adiabatic axisymmetric perturbations (e.g.~\citealt{Tassoul1978}).

For illustration, in Fig.~\ref{0a} we plot the angular velocity
$\Omega$, vertical shear $\partial_{z}(R\Omega)$ and density $\rho$ on
the $(R,z)$ plane for a disc with $c_{0}=0.05, p=-1.5$ and $q=-1$. This shows that the angular velocity depends on $z$, and that the vertical shear increases monotonically with $z$ until its reaches a maximum far away from the mid-plane, where the density is negligible.

\subsubsection{Locally polytropic disc with a radial power law in entropy}
\label{polybasic}

The second model is a locally polytropic model in
which $P=K_s(R) \rho^{\gamma}$ (where $\gamma$ is the adiabatic index)
with an entropy function 
\begin{eqnarray}
K_{s}(R)=P\rho^{-\gamma}=K_{0}\left(\frac{R}{R_{0}}\right)^{q_s},
\end{eqnarray}
with $q_s$ and $K_0$ constants. Hence $K_s$ is only a function of cylindrical radius, so that $S=S(R)\propto \ln K_{s}(R)$. This disc model is neutrally stratified in the vertical direction. The corresponding density profile which satisfies Eq.~\ref{forcebalance} is
\begin{eqnarray}
\rho(R,z)&=&\rho_{m}(R)\left(1+\frac{(1+m)}{K_{s}(R)}\left[\Phi_{0}(R)-\Phi(R,z)\right]\right)^{m}
\\ 
&\approx & \rho_{m}(R)\left(1-\frac{z^2}{H_{0}^2(R)}\right)^{m},
\end{eqnarray}
where $m=1/(\gamma-1)$ is the polytropic index. For the last line, the potential has been expanded for a thin disc, and we have defined the local disc thickness 
\begin{eqnarray}
H_{0}(R) = \sqrt{\frac{2 (1+m) K_{s}(R)}{\Omega_{0}^2}},
\end{eqnarray}
where $\Omega_{0}$ was defined in \S~\ref{isobasic}. This disc model possesses a surface at which $\rho=0$ when $z=H_{0}(R)$. The adiabatic sound speed is $a(R,z) =\sqrt{\gamma P(R,z)/\rho(R,z)}$, which becomes $a_m(R)$ at the mid-plane (taking the value $a_{0}$ at $R=R_{0}$), and we define $M(R,z)=R\, \Omega_{0}(R)/a(R,z)$ and $M_m(R)=R\, \Omega_{0}(R)/a_m(R)$ to be the Mach number and mid-plane Mach number of the flow.
The angular velocity that satisfies Eq.~\ref{forcebalance} is
\begin{eqnarray}
\Omega(R,z)=\Omega_{0}(R)\left(1+q_s+\frac{p}{M_{m}^2}+\frac{q_s}{\gamma M^{2}}-\frac{q_s R}{\sqrt{R^2+z^2}}\right)^{\frac{1}{2}}.
\end{eqnarray}
As earlier, 
the angular velocity depends on $z$ whenever $q_s\ne 0$. Since this model is neutrally stratified in the vertical direction, it can become unstable to adiabatic disturbances for certain choices of $p$ and $q_s$, whenever one of the Solberg-H\o iland criteria are violated. However, we do not set out to study such instabilities in this work, and we instead focus on discs that would be stable to adiabatic perturbations. 

For illustration, in Fig.~\ref{0b} we plot the vertical shear
$\partial_{z}(R\Omega)$ and density $\rho$ on the $(R,z)$ plane for a disc with $a_{0}=0.05, \gamma =1.4, p=-1.5$ and $q_s=-1$. This shows that the angular velocity depends on $z$, and that the magnitude of the vertical shear increases monotonically with $|z|$ until its reaches a maximum just below the disc surface.

In the next three sections we work through the stability of these two
equilibria. Because the full global analysis is challenging, we first
treat their quasi-local approximations in a reduced model first outlined in \cite{Nelson2013}. This helps tease out the most important features.
Once this is done we perform the full global analysis on the locally isothermal
model only.

\section{Linear stability of the locally isothermal disc: reduced model}
\label{VSIiso}
\label{VSIisoreduced}
In this section we revisit the reduced model of \cite{Nelson2013}, which describes the dynamics of the VSI in locally isothermal discs with a radial power law in temperature. We begin by outlining the derivation of this model (a similar derivation is presented in detail in Appendix \ref{derivation} for the locally polytropic disc) and go on to analyse its most important properties.

We define $\epsilon=H/R$ and consider a thin disc in which $\epsilon
\ll 1$.
We are interested in slow modes with frequencies and growth rates that
are each $O(\epsilon\Omega)$, 
with vertical scales $\ell_z$ that are comparable with the thickness of the
disc ($\ell_z \sim \epsilon R$) and radial scales $\ell_R$ 
that are much smaller
($\ell_R \sim \epsilon^2 R$), cf. Eq.~\ref{lengths}. 
If we also consider vertical velocities that are
mildly subsonic or transonic (by a factor $\epsilon$) and radial
velocities that are very subsonic (by a factor $\epsilon^2$),
 then we require $O(\epsilon)$ density perturbations,
for consistency. 
On such small radial scales, the curvature of the disc can be
neglected and the geometry is 
locally Cartesian, similar to the classical shearing box
\citep{GoldreichLyndenBell1965,UmurhanRegev2004}.
 In this limit, we are looking at low frequency (inertial) dynamics
 that 
are anelastic and in approximate radial geostrophic balance. 

The linearised reduced equations for the rescaled velocity perturbation
$(u,\,v,\,w)$ and the fractional density perturbation $\Pi=\rho^{\prime}/\overline{\rho}$
are \citep{Nelson2013}
\begin{eqnarray}
\label{reducedisothermal1} 
0&=&2v -\partial_{x}\Pi, \\
\label{reducedisothermal2} 
\partial_{\tau}v&=&-\frac{u}{2}-\frac{qzw}{2}, \\
\partial_{\tau}w&=&-\partial_{z}\Pi, \\
0&=&\partial_{x}(\bar{\rho} u)+\partial_{z}(\bar{\rho} w).
\label{reducedisothermal4}
\end{eqnarray}
Here $\tau$ is rescaled time, and $x$ is a local
radial variable. The background density is
$\overline{\rho}=\mathrm{e}^{-\frac{z^2}{2}}$,
 after appropriate normalisation. The crucial term for the appearance of the VSI is the last one in Eq.~\ref{reducedisothermal2}, which arises from thermal wind balance if there is radial variation in temperature (see Eq.~\ref{TWE}).

We seek solutions of this system of the form
\begin{eqnarray}
\Pi = \mathrm{Re}\left[\hat{\Pi}(z) \mathrm{e}^{\mathrm{i}\left(kx-\omega \tau\right)}\right],
\end{eqnarray}
and so on for other variables, where we subsequently drop the hats for
clarity. This 
allows us to reduce the system to a single ODE that can be written
most simply 
by defining new (complex) coordinates $\zeta=z \sqrt{1+\mathrm{i} kq}$, as
\begin{eqnarray}
\frac{\mathrm{d}^2 \Pi}{\mathrm{d}\zeta^2} -\zeta \frac{\mathrm{d}\Pi}{\mathrm{d}\zeta} +\lambda \Pi=0,
\end{eqnarray}
where we have set
\begin{eqnarray} \label{dispreln}
\lambda=\frac{\omega^2 k^2}{1+\mathrm{i} k q}.
\end{eqnarray}
This is the well-known Hermite differential equation (the probabilist's version), which has solutions
\begin{eqnarray}
\label{fullsoln}
\Pi=a_1\,\, \mathrm{He}_{\lambda}(\zeta) + a_2\, \,_{1}F_{1}\left(-\frac{\lambda}{2},\frac{1}{2},\frac{\zeta^2}{2}\right).
\end{eqnarray}
The first function is the Hermite function and the second
function is a 
confluent hypergeometric function of the first kind, with $a_1,\,a_2$
arbitrary constants.
 Solutions that are polynomially bounded as $|\zeta|\rightarrow\infty$ 
(so that $|\rho^{\prime}|\rightarrow 0$ as $|\zeta|\rightarrow\infty$) 
require $\lambda=n\in \mathbb{N}$ and $a_2=0$ (see \citealt{Okazaki1987,Kato2001}). The regular solutions are therefore
\begin{eqnarray}
\label{IWsoln}
\Pi \propto \mathrm{He}_{n}(\zeta),
\end{eqnarray}
with $\mathrm{He}_n$ a Hermite polynomial of order $n$. The
corresponding $\omega$ can be obtained from the dispersion
relation Eq.~\eqref{dispreln}.
These solutions describe the vertical structure of modes in the low
frequency limit.
 Since they are (complex) polynomials in $\zeta$, 
they describe global modes which are not localised in the vicinity of any particular location $\zeta\ne0$. 

\subsection{Non-vertically shearing case, $q=0$}
\label{VSInoshear}

Before treating the VSI we examine the case when $q=0$ in order to make contact
with the existing literature on wave modes in vertically stratified
isothermal discs (namely \citealt{LP1993}). This then can assure us
of the validity of the reduced model and the regime of its
applicability.

Since $\lambda$ is quantised, we obtain a discrete set of frequencies. When $q=0$, these are real and 
\begin{eqnarray}
\omega=\pm \frac{\sqrt{n}}{k},
\label{reducedfreq}
\end{eqnarray}
which represents a pair of low frequency inertial waves\footnote{Note that these are sometimes referred to as $r$-modes in the literature (and sometimes inappropriately as g-modes even when buoyancy forces are absent).} travelling in opposite directions.
The full isothermal disc allows axisymmetric waves with frequencies that satisfy the following dispersion relation \citep{LP1993}
\begin{eqnarray}
(-\omega^{2}+n c_{s}^2)(-\omega^{2}+1)-c_{s}^2 k^2\omega^{2}=0.
\end{eqnarray}
There are two branches of solutions, which represent either high frequency acoustic waves or low frequency inertial waves. Note that there are no surface modes in the isothermal disc, since the model lacks a surface.
The inertial waves have $\omega^2 \approx n/(k^2+n)$, therefore
Eq.~\ref{reducedfreq} is a good approximation for modes with $k\gg
\sqrt{n}$, as expected from the
assumption $\ell_R/\ell_z \sim \epsilon$.
When this is violated, physically incorrect
solutions appear with frequencies larger than one. 
The restriction on the size of the vertical quantum number $n$
means that the model can never capture purely local modes (for which
both $n$ and $k$ are large).

\subsection{Vertically shearing case, $q\ne 0$}
\label{trueisothermalVSI}

When $q\ne 0$ the disc is unstable to the VSI. The frequencies are now
\begin{eqnarray}
\omega=\pm \frac{\sqrt{n}}{k}\left(1+\mathrm{i}k q\right)^{\frac{1}{2}},
\label{growthprediction}
\end{eqnarray}
which are complex. The growth rate $\sigma=\mathrm{Im}[\omega]$ may be computed from the positive root of
\begin{eqnarray}
\sigma^2 = \frac{n}{2k^2}\left(\sqrt{1+k^2 q^2} -1\right).
\end{eqnarray}
 The unstable modes in this case are the classic inertial waves of the
 vertically stratified disc, destabilised by the vertical
 shear. They are all global ``body modes" that have no localisation
 near to any $\zeta\ne 0$.

In the limit $|kq|\ll 1$,
\begin{eqnarray}
\omega = \pm \frac{\sqrt{n}}{k}\left(1+\mathrm{i}\frac{k q}{2} + O(k^2 q^2)\right),
\end{eqnarray}
which possesses the same frequency as the classical modes, cementing
the identification. They grow, however,
at the rate $\sigma \approx \sqrt{n}|q|/2$. In the opposite limit $|kq| \gg 1$, 
\begin{eqnarray}
\omega = \pm \sqrt{\frac{n q}{2k}} \left(1+\mathrm{i}\right) + O\left(\sqrt{\frac{n}{k^3 q}}\right),
\end{eqnarray}
therefore the growth rate is $\sigma=\sqrt{n|q|/(2k)}$, which is no longer linear in the shear.

Note that the growth rate increases with polynomial order as
$\sqrt{n}$, 
which may seem somewhat surprising but can be understood by
considering the vertical shear profile. An 
isothermal thin-disc lacks a surface, yet
$|\partial_{z}(R\Omega)|\propto |q z/2|$,
 which increases without bound as $|z|\rightarrow \infty$.
Since modes with large $n$ extend over a greater vertical range, they
can better 
tap into the energy associated with the larger shear at
larger $z$. Hence it follows that the growth rate should
increase 
with $n$ (the square root comes from the fact that $n$ has units of
inverse length squared).
It should be remembered, however, that these 
expressions apply only when the reduced model is valid, which requires
$\sqrt{n} \ll k$.

Finally, note that the unstable modes here
have a wave character, whereas those issuing from the local Boussinesq
calculation grow monotonically
\citep{UrpinBrand1998,Urpin2003}. Their scalings with wavenumber also differ.
 In the isothermal limit with $k/k_z \gg 1$ and $k \gg 1$,
local Boussinesq modes grow at a rate $\sigma \sim \sqrt{k_z q/k}$, where 
$k_z$ is the vertical wavenumber. 
The system, being scale free, only depends on the ratio of
  wavenumbers, $k_z/k$. In contrast, the quasi-global model separates out
  the vertical and radial scales, forcing the former to be near $H$.
 Consequently, it will always struggle to reproduce the
local limit. Only if we force modes to localise at a fixed $z$ (such as at a
boundary) and then undertake 
a plane-wave analysis, can the reduced model produce
the local Boussinesq results \citep{Nelson2013}.

\subsection{Vertically shearing case with imposed boundaries: appearance of surface modes}
\label{VSIisoboundaries}

To connect with recent numerical simulations of a locally isothermal
disc \citep{Nelson2013,StollKley2014}, we
adopt an artificial boundary at a finite height in the vertical
direction. In realistic discs there is surely some form of transition
at the photosphere, but whether a numerical boundary adequately 
 mimics this
feature is unclear. We emphasise that within the confines
of a strict isothermal model, a numerical boundary is an artificial
addition, yet it has important effects.

We consider boundary conditions such that the vertical momentum
$\rho_{0}w=0$ 
at $|z|=H$, where $H$ is free parameter
(other choices would lead to similar behaviour for the VSI). 
To illustrate the unstable modes in this case, we solve
Eqs.~\ref{reducedisothermal1}--\ref{reducedisothermal4} numerically
using a Chebyshev collocation method on $N+1$ points of a
Gauss-Lobatto grid. This results in a matrix eigenvalue problem that
can be solved using a QZ method \citep{Boyd2001,GolubvanLoan1996}. 
Numerical convergence for the eigenvalues is verified by varying the
resolution and comparing eigenvalues.

\begin{figure*}
  \begin{center}
     \subfigure[$\omega=-0.2547+2.3109\mathrm{i}$]{\includegraphics[trim=7cm 1cm 8cm 2cm, clip=true,width=0.24\textwidth]{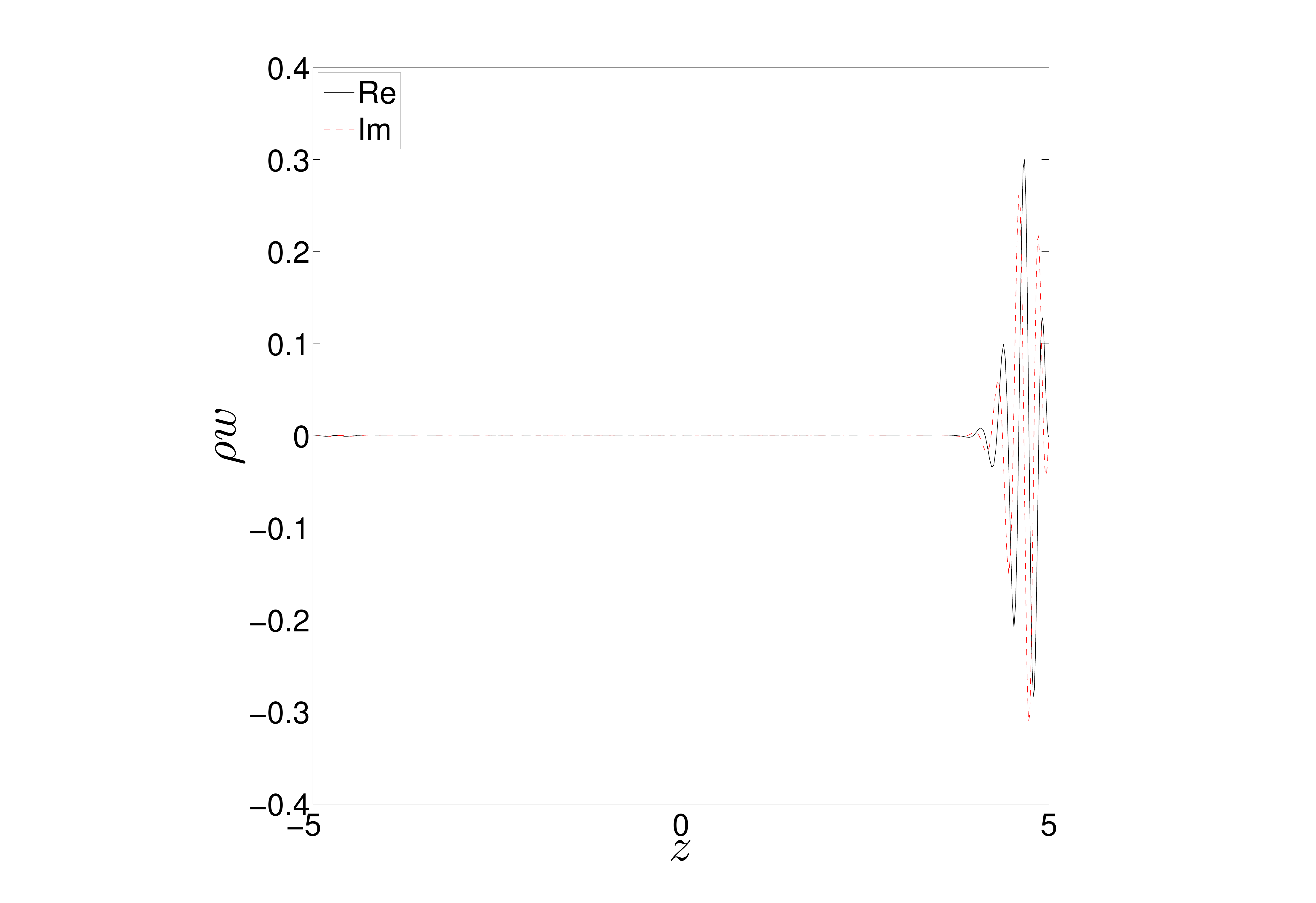} }
       \subfigure[$\omega=-0.2485+2.041\mathrm{i}$]{\includegraphics[trim=7cm 1cm 8cm 2cm, clip=true,width=0.24\textwidth]{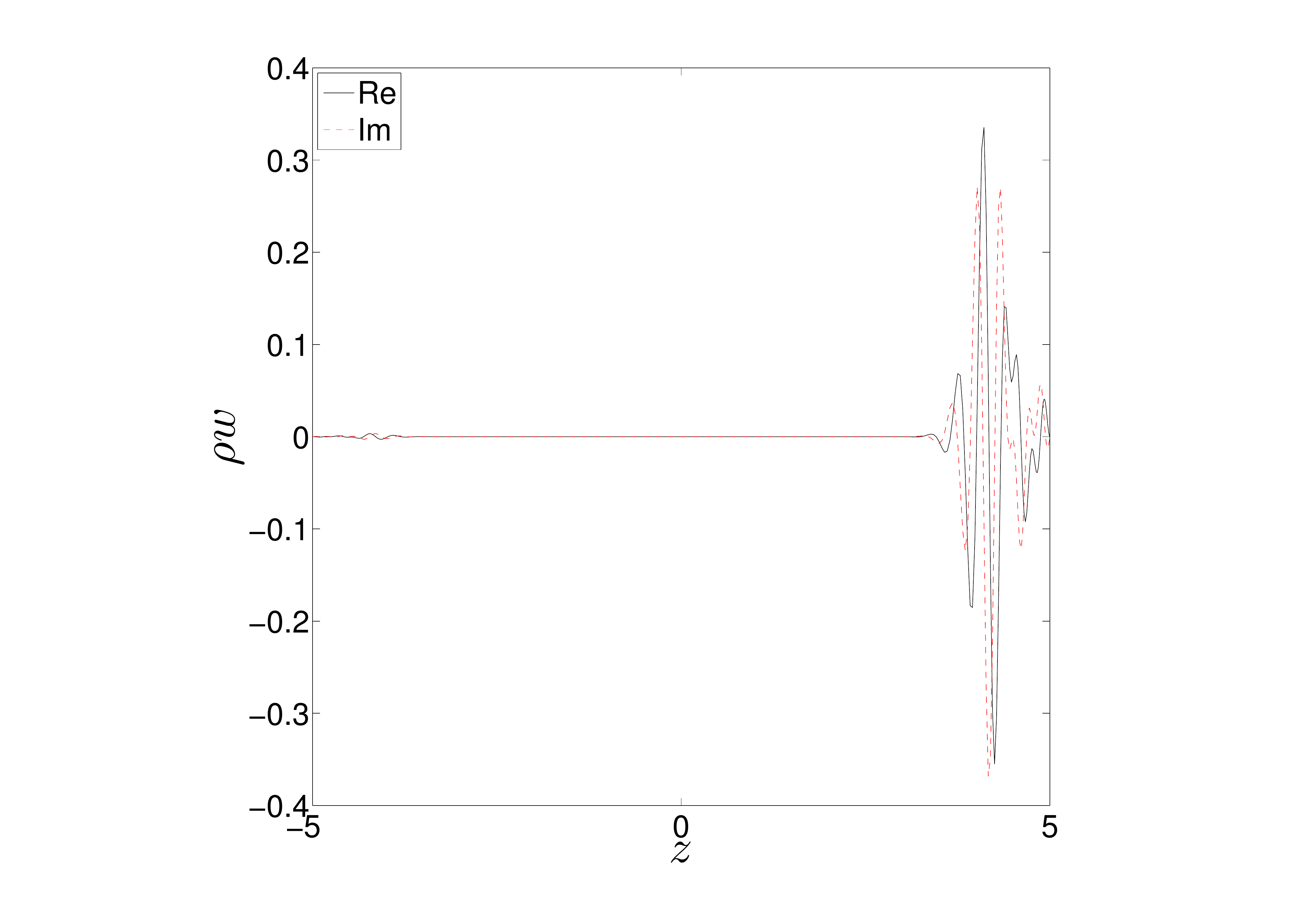} }
        \subfigure[$\omega=-0.2494+1.1505\mathrm{i}$]{\includegraphics[trim=7cm 1cm 8cm 2cm, clip=true,width=0.24\textwidth]{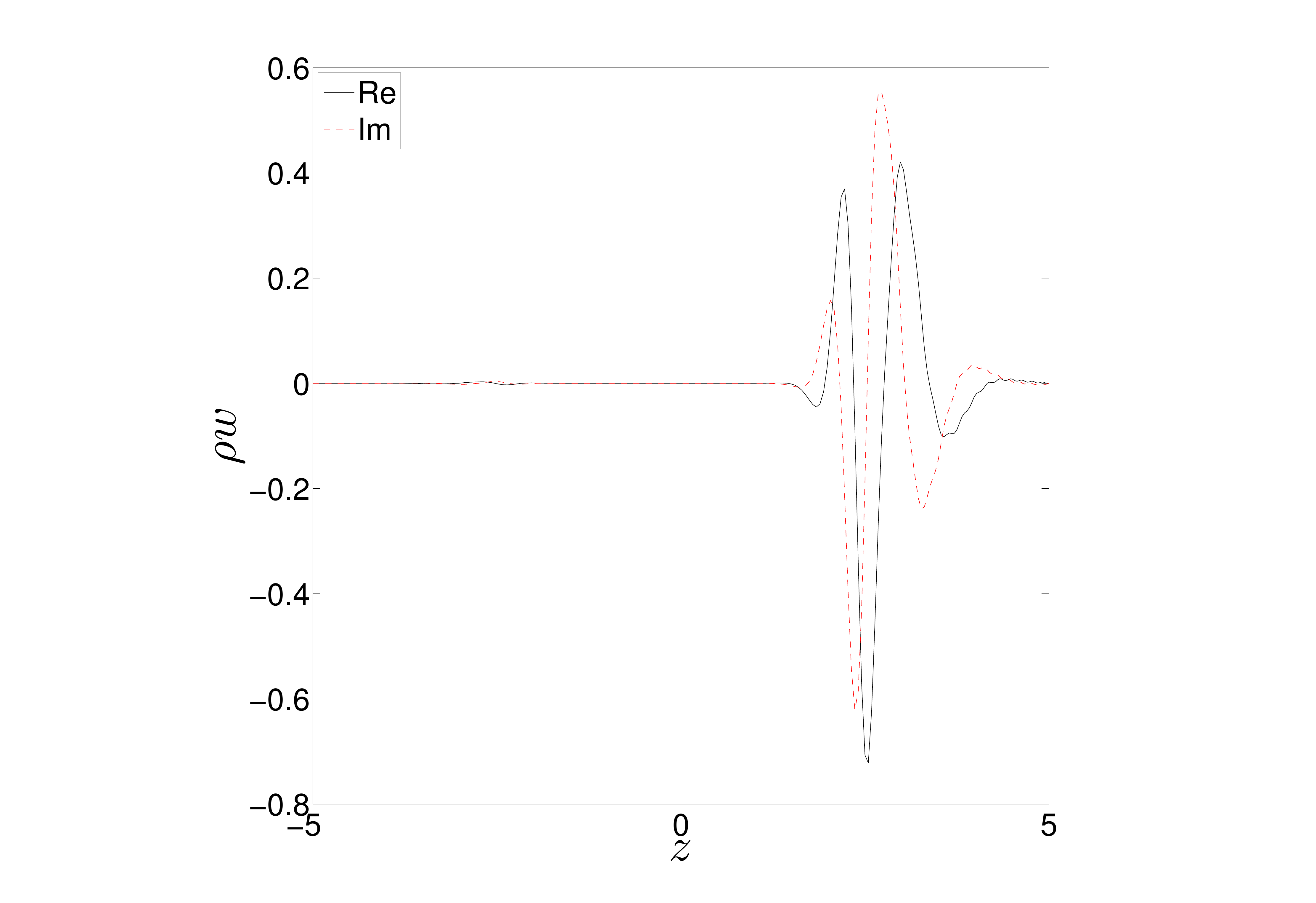} }
        \subfigure[$\omega=-0.2676+0.8557\mathrm{i}$]{\includegraphics[trim=7cm 1cm 8cm 2cm, clip=true,width=0.24\textwidth]{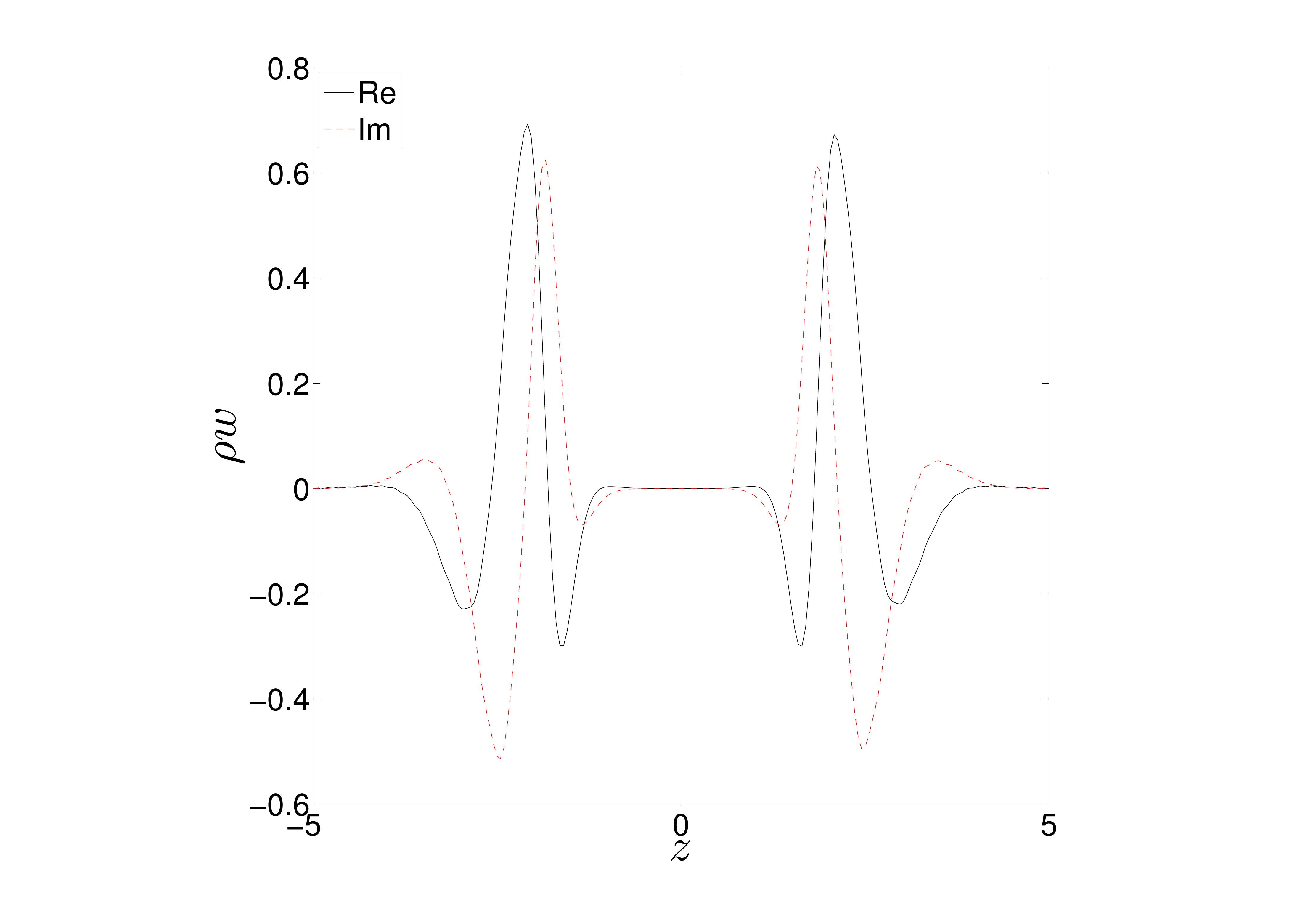} }
        \subfigure[$\omega=-0.4419+0.3212\mathrm{i}$]{\includegraphics[trim=7cm 1cm 8cm 2cm, clip=true,width=0.24\textwidth]{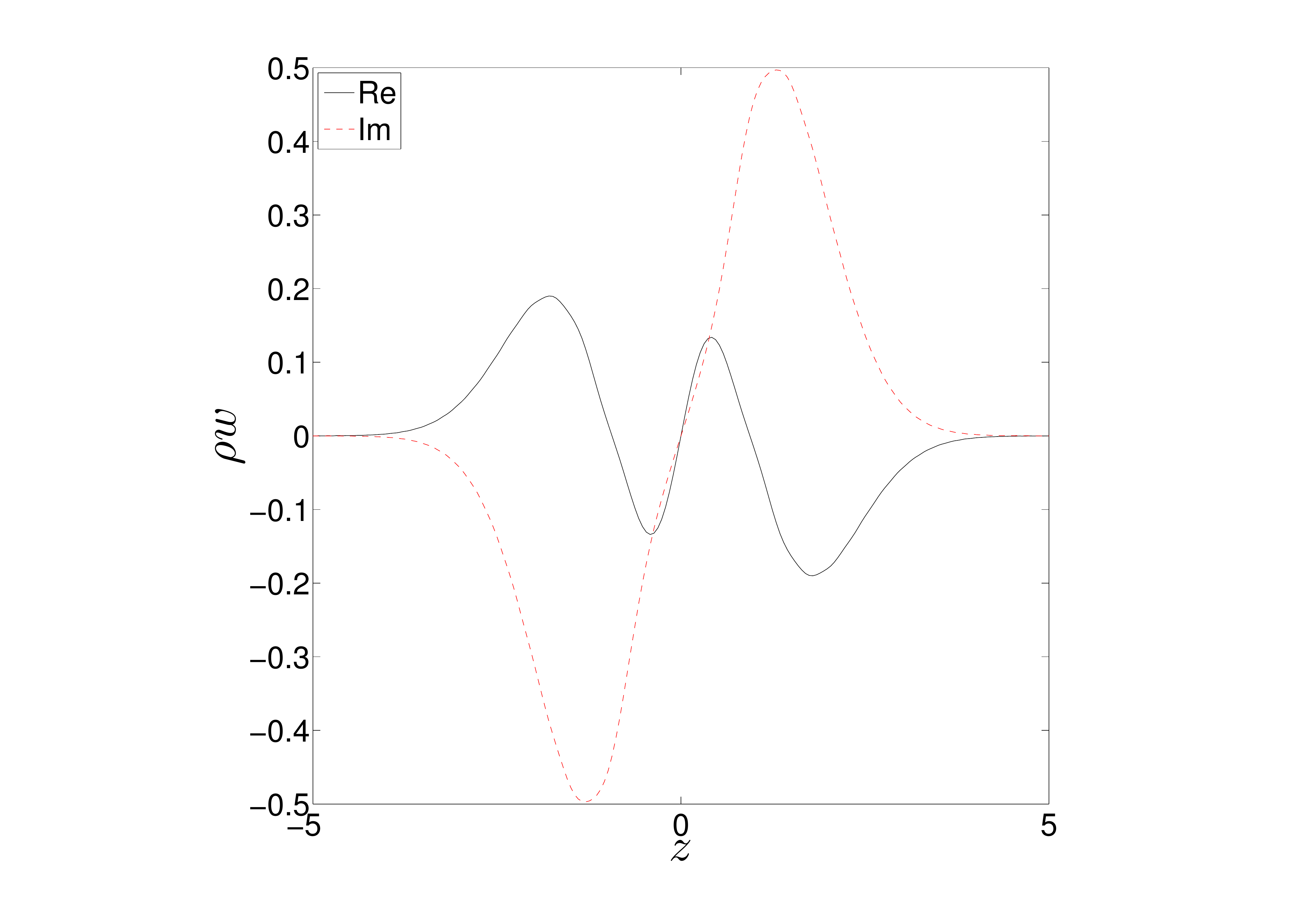} }
        \subfigure[$\omega=-0.4961+0.2950\mathrm{i}$]{\includegraphics[trim=7cm 1cm 8cm 2cm, clip=true,width=0.24\textwidth]{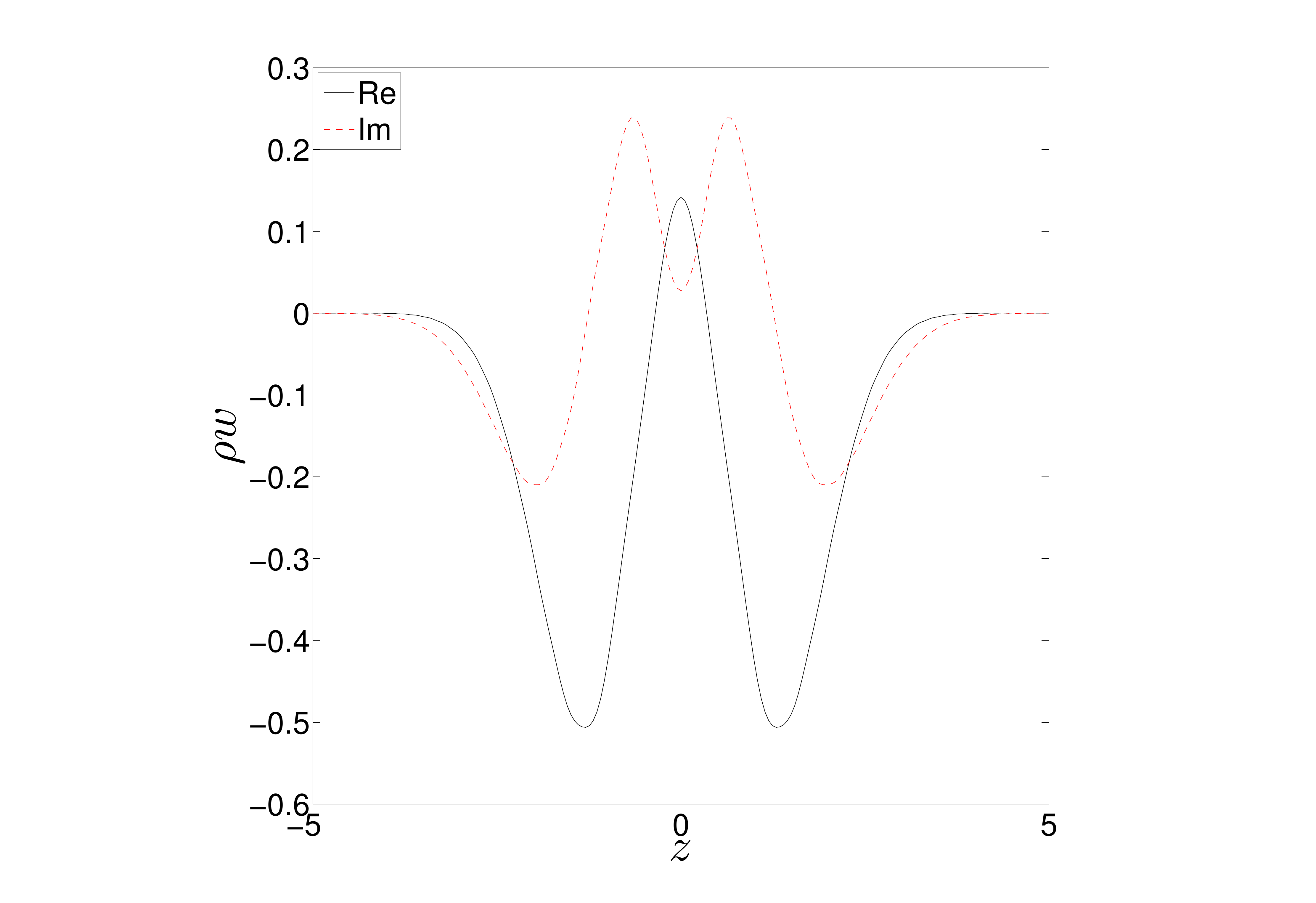} }
        \subfigure[$\omega=-0.6052+0.2579\mathrm{i}$]{\includegraphics[trim=7cm 1cm 8cm 2cm, clip=true,width=0.24\textwidth]{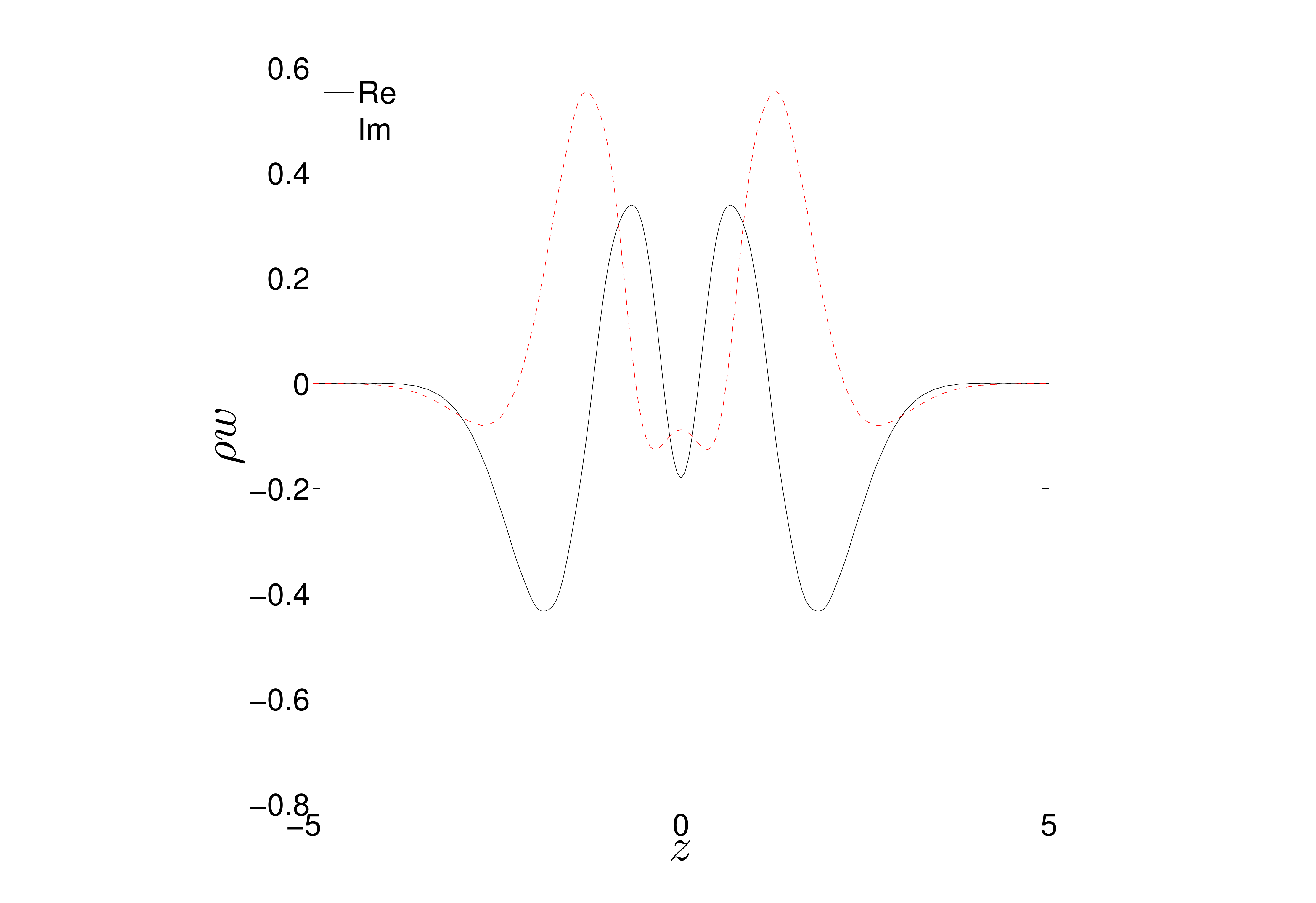} }
     \subfigure[$\omega=-0.2350+0.2127\mathrm{i}$]{\includegraphics[trim=7cm 1cm 8cm 2cm, clip=true,width=0.24\textwidth]{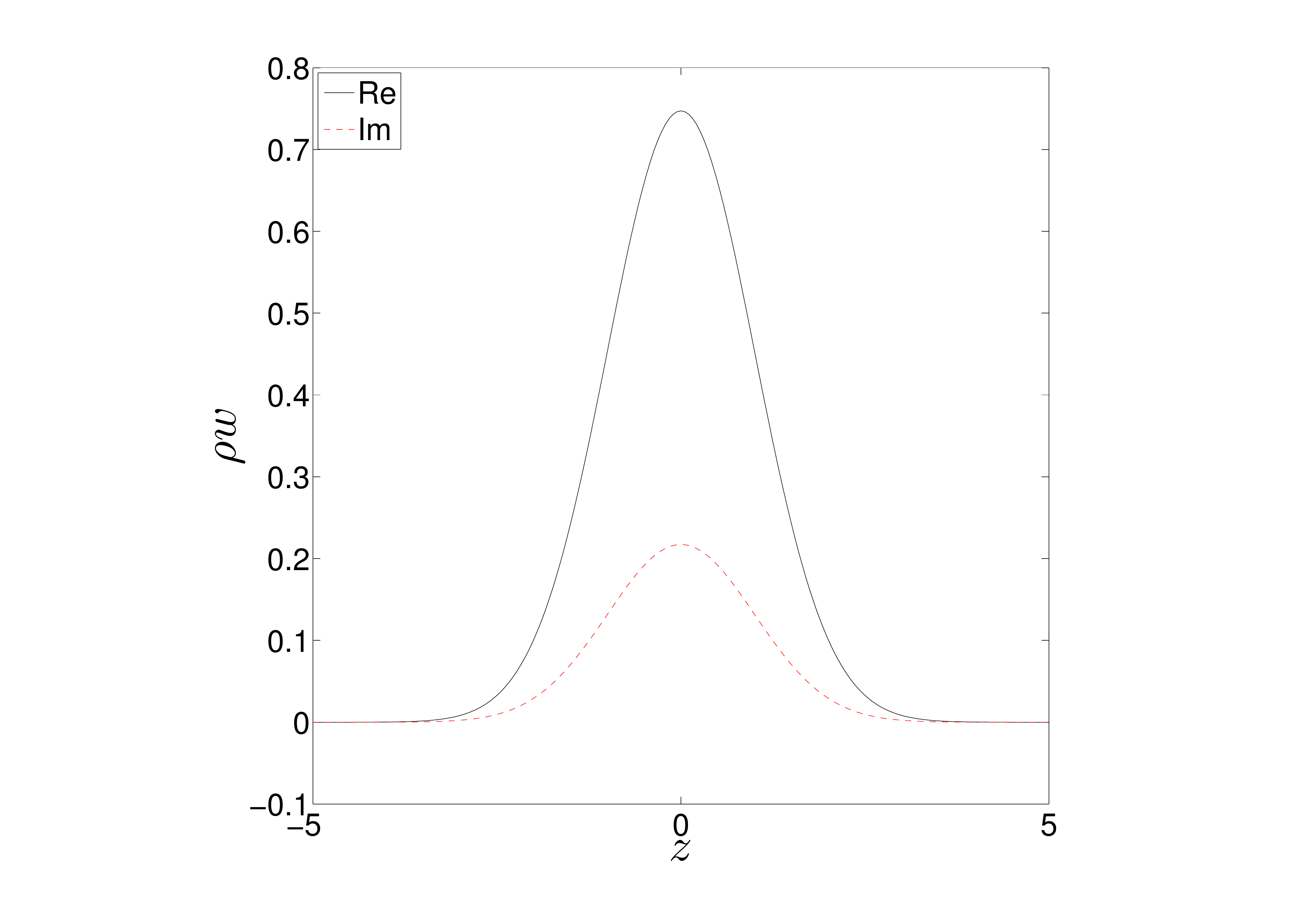} }
    \end{center}
  \caption{Illustration of the real and imaginary parts of the
    vertical momenta for a representative selection of the two types
    of modes in an isothermal disc with imposed numerical boundaries. Here $k=10$, $q=-1$, $H=5$ and $N=300$. The figure labels indicate the complex frequencies of the modes. Surface modes are shown in the top row (the fastest growing mode is in the top left panel), and body modes are shown in the bottom row. Similar surface modes localised at the bottom boundary are also obtained. The lowest frequency body mode is plotted in the bottom right panel and is an $n=1$ ($n=0$ for $w$) inertial wave, i.e. the fundamental ``corrugation mode", which is well described by an $n=1$ ($n=0$ for $w$) Hermite polynomial in the absence of vertical boundaries, and its complex frequency is that predicted by Eq.~\ref{growthprediction}.}
  \label{2}
\end{figure*}

In the presence of an artificial boundary, a new class of ``surface
modes" appears
 that localise near to the boundaries.
These are in addition to the ``body modes" obtained previously.
 We show the vertical momenta for several examples
of each type
 of mode in Fig.~\ref{2} for $k=10$ and $q=-1$: the top panel shows a selection of
 typical surface modes, and the bottom
 panel shows typical body modes. The mode plotted in the top right panel is one of the slowest growing surface modes, and its character is intermediate between the two classes of mode. The
appearance of these two classes of modes was also found by \cite{Nelson2013} and \cite{McNallyPessah2014}.

Mathematically, the emergence of surface modes can be understood by examining
Eq.~\ref{fullsoln}. In the presence of a boundary at a finite height,
$\lambda$ is complex and $a_{1}$ and $a_{2}$ are both nonzero, in
general. New modes, localised near to $|z|=H$, appear due to the need
to match boundary conditions at this location, rather than at infinity 
(this can be verified by plotting $\mathrm{He}_{\lambda}(\zeta)$ for $\lambda\in \mathbb{C}$).
Thus the new surface modes rely entirely on the boundary for their existence.

\begin{figure}
  \begin{center}
     \subfigure{\includegraphics[trim=7.0cm 0cm 9cm 1cm, clip=true,width=0.42\textwidth]{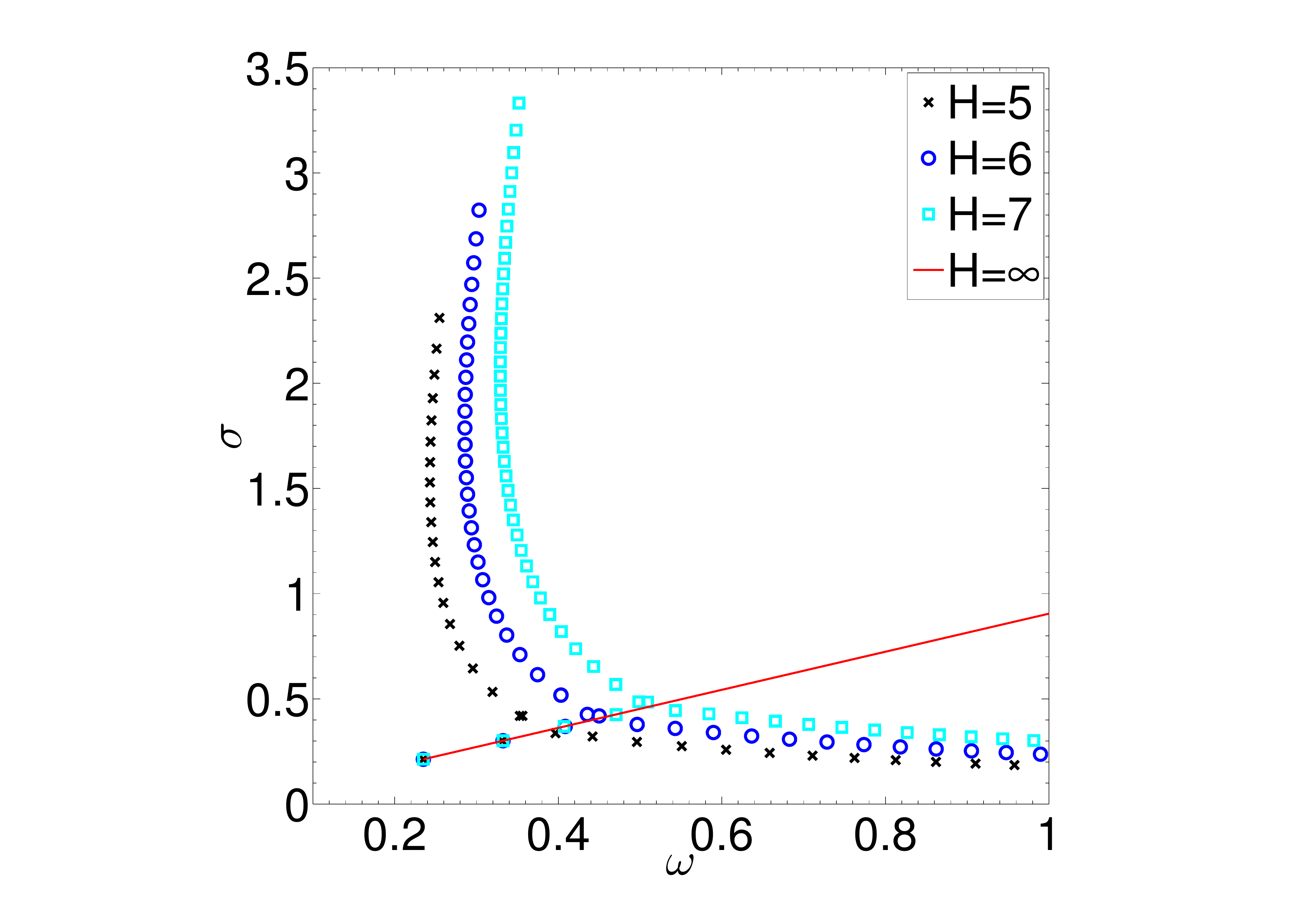} }
    \end{center}
  \caption{Growth rate of the unstable modes versus their (real)
    frequencies in an isothermal disc with artificial boundaries at
    $|z|=H$, illustrating the lack of convergence as $H$ is varied for
    cases with $k=10$ and $q=-1$ (converged results were obtained with
    $N=300$). Note that the maximum growth rate in this case is
    approximately $|qH|/2$. We have also indicated the analytical
    prediction of Eq.~\ref{growthprediction} with the solid red line
    to guide the eye (however, it should be remembered that this
    represents a set of discrete modes and not a continuum). The
    branch that extends approximately vertically represents surface
    modes.}
  \label{3}
\end{figure}

In Fig.~\ref{3} we show the dependence of the spectrum on disc height
$H$. Here the numerical growth rates of the unstable modes are plotted 
in the complex frequency plane
for three different heights, $H=5,6,7$. The remaining parameters are $k=10$
and $q=-1$. We also plot the $H=\infty$ case
Eq.~\ref{growthprediction},
which is given by the solid red line in the figure. Roughly, surface
modes correspond to the more ``vertical'' segment of the spectrum, and the
inertial waves to the more ``horizontal'' segment.
 
Obviously, there is no convergence with increasing $H$, especially for
the surface modes.
As the height of the domain is increased, the maximum growth
 rate increases in direct proportion with $H$. This is what we would 
expect, since $\sigma_{\mathrm{max}} \propto
\mathrm{max}|\partial_{z}(R\Omega)| =|q|H/2$
 (even larger growth rates than shown in Fig.~\ref{3} 
are obtained for modes with $k\gg 10$, which asymptotically attain
this value). The presence of a boundary also strongly influences
the inertial waves, only the lowest $n$ of
which 
are well described by Eq.~\ref{growthprediction}.
As $H$ is increased, however,
modes with increasingly larger $n$ converge to the 
$H=\infty$ analytical prediction.

 The appearance
 and lack of convergence of the surface modes is a special pathology of the
isothermal model with imposed boundaries. It makes the
interpretation of these surface modes, and their role in
any ensuing turbulence,
especially problematic.
Are they merely numerical artefacts? In the next section we argue that
they are, in fact, more than that and that it is the vertically isothermal model
itself that is the problem.

\section{Linear stability of the locally polytropic disc: reduced model}
\label{VSIpoly}

In this section we analyse a reduced model of the VSI in discs with a
radial power law in entropy. The vertical structure of this locally
polytropic model is a good approximation for an optically thick disc,
and is more realistic than the locally isothermal model because it
possesses upper and lower surfaces. In addition, there is a
well-defined maximum vertical shear rate, which occurs at the disc
surface. Therefore this model is better defined mathematically and physically.

A formal derivation of the reduced model is relegated to Appendix
\ref{derivation}. The same scalings as in \S~\ref{VSIiso} are adopted,
corresponding to anelastic radially geostrophic phenomena. In
addition, thermal diffusion is likely to dominate the thermodynamics
of such slow and short-scale modes. Consequently, we may neglect the
entropy perturbation entirely,
as well as the stabilising influence of stratification.

The resulting linearised equations for such perturbations are
\begin{eqnarray}
\label{reducedpoly1a}
0&=&2v-\partial_{x} h, \\
\partial_{\tau} v &=& -\frac{u}{2} - w \frac{q_s z}{2\gamma}, \\
\partial_{\tau} w &=& -\partial_{z} h, \\
0&=&\partial_{x}u + \partial_{z}w - \frac{2mzw}{1-z^2},
\label{reducedpoly2a}
\end{eqnarray}
where the pseudo-enthalpy perturbation is $h= P^{\prime}/(1-z^2)^m$ and
$P^\prime$ is the (scaled) pressure perturbation. Rapid thermal 
diffusion means that the perturbations evolve isothermally, hence the
equations are similar in form to Eqs \ref{reducedisothermal1}-\ref{reducedisothermal4}.

We seek solutions of the form
\begin{eqnarray}
u= \mathrm{Re}\left[\tilde{u}(z) \mathrm{e}^{\mathrm{i}\left(kx-\omega \tau\right)}\right],
\end{eqnarray}
and so on for other variables.
The resulting system of equations is equivalent to the single ODE
\begin{eqnarray}\label{Polyredeqn}
\frac{\mathrm{d}^2 h}{\mathrm{d}z^2} -
z\left[\frac{2m}{1-z^2}+\frac{\mathrm{i}kq_s}{\gamma}\right]\frac{\mathrm{d}h}{\mathrm{d} z} +\omega^2k^2 h=0.
\end{eqnarray}
Unfortunately, this cannot be solved in closed form analytically. 
However, we can attack the problem using a matched WKBJ approach, or
numerically with the same method
as \S~\ref{VSIisoboundaries}: Chebyshev collocation on $N+1$ points,
followed by the eigensolution of the resulting matrix equation using
the QZ method. 
No explicit boundary conditions are imposed, but implicit regularity 
is assumed at $|z|=1$ (this gives identical results to imposing a free-surface condition explicitly).

Finally, we note that the resulting system can be written in a
scale-free manner by transforming our
 variables to hatted quantities, as follows:
\begin{eqnarray}
\label{rescaling}
v=k \hat{v}, \;\;\; w=k\hat{w}, \;\;\; \omega=\hat{\omega}/k, \;\;\; q_s=2\hat{q}\gamma/k.
\end{eqnarray} 
A similar rescaling also applies to the isothermal model. Under this
rescaling the governing equation depends solely on the two parameters
$\hat{q}$ and $m$. Each, discrete, growth rate then has the following
behaviour:
\begin{eqnarray}
\omega_n = \frac{1}{k}\hat{\omega}_n(k q_s,\,\gamma), 
\end{eqnarray}
where the function $\hat{\omega}_n$ is determined numerically.

\subsection{Non-vertically shearing case, $q_s=0$}

We first demonstrate that the reduced model correctly 
captures the inertial waves in a polytropic disc in the limit of large
$k$, in order
to gain confidence in the reduced model.

\begin{figure}
  \begin{center}
 \subfigure{\includegraphics[trim=8cm 0cm 8.5cm 1cm, clip=true,width=0.4\textwidth]{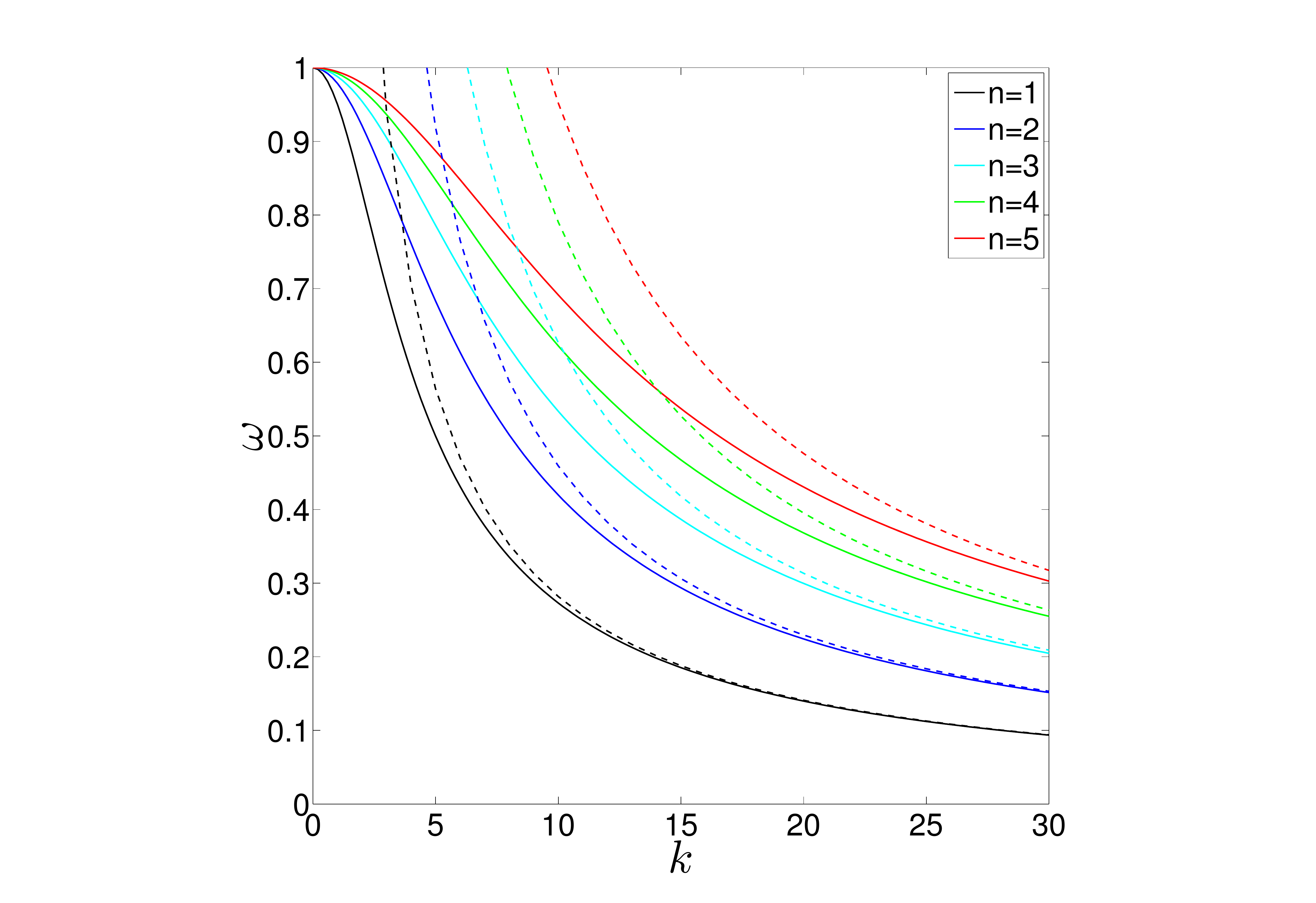}}
    \end{center}
  \caption{Comparison of the dispersion relation computed numerically from the locally polytropic reduced model (Eqs.~\ref{reducedpoly1a}--\ref{reducedpoly2a}) with $q_s=0$ (dashed lines) together with the numerically computed predictions for inertial modes in a polytropic disc \citep{KP1995} (solid lines) with $n=1,2,3,4,5$ vertical nodes adopting $\gamma=1.4$. The reduced model is only valid when $k\gtrsim O(\sqrt{n})$.}
  \label{4}
\end{figure}

The full polytropic disc model without vertical shear 
($q_s=0$) possesses three different classes of neutrally stable
axisymmetric modes: 
high-frequency acoustic modes, low-frequency inertial modes, 
and surface gravity modes of intermediate frequency
\citep{KP1995,Ogilvie1998}.
 The reduced model is designed to capture only the low frequency inertial modes.

In Fig.~\ref{4} we compare the
frequencies of five inertial modes with vertical modenumbers (vertical nodes)
$n=1,2,3,4,5$ and $q_s=0$ for several $k$ as predicted by the reduced
model and the full polytropic model of
\cite{KP1995} (solving their Eqs.~4-8). The latter are represented by
solid and the former by dashed lines.
Unlike the
isothermal disc, 
these frequencies must generally be obtained numerically.
There is general agreement for 
$k\gtrsim 20$, which illustrates that the reduced model correctly captures
the 
frequencies of these inertial modes for sufficiently large
$k$. However,
 only modes with $k\geq O(\sqrt{n})$ are correctly captured.

In the limit of large $k$, a WKBJ analysis shows that the frequencies 
take a simple form, 
\begin{equation}\label{wkbjfreq}
\omega= \pi(2n+m)/(4k),
\end{equation}
 where $n$ is an integer. 
Though the formula is most accurate in the limit of
large $n$, it does well across the whole
range of frequencies. Details of this calculation can be found in
Appendix B.

\subsection{Vertically shearing case, $q_s\neq 0$}
\label{instabilitypoly}

\begin{figure}
 \begin{center}
\subfigure{\includegraphics[trim=6.5cm 2.5cm 7.5cm 3.5cm, clip=true,width=0.42\textwidth]{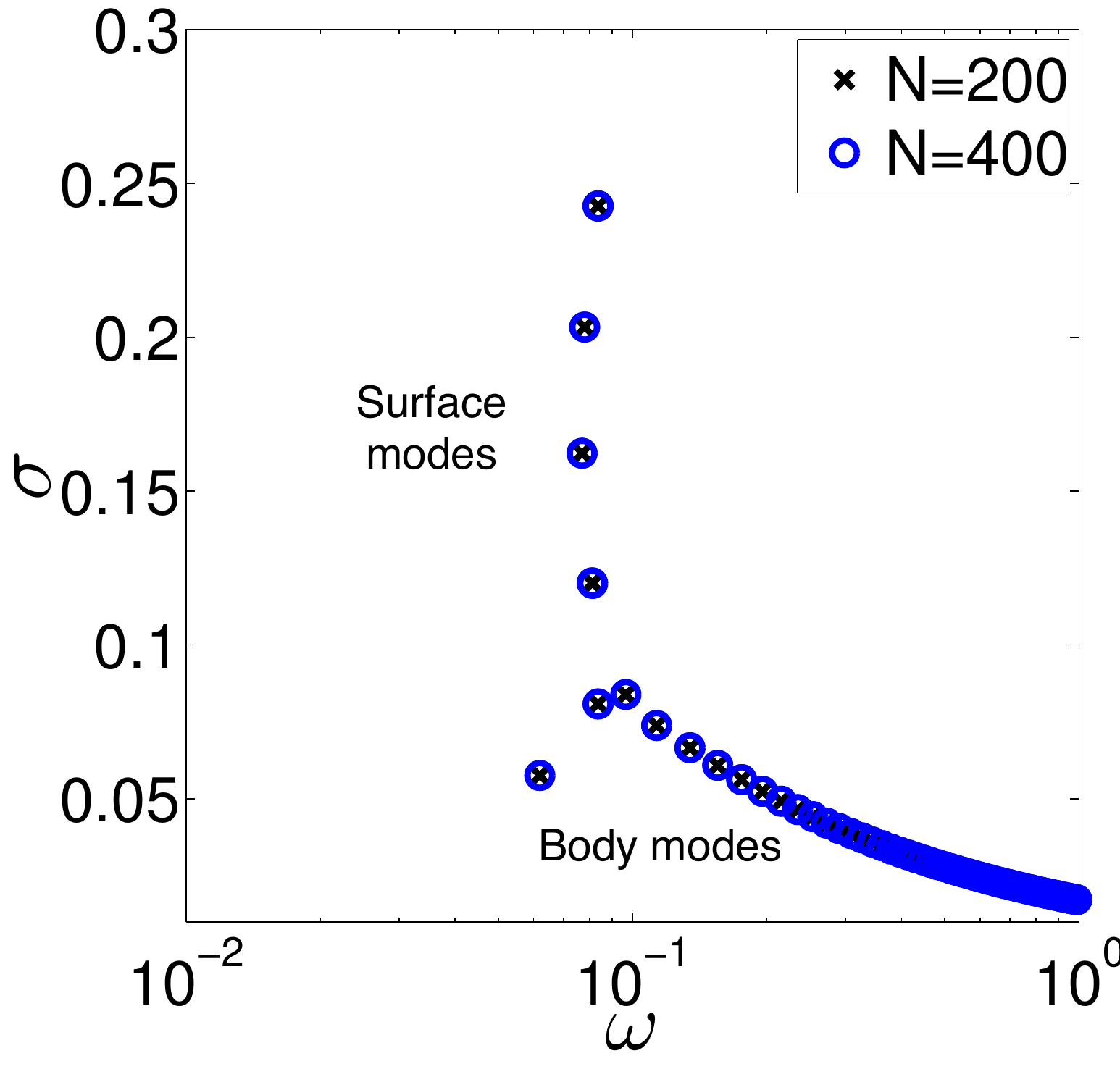}}
    \end{center}
  \caption{Growth rates of the unstable modes versus their (real) frequencies in a polytropic disc with $k=100$, $q_s=-1$, and $\gamma=1.4$ computed using $N=200$ and $N=400$ points. This shows that the modes are well captured using $N=200$, and the distribution is similar to the isothermal case in Fig.~\ref{3}.}
  \label{5}
\end{figure}

When $q_s\ne 0$, the disc is unstable to the VSI. 
As an illustrative example, we plot in Fig.~\ref{5} the
complex frequencies of the unstable modes computed from the
eigenvalue problem
Eqs.~\ref{reducedpoly1a}--\ref{reducedpoly2a} when
$k=100$, $q_s=-1$ and $\gamma=1.4$ for two different vertical
resolutions $N=200$ and $N=400$. The unstable modes (for this $k$) are
shown to be well resolved using $N=200$. 

\begin{figure*}
  \begin{center}
     \subfigure[$\omega=-0.0824+0.2428\mathrm{i}$]{\includegraphics[trim=6cm 0cm 8.5cm 1cm, clip=true,width=0.24\textwidth]{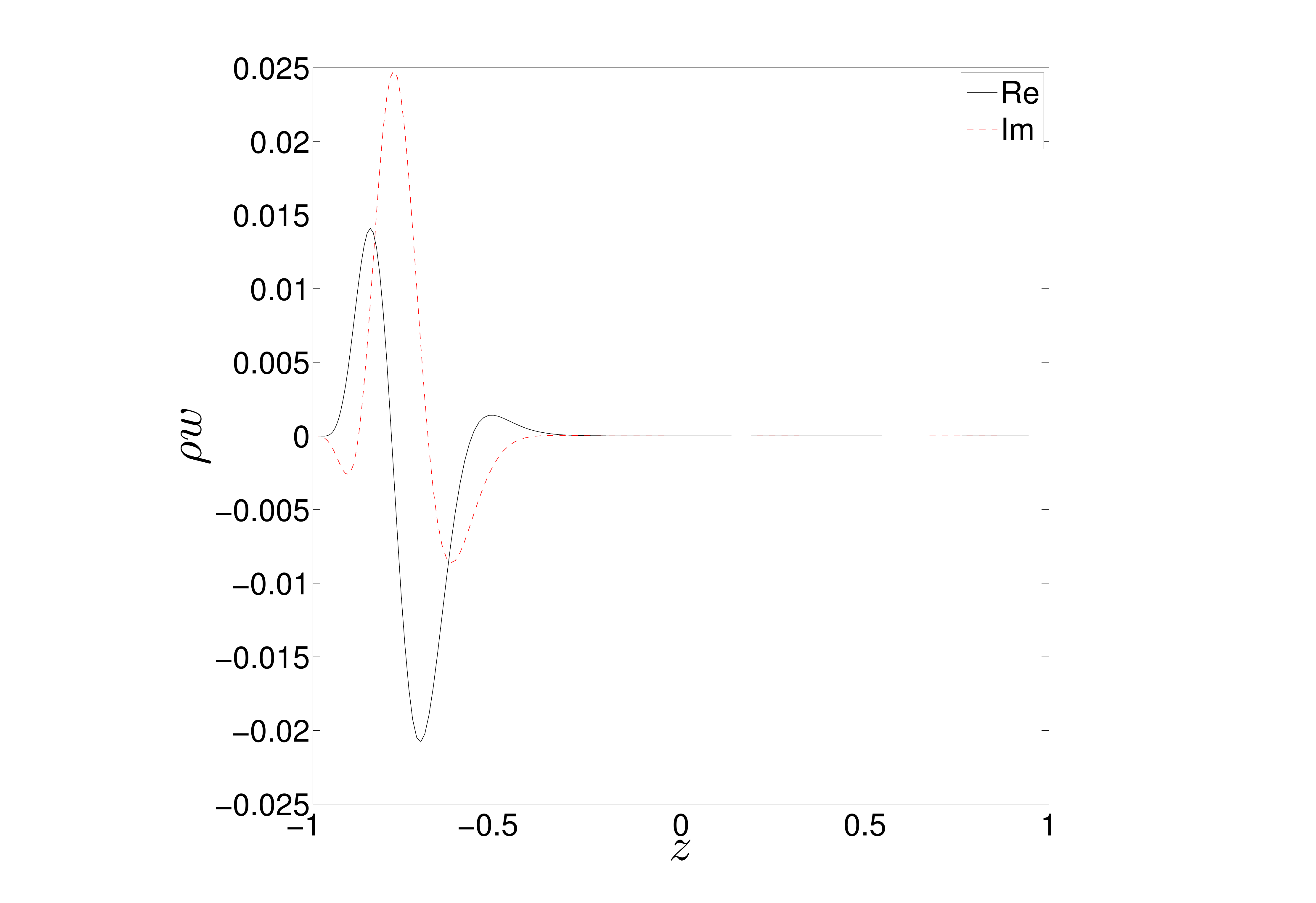}}
      \subfigure[$\omega=-0.0824+0.2428\mathrm{i}$]{\includegraphics[trim=6cm 0cm 8.5cm 1cm, clip=true,width=0.24\textwidth]{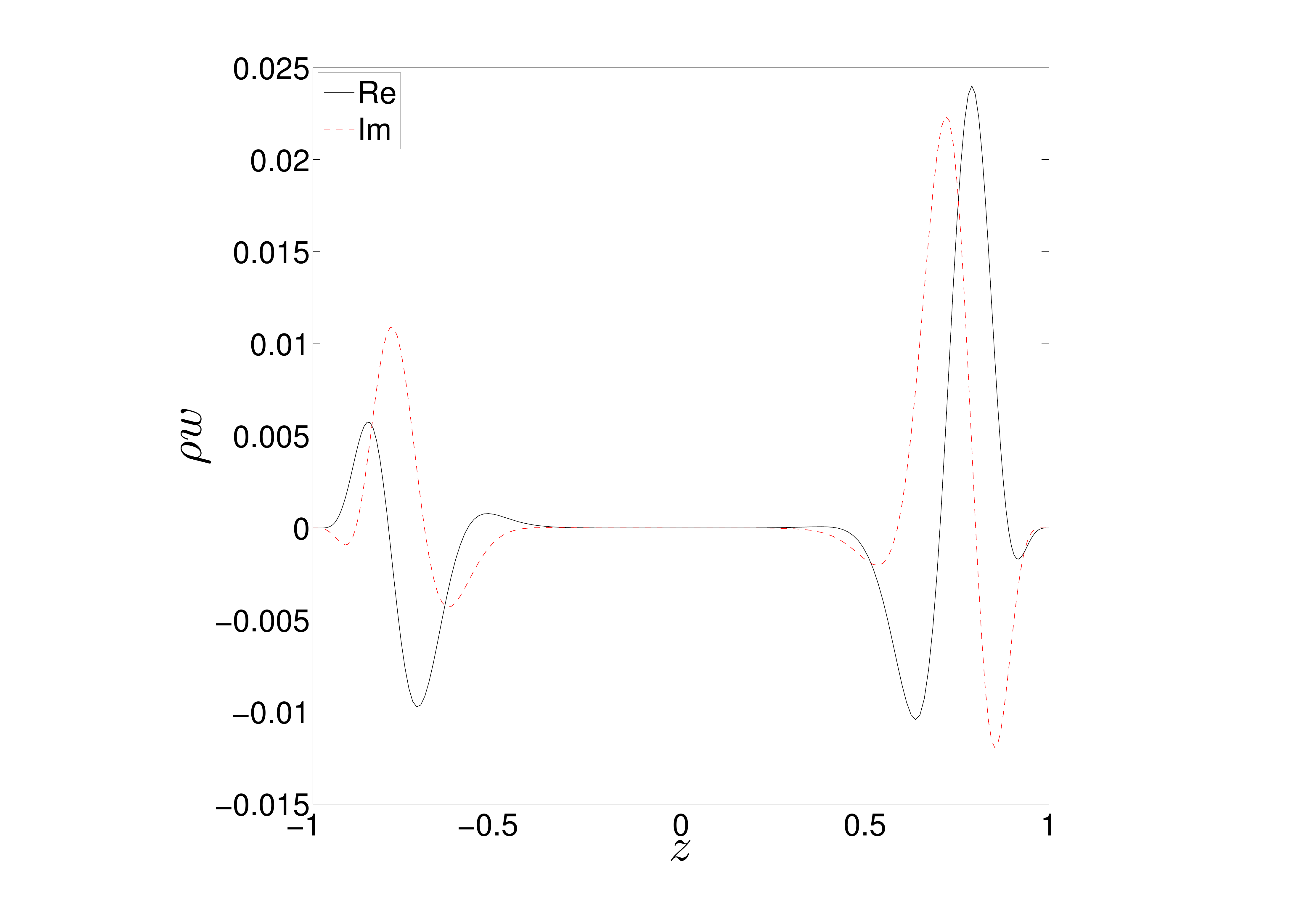}} 
      \subfigure[$\omega=-0.0768+0.2035\mathrm{i}$]{\includegraphics[trim=6cm 0cm 8.5cm 1cm, clip=true,width=0.24\textwidth]{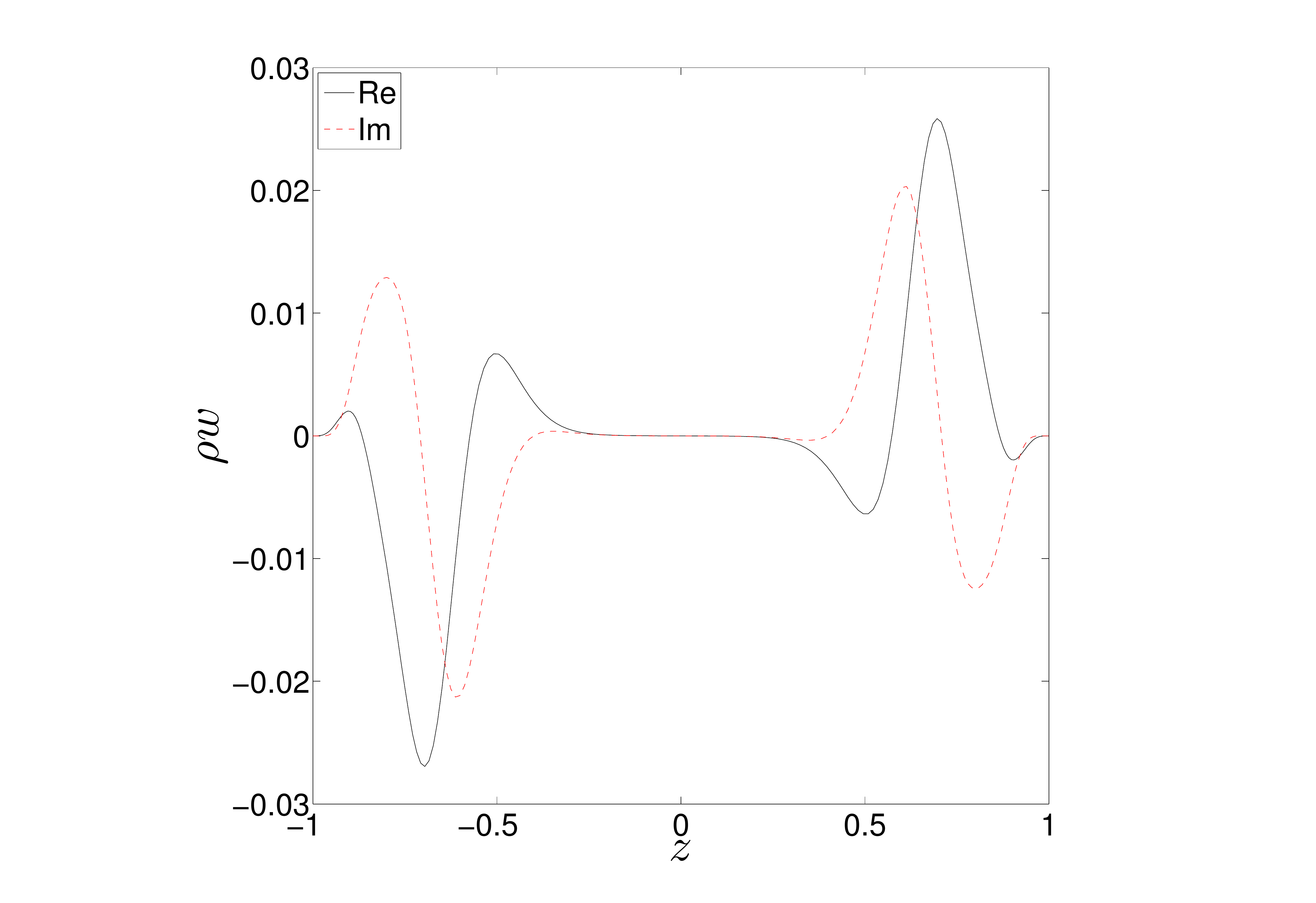}}
      \subfigure[$\omega=-0.0758+0.1624\mathrm{i}$]{\includegraphics[trim=6cm 0cm 8.5cm 1cm, clip=true,width=0.24\textwidth]{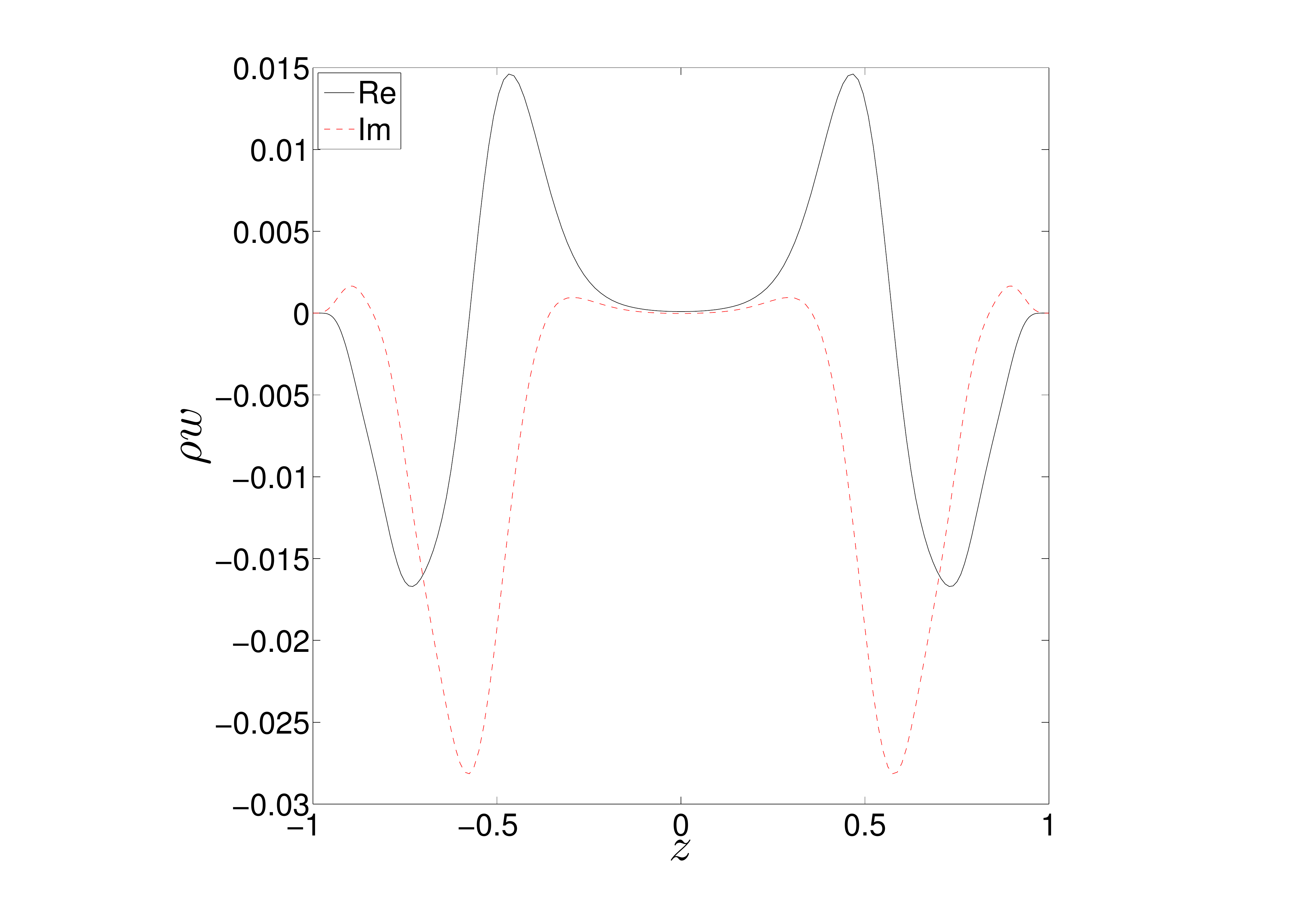}}
      \subfigure[$\omega=-0.1751+0.0556\mathrm{i}$]{\includegraphics[trim=6cm 0cm 8.5cm 1cm, clip=true,width=0.24\textwidth]{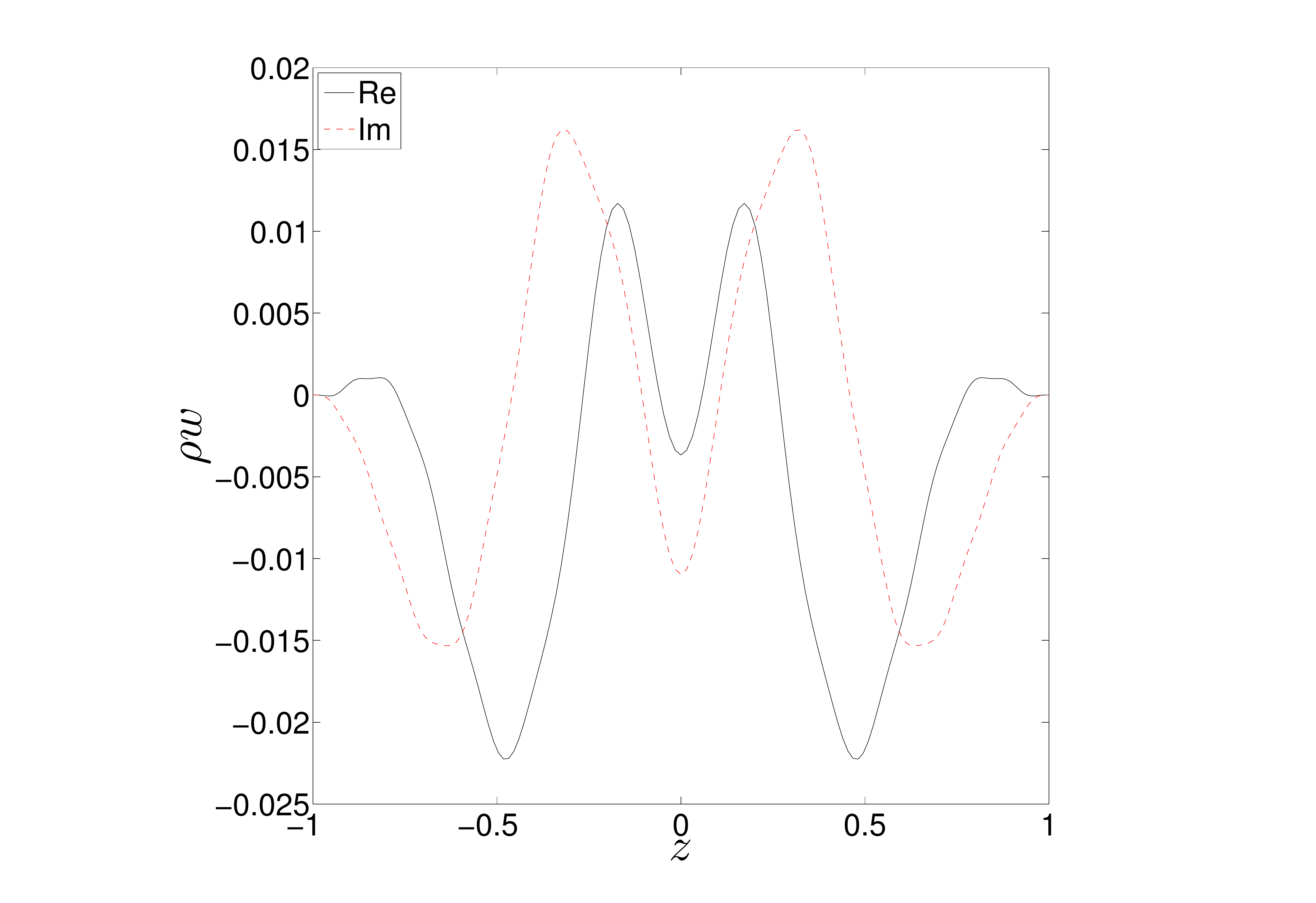}}
      \subfigure[$\omega=-0.1951+0.0518\mathrm{i}$]{\includegraphics[trim=6cm 0cm 8.5cm 1cm, clip=true,width=0.24\textwidth]{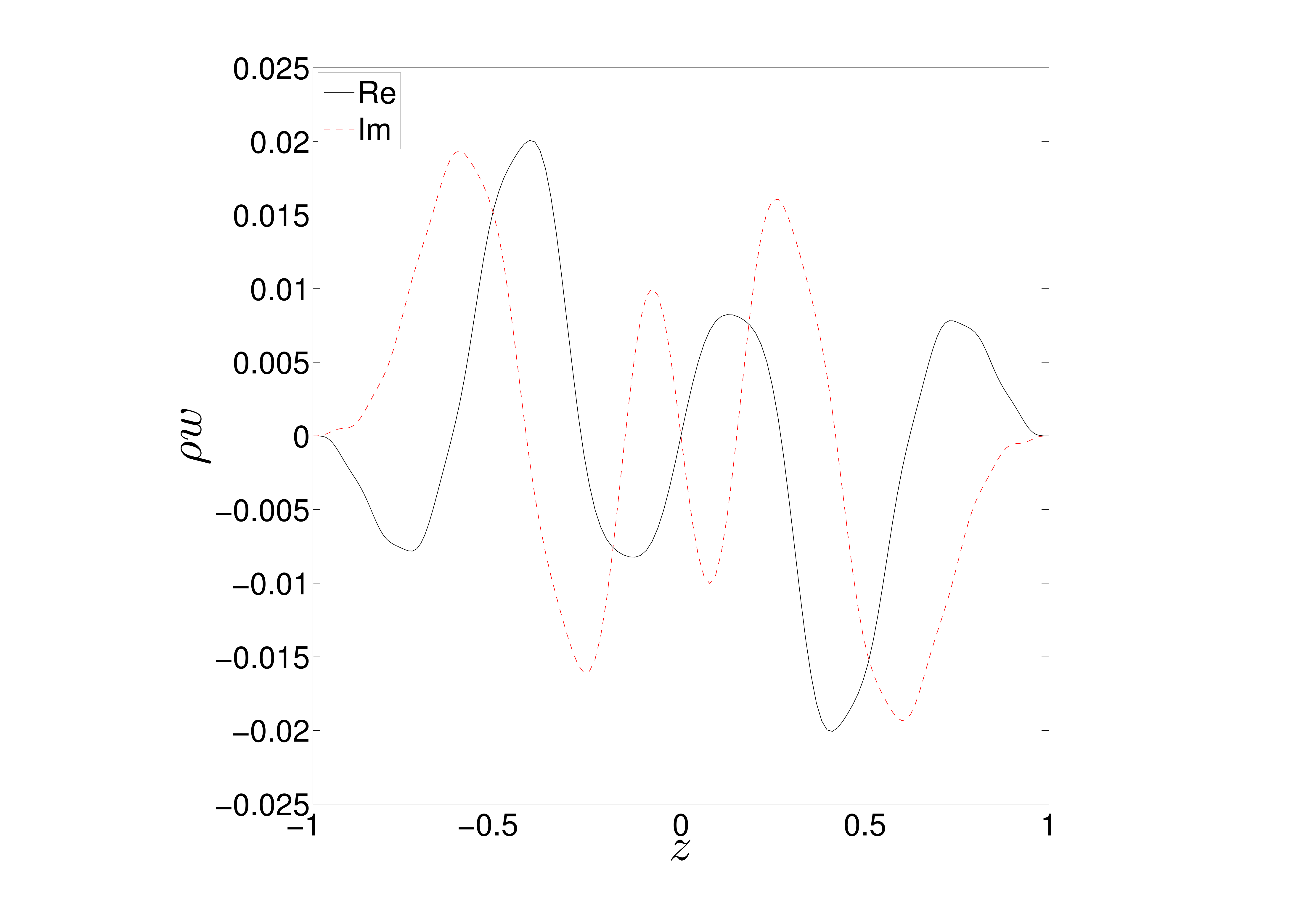}}
      \subfigure[$\omega=-0.2148+0.0487\mathrm{i}$]{\includegraphics[trim=6cm 0cm 8.5cm 1cm, clip=true,width=0.24\textwidth]{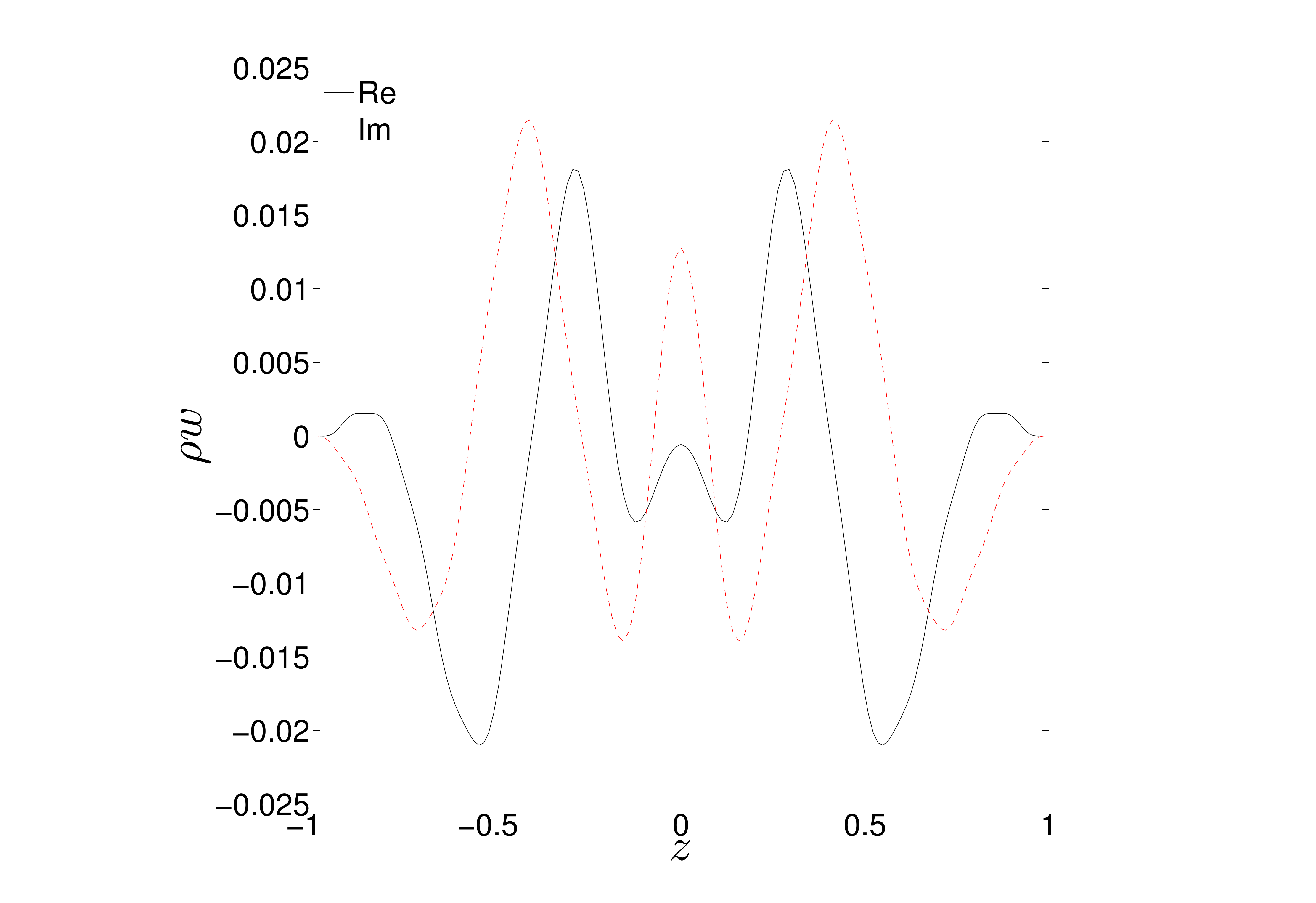}}
       \subfigure[$\omega=-0.0618+0.0577\mathrm{i}$]{\includegraphics[trim=6cm 0cm 8.5cm 1cm, clip=true,width=0.24\textwidth]{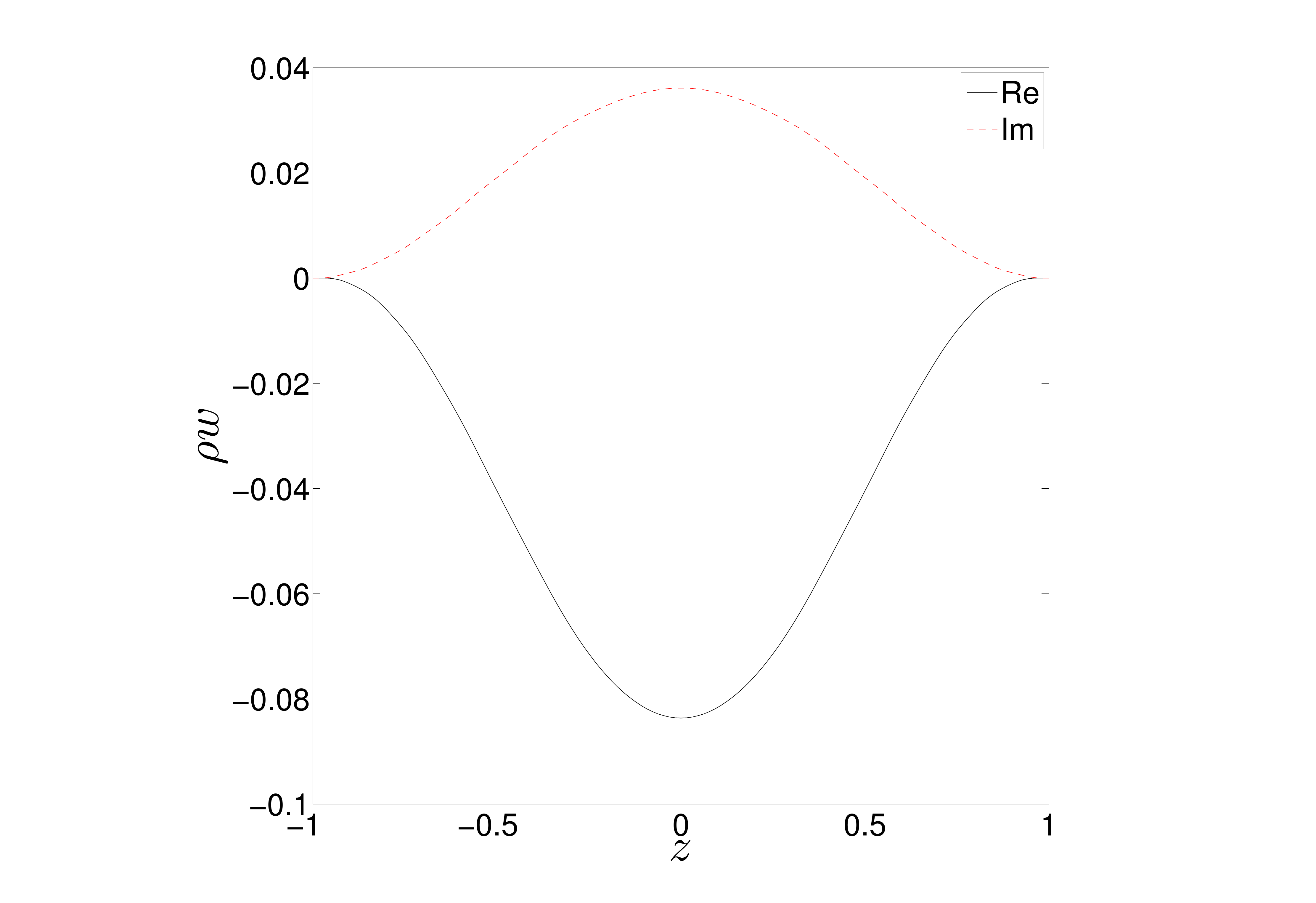}}
    \end{center}
  \caption{The real and imaginary parts of the
    vertical momenta for several representative examples of both types of modes in a polytropic disc with $k=100$, $q_s=-1$, $\gamma=1.4$
    using $N=200$, where the figure labels indicate the complex frequencies of the modes. In the top row we show several examples of surface
    modes, with the fastest growing modes plotted in the two top left panels. In the bottom row
    we show several examples of body modes. The lowest frequency mode is
    plotted in the bottom right panel, and is an $n=1$ ($n=0$ for $w$)
    inertial wave, i.e. the fundamental ``corrugation mode". The other bodes modes are higher order body modes.}
  \label{6}
\end{figure*}

\begin{figure}
  \begin{center}
 \subfigure{\includegraphics[trim=5cm 1cm 7cm 2cm, clip=true,width=0.42\textwidth]{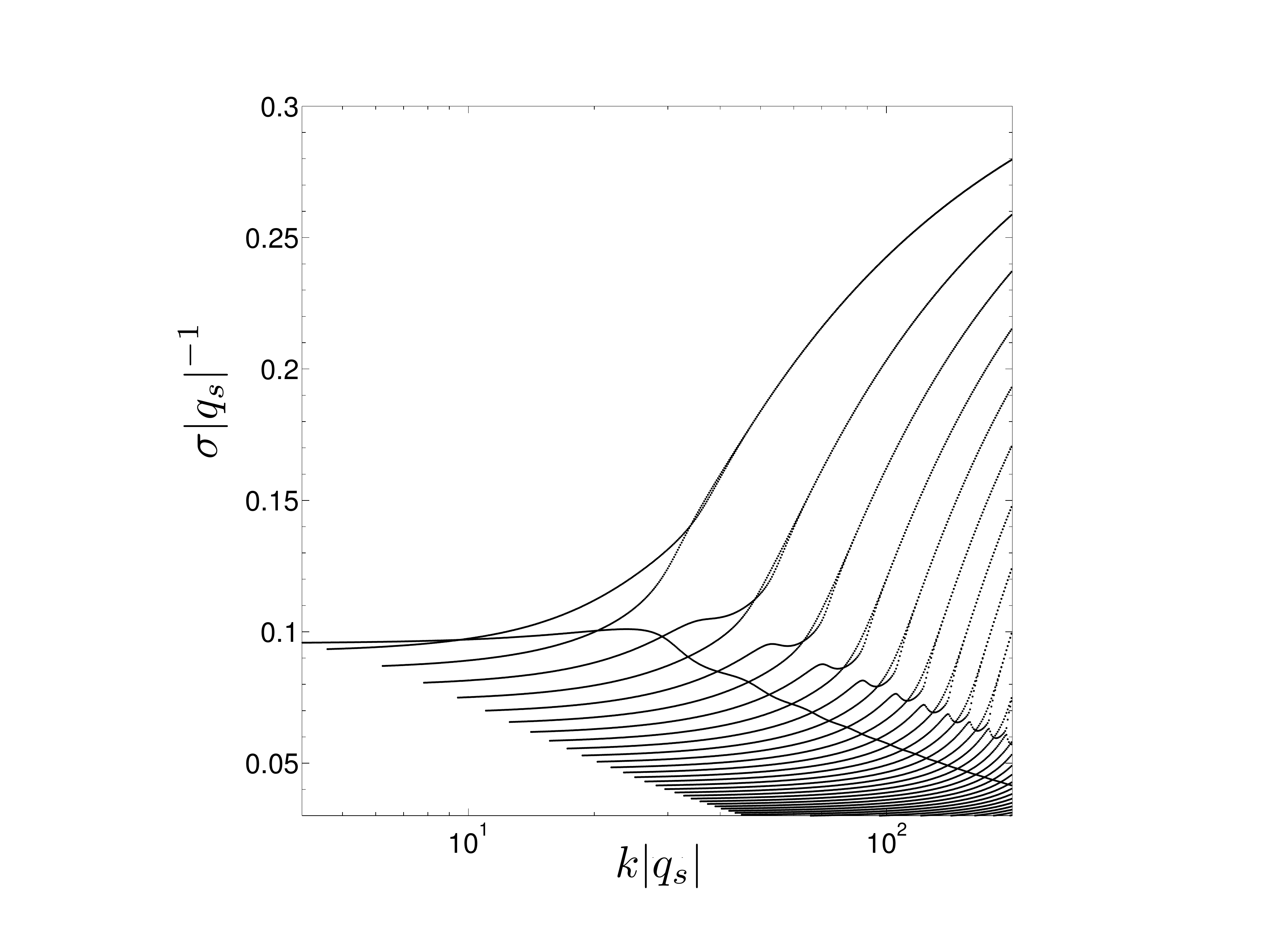}}
    \end{center}
  \caption{Normalised growth rates of all unstable modes as a function of the scaled radial wavenumber $k |q_s|$ in a polytropic disc with $\gamma=1.4$, computed using $N=200$. The absence of unstable modes in the lower left of the figure results from our elimination of modes with frequencies larger than $1$. Surface modes appear in pairs for $k|q_s|\gtrsim 30$, with the fastest growing arising from the $n=2$ ``breathing mode" and the $n=3$ inertial mode. The $n=1$ fundamental ``corrugation mode" lies along the line that ends up at the bottom right. The scaled growth rate of the fastest growing surface mode asymptotically approaches the maximum vertical shear rate (0.357) as $k|q_s|\rightarrow \infty$.}
  \label{8}
\end{figure}

Fig.~\ref{5} also shows clearly
that a locally polytropic disc can be
unstable to 
two different types of modes: (a) modestly growing inertial waves
(``body modes''; these occur
on longer radial scales, as we will show in Fig.~\ref{8}), and (b) rapidly growing short-wavelength
surface modes (these only occur
 when $|kq_s|\gtrsim 30$, as we will also show in Fig.~\ref{8}). The growth rates of the two classes of modes
 are labelled in Fig.~\ref{5}.
 Examples for each type of 
mode are plotted in Fig.~\ref{6}. Hence we find that the VSI in the locally polytropic
disc, with well-defined physical surfaces, is
 qualitatively similar to the VSI in the isothermal disc with an 
imposed artifical surface at a finite height, as discussed in \S~\ref{VSIisoboundaries}. 

\begin{figure}
  \begin{center}
 \subfigure{\includegraphics[trim=7.5cm 0cm 8.5cm 1cm, clip=true,width=0.42\textwidth]{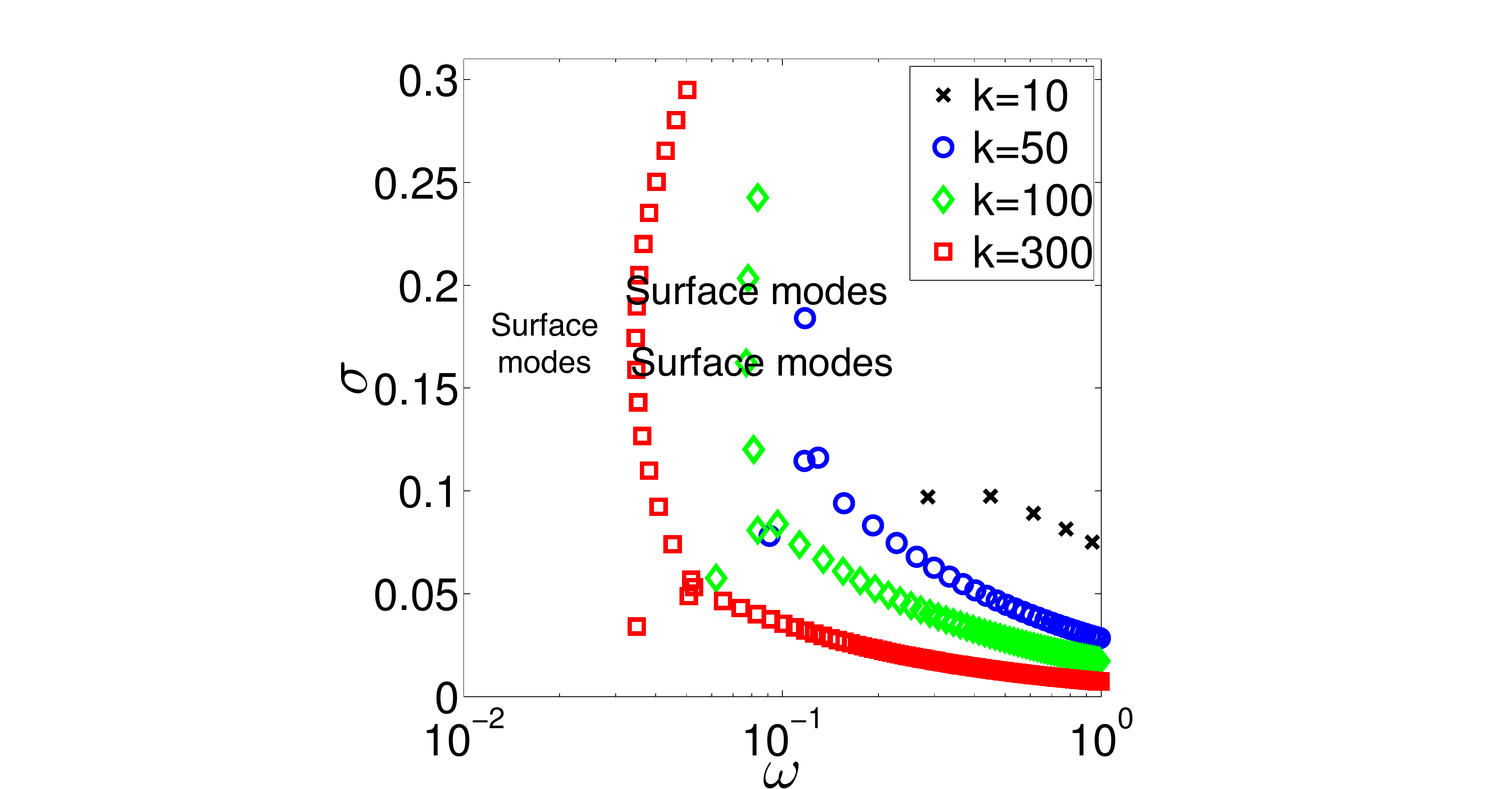}}
    \end{center}
  \caption{Growth rates of the unstable modes versus their (real)
    frequencies for several values of $k$ in a polytropic disc with
    $q_s=-1$ and $\gamma=1.4$, computed using up to $N=400$ points
    (which was found to be sufficient for all $k$). This illustrates
    the dependence of the unstable modes on $k$.}
  \label{7}
\end{figure}

In Fig.~\ref{7}, we plot the complex frequencies of the unstable modes
when $q_s=-1$ and $\gamma=1.4$ for several values of $k$. To accurately
capture the fastest growing modes for larger $k$ requires increasing
the vertical resolution ($N=400$ was required to obtain convergence
when $k=300$). Fig.~\ref{7} shows that the growth rate of the fastest
growing mode increases, and its frequency decreases, as we consider
smaller horizontal lengthscales (larger $k$). For a given $|q_s|$, the
number of rapidly growing surface modes increases with $k$. All
unstable modes are body modes when $k=10$, but as we increase $k$
surface modes appear, and by $k=300$ there is a large population of
them. The maximum growth rate is on track to approach the maximum 
vertical shear rate $|\partial_{z}(R\Omega)| \sim |q_s|H/(2\gamma)
\approx 0.357$
 as $k\rightarrow \infty$, which we expect to provide an upper bound on the growth rate of the VSI.

In Fig.~\ref{8}, we have exploited the rescaling of Eq.~\ref{rescaling} to illustrate the general dependence of the scaled growth rate $\sigma/|q_s|$ of the VSI as a function of $k |q_s|$ for all unstable modes. We do not consider modes with real frequencies that are larger than unity 
--  their omission results in a region where the VSI is absent in the bottom left of the figure. Pairs of surface modes appear for sufficiently large $k|q_s|$. The first, and fastest growing, pair of surface modes appear when $k|q_s|\gtrsim 30$, and corresponds to the $n=2$ ``breathing mode" and the $n=3$ inertial mode changing their character. The $n=1$ fundamental ``corrugation mode" is the first to become unstable when $k|q_s|\gtrsim 3$, but its growth becomes weaker for larger $k|q_s|$. This is represented by the curve that crosses all others and ends up at the bottom right of the figure. As $k|q_s|\rightarrow \infty$, there are more unstable modes, and their scaled growth rates approach the scaled maximum vertical shear rate $1/(2\gamma) \approx 0.357$, as expected.

The fastest growing body mode is the $n=1$ corrugation mode, and has a maximum growth rate of approximately $0.1|q_s| (1.4/\gamma)$ when $|kq_s|\lesssim 10$. The fastest growing mode is a surface mode, with a maximum growth rate that approaches the maximum vertical shear rate $|\partial_{z}(R\Omega)|\approx |q_s|/(2\gamma)$ on the smallest scales. (We have confirmed the dependence of the fastest growing surface mode on $\gamma$, though we omit this for brevity -- this arises because the basic disc structure varies with $\gamma$.) Note that the full polytropic disc possesses a class of surface modes that are restored by gravity \citep{KP1995,Ogilvie1998}. However, these are unrelated to the class of low frequency surface modes that we have discussed in this section, and occur in a different frequency range.

Why do surface modes appear when $|kq_s|\gtrsim 30$? 
The vertical shear rate takes its maximum magnitude at the disc
surface,
so it makes sense that if localised modes were to appear, they should do so just below the disc surface. 
But when $|kq_s|\lesssim 30$, it is not possible for a mode (with an inertial wave character) to become
sufficiently
localised close to the disc surface (i.e. $\ell_{z}$ cannot be made
sufficiently small). 
The criterion for the appearance of surface modes is therefore
 tied to the shortest vertical lengthscale permitted.

Finally, in Appendix B we show that the body modes' growth rate
can be obtained analytically in the limit of small $q_s$:
\begin{equation}
\sigma \approx \frac{m q_s}{\pi\gamma(2n+m)}\log\left[\tfrac{1}{2}\pi(2n+m)\right],
\end{equation}
where $n$ is a (large) integer. The corresponding wave frequency is
given by Eq.~\ref{wkbjfreq}. It should be stressed that 
the growth rate $\sigma$ being at subdominant
order is only a rough
estimate, because errors arising from the WKBJ
method itself enter at the same order.
Note that the growth rate is linear in the shear, but the larger $n$, the
smaller $\sigma$, in contrast to the isothermal case, but in agreement
with Figs.~\ref{5} and \ref{7}. 
For general $q_s$ an estimate for the surface modes' maximum growth rate is 
$\sigma \approx q_s/(2\gamma)$, while its
wave frequency scales as $\ln k/k$.
Being leading order, this is a more robust estimate, which is in approximate agreement with our results.

In summary, the VSI in the locally polytropic model is very similar in character
to the VSI in the locally isothermal model with artificially imposed
vertical boundaries
(\S~\ref{VSIisoboundaries}).
The presence of a surface, be it physically justified
 or a numerical convenience, provides a
special location upon which modes can affix themselves and localise.
The polytropic disc however yields solutions that
 are clearer to interpret because the location of the surface
is specified (not an adjustable parameter). Consequently,
there exists a well-defined maximum
growth rate for the instability, set by the vertical shear rate
at the disc surface.

\section{Global calculations in the locally isothermal disc}
\label{2D}

We have so far analysed the VSI using reduced models of locally isothermal and polytropic discs. Here we present the first two-dimensional stability calculations of the VSI in a locally isothermal disc. One motivation for doing so is to verify the validity of the model analysed in \S~\ref{VSIisoreduced}, another is to reproduce the instability in a setup that more directly matches that of recent global simulations \citep{Nelson2013,StollKley2014}. An issue with global stability calculations is that they are computationally demanding (primarily regarding their memory usage), so we are limited to studying the VSI with relatively low resolutions. This is primarily a problem for capturing the surface modes, since they occur on very short lengthscales. However, the lowest order body modes have less vertical structure and are better captured in these calculations.

We consider axisymmetric perturbations
($u_{R},u_{\phi},u_{z},\rho^{\prime}$) to the
global disc model of Section 3.1.1, assuming their time-dependence to
be $\propto \text{e}^{-\text{i}\omega t}$. The equations governing their
linear evolution are then
\begin{eqnarray}
\label{linsystem1a}
-\mathrm{i}\omega u_{R}  &=& 2 R\Omega u_{\phi}-\frac{1}{\rho}\partial_{R}(c_{s}^2\rho^{\prime}) +\frac{\rho^{\prime}}{\rho^2}\partial_{R}(c_{s}^2 \rho) +f_{R}, \\
-\mathrm{i}\omega u_{\phi} &=&-u_{R}\frac{1}{R}\partial_{R} (R^2\Omega) - u_{z}\partial_{z}(R\Omega) +f_{\phi}, \\
-\mathrm{i}\omega u_{z} &=& -\frac{1}{\rho}\partial_{z}(c_{s}^2\rho^{\prime})+\frac{\rho^{\prime}}{\rho^2}\partial_{z}(c_{s}^2\rho)+f_{z}, \\
-\mathrm{i}\omega \rho^{\prime} &=& -u_{R}\partial_{R}\rho-u_{z}\partial_{z}\rho-\frac{\rho}{R}\partial_{R}(Ru_{R}) - \rho\partial_{z}u_{z},
\label{linsystem2a}
\end{eqnarray}
where $P^{\prime}=c_{s}^2(R) \rho^{\prime}$, and the background state
quantities $c_{s}(R), \Omega(R,z)$ and $\rho(R,z)$ are defined in
\S~\ref{isobasic}. To regularise the solutions in
some calculations we
 include a Navier-Stokes shear viscosity with constant kinematic viscosity $\nu$, with the extra terms
\begin{eqnarray}
\nonumber
\boldsymbol{f} &=& \nu\rho \left[\nabla^2\boldsymbol{u}+\frac{1}{3}\nabla (\nabla \cdot \boldsymbol{u})\right]+ \nu \rho^{\prime}\nabla^2\boldsymbol{U}_{0} \\
&& +\nu\nabla \rho\cdot \left[\nabla \boldsymbol{u}+(\nabla \boldsymbol{u})^T - \frac{2}{3}(\nabla \cdot \boldsymbol{u})\mathrm{I}\right] \\
&& \nonumber +\nu \nabla \rho^{\prime}\cdot \left[ \nabla \boldsymbol{U}_{0}+(\nabla \boldsymbol{U}_{0})^T\right],
\end{eqnarray}
where $\boldsymbol{U}_{0}=R\Omega \boldsymbol{e}_{\phi}$.
We adopt a 2D cylindrical domain with $R\in [R_{0},R_{1}]$ and $z\in [-z_{0},z_{0}]$, where $z_{0}$ is some multiple of $H_{0}=\epsilon$ (this differs from the spherical wedge considered by \cite{Nelson2013} and \cite{StollKley2014}, but this difference is probably unimportant). Our units of length and time are chosen such that $R_{0}=1$ and $\Omega(R_{0})=1$. If $\boldsymbol{f}=0$,
boundary conditions are enforced such that
\begin{eqnarray}
&& \rho u_{R} =0    \;\;\;\;\; \mathrm{at} \;\;\; R=R_{0} \;\;\; \& \;\;\; R=R_{1}, \\
&& \rho u_{z} =0    \;\;\;\;\; \mathrm{at} \;\;\; z=-z_{0} \;\;\; \& \;\;\; z=z_{0},
\end{eqnarray}
otherwise we supplement these with stress-free conditions (no tangential stresses) on each boundary. 
This system is solved (after writing the equations in terms of momenta
rather than velocities) numerically using a Chebyshev collocation
method in both $R$ and $z$, with $N_{R}+1$ and $N_{z}+1$ points on a
Gauss-Lobatto grid, respectively. The resulting generalised eigenvalue
problem involves matrices of size $L\times L$, where $L=4
(N_{R}+1)(N_{z}+1)$, and is solved using one of two methods: a QZ
algorithm to obtain an approximation of the full spectrum, or an
Arnoldi iterative method to obtain an approximation to a desired part
of the spectrum. The QZ method is computationally expensive when
$L\gtrsim 20,000$, so our results are then supplemented by the Arnoldi
method, once we determine which modes to focus on. In what follows we set $p=0$, since preliminary investigation suggested this parameter to be unimportant.

The code has been tested in several ways. First, without viscosity and with a small radial domain (e.g.~$R_{1}=1.001$), we have confirmed that our code accurately reproduces the inertial and acoustic modes of the vertically unbounded isothermal disc when $q=0$ \citep{LP1993} -- except for modes with large $n$, for which the confining effect of the vertical boundaries modifies the solutions. Using a similarly small radial domain, we have verified that viscous damping produces the expected decay rate for a mode with a given short radial wavelength ($\nu k^2$ when $k\gg 1$). In all calculations, the discretisation inevitably produces many junk modes, that involve oscillations on the grid-scale. Where possible, we eliminate these modes by comparing eigenvalues for several different resolutions, and by inspecting the spatial structure of the eigenfunctions, discarding those that oscillate on the grid-scale.

\subsection{Inviscid calculations}

We first illustrate the properties of the VSI in the absence of
viscosity. As in the reduced model in \S~\ref{VSIisoreduced}, we
obtain two classes of modes: modestly growing body modes, and rapidly
growing surface modes localised near the numerical boundaries
in $z$. As in Section 4.3, these modes would vanish if the
boundaries could be taken to infinity. However, we do not discard
these modes entirely as they still probably
represent physical solutions in some sense (see \S~\ref{VSIpoly}).

\subsubsection{Body modes}

\begin{figure*}
 \begin{center}
    \subfigure[$\omega=0.1028+0.0014\mathrm{i}$]{\includegraphics[trim=5cm 0cm 5cm 1cm, clip=true,width=0.38\textwidth]{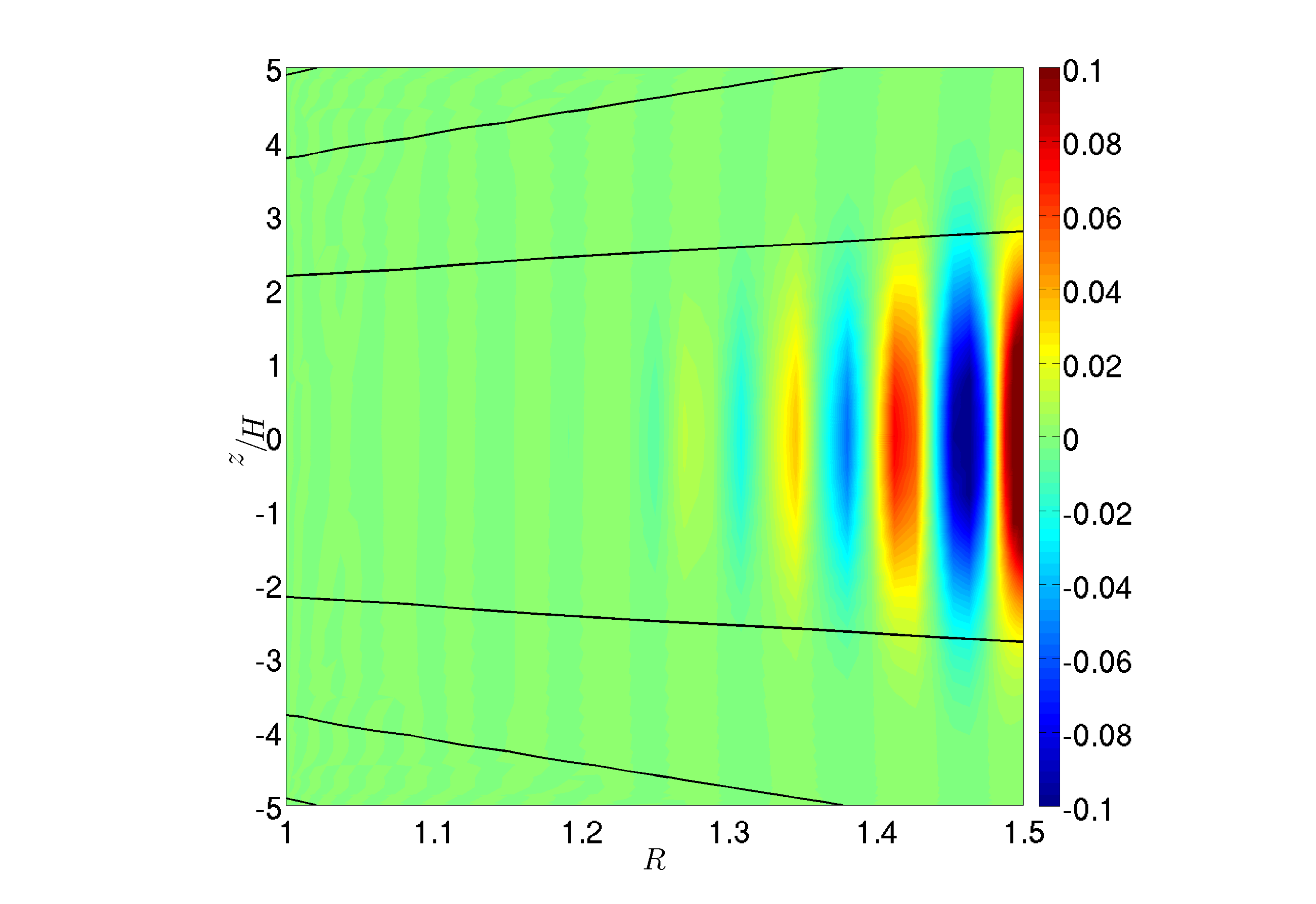} } \hspace{1.5cm}
     \subfigure[$\omega=0.1587+0.0016\mathrm{i}$]{\includegraphics[trim=5cm 0cm 5cm 1cm, clip=true,width=0.38\textwidth]{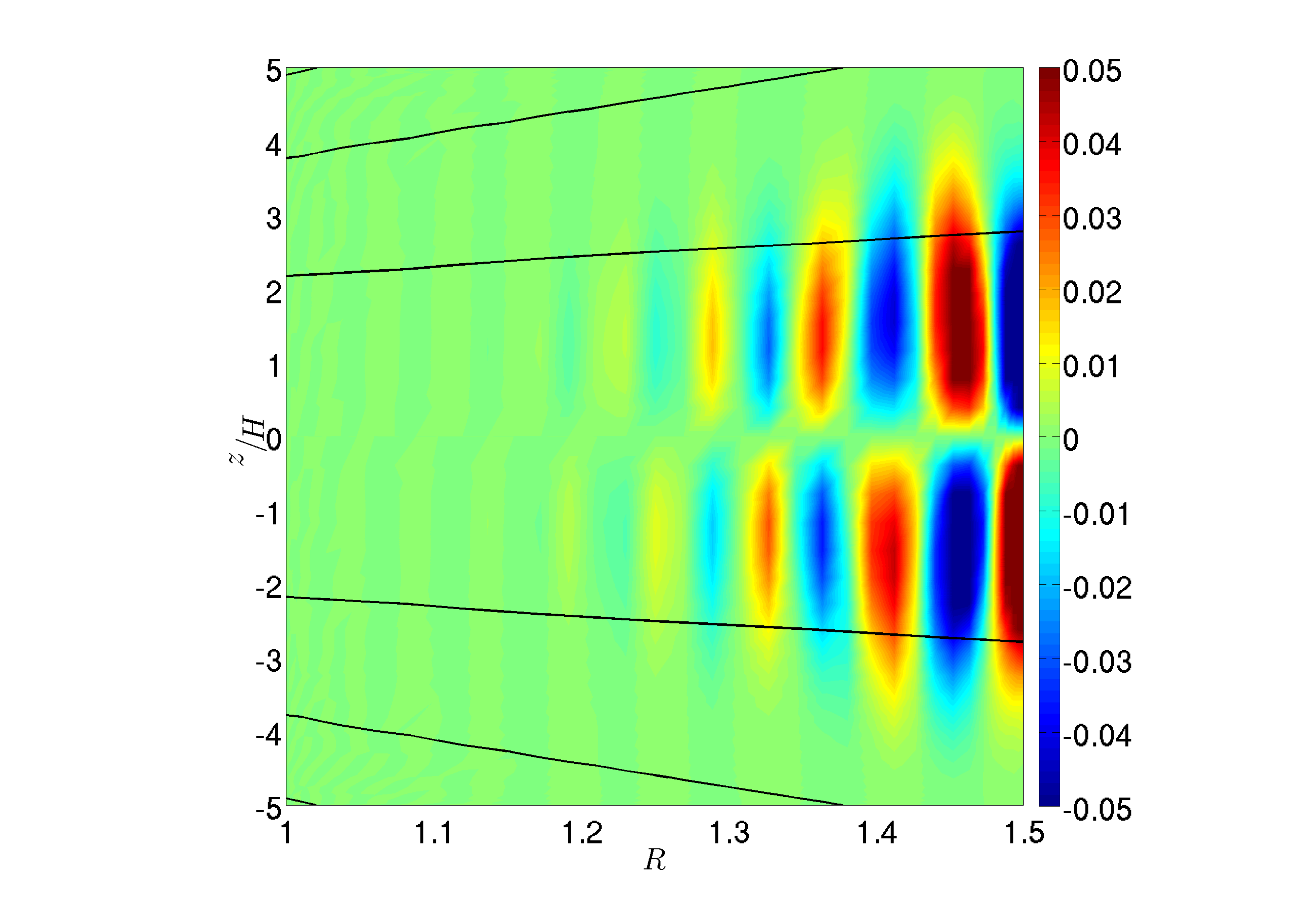} }\\
     \subfigure[$\omega=0.2019+0.0034\mathrm{i}$]{\includegraphics[trim=5cm 0cm 5cm 1cm, clip=true,width=0.38\textwidth]{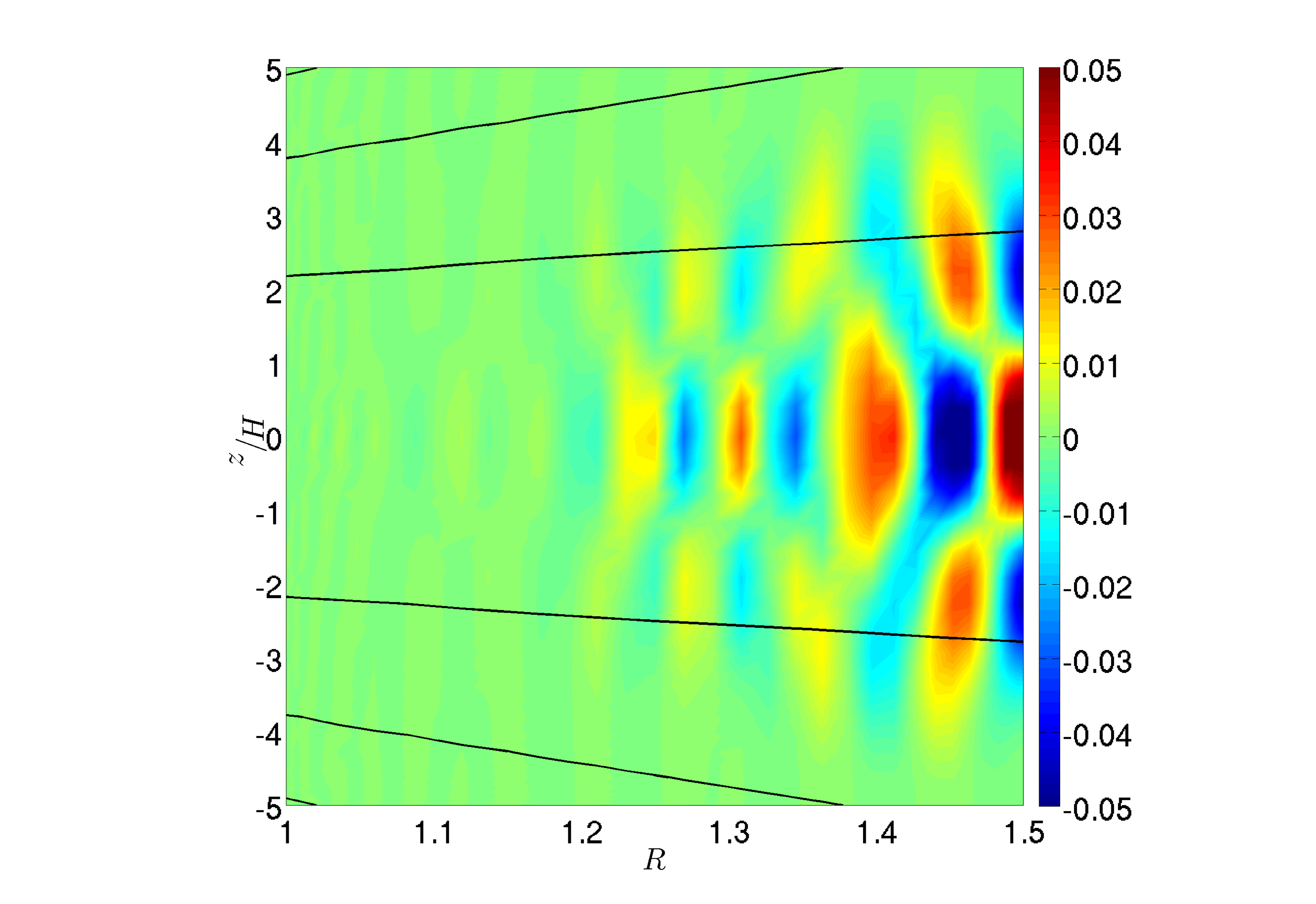} } \hspace{1.5cm}
     \subfigure[$\omega=0.2209+0.0035\mathrm{i}$]{\includegraphics[trim=5cm 0cm 5cm 1cm, clip=true,width=0.38\textwidth]{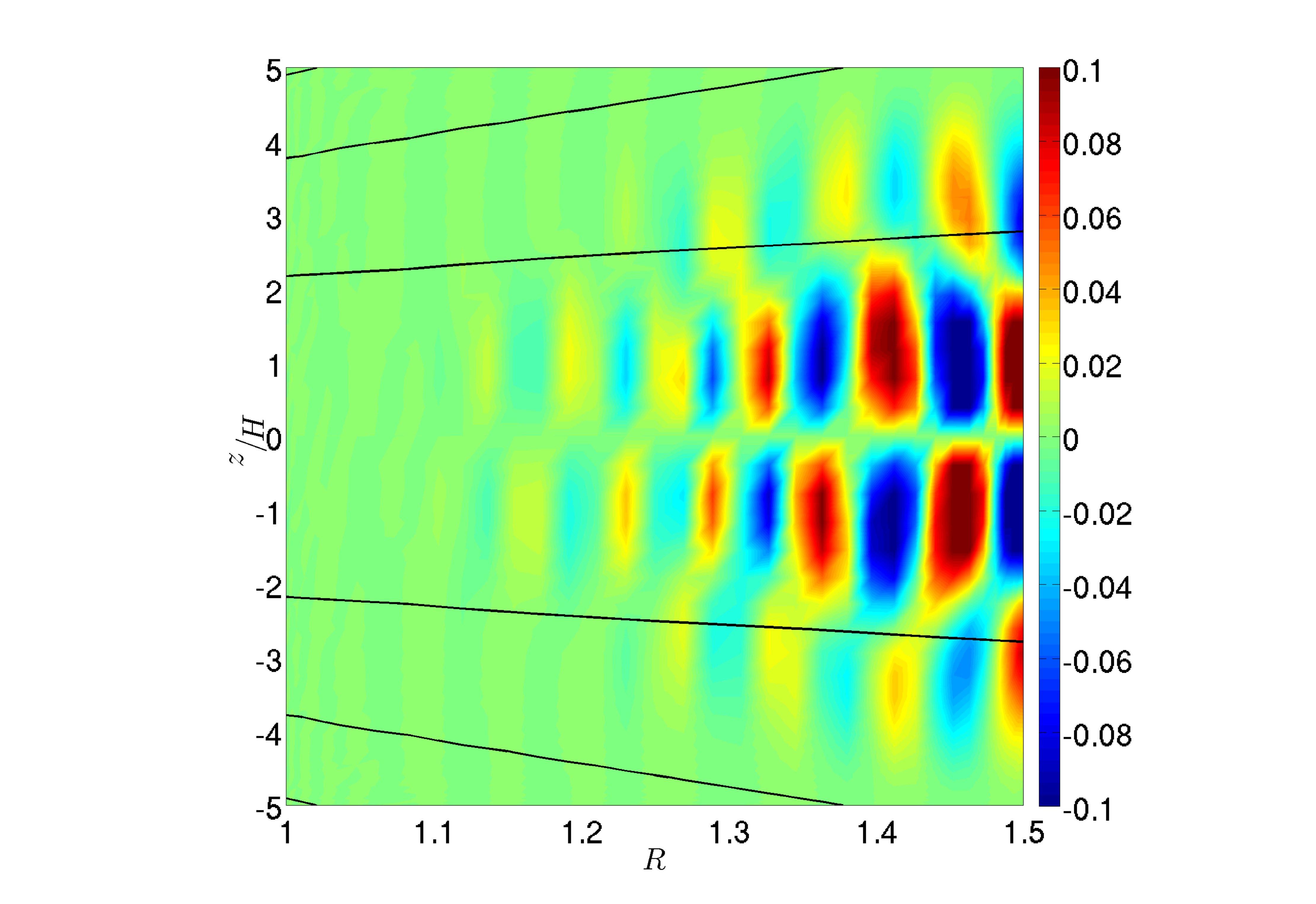} }
    \end{center}
  \caption{Illustration of the vertical momenta on the $(R,z)$-plane
    for several body modes in the locally isothermal disc without viscosity with
    $\epsilon=0.05$, $q=-1$, $z_{0}=5\epsilon$ and $R_1=1.5$, obtained using
    $N_{R}=N_{z}=40$. These modes have vertical structures for $u_z$ that are
    well described by Hermite polynomials with $n=0,1,2$ and $3$. Many modes with similar vertical structures (but
    different radial structures) to each of these are also unstable. 
    Also over-plotted are three contours of
    constant density (solid black lines) with
    $\rho=10^{-5},10^{-3},10^{-1}$ for reference. The amplitude in
    each panel is arbitrary.}
  \label{2Da}
\end{figure*}

As in \S~\ref{trueisothermalVSI}, we obtain a set of body modes,
essentially classical inertial waves that grow in the presence of
vertical shear.  We plot the vertical momenta for several examples
with either even or odd symmetry in Fig.~\ref{2Da} in a domain with $R_1=1.5$ and $z_0=5\epsilon$. These modes have
vertical structures with $n=1,2,3$ and $4$ ($n=0,1,2,3$ for $\rho
w$), which correspond with those obtained in the isothermal model
without boundaries in \S~\ref{trueisothermalVSI} (and the bottom panels
in Fig.~\ref{2}) for various radial structures. The vertical shear is
strongest at the inner radial boundary of the domain, whereas these
modes are concentrated near the outer boundary. We might
  expect this to be the case because axisymmetric inertial waves are
  localised at and within their turning surfaces, defined by $\kappa=\omega$ (where $\kappa\approx\Omega$ is the epicyclic frequency). The
  best resolved modes in our calculations 
  have low frequencies, and thus their turning
  surfaces lie near or beyond the outer boundary $R_1$. Consequently such
  modes prefer the largest radii possible in the
  computational domain. Those plotted here were chosen to illustrate that the global model exhibits body modes with the same vertical structure as the reduced model in \S~\ref{trueisothermalVSI}. The body modes are modestly growing, with a growth rate no larger than a third of the maximum vertical shear rate. These are the unstable modes with the longest wavelengths, and have a radial scale that is shorter than their vertical scale by a factor $O(\epsilon)$.

\subsubsection{Surface modes}

\begin{figure}
  \begin{center}
     \subfigure{\includegraphics[trim=8cm 0cm 8cm 1cm, clip=true,width=0.38\textwidth]{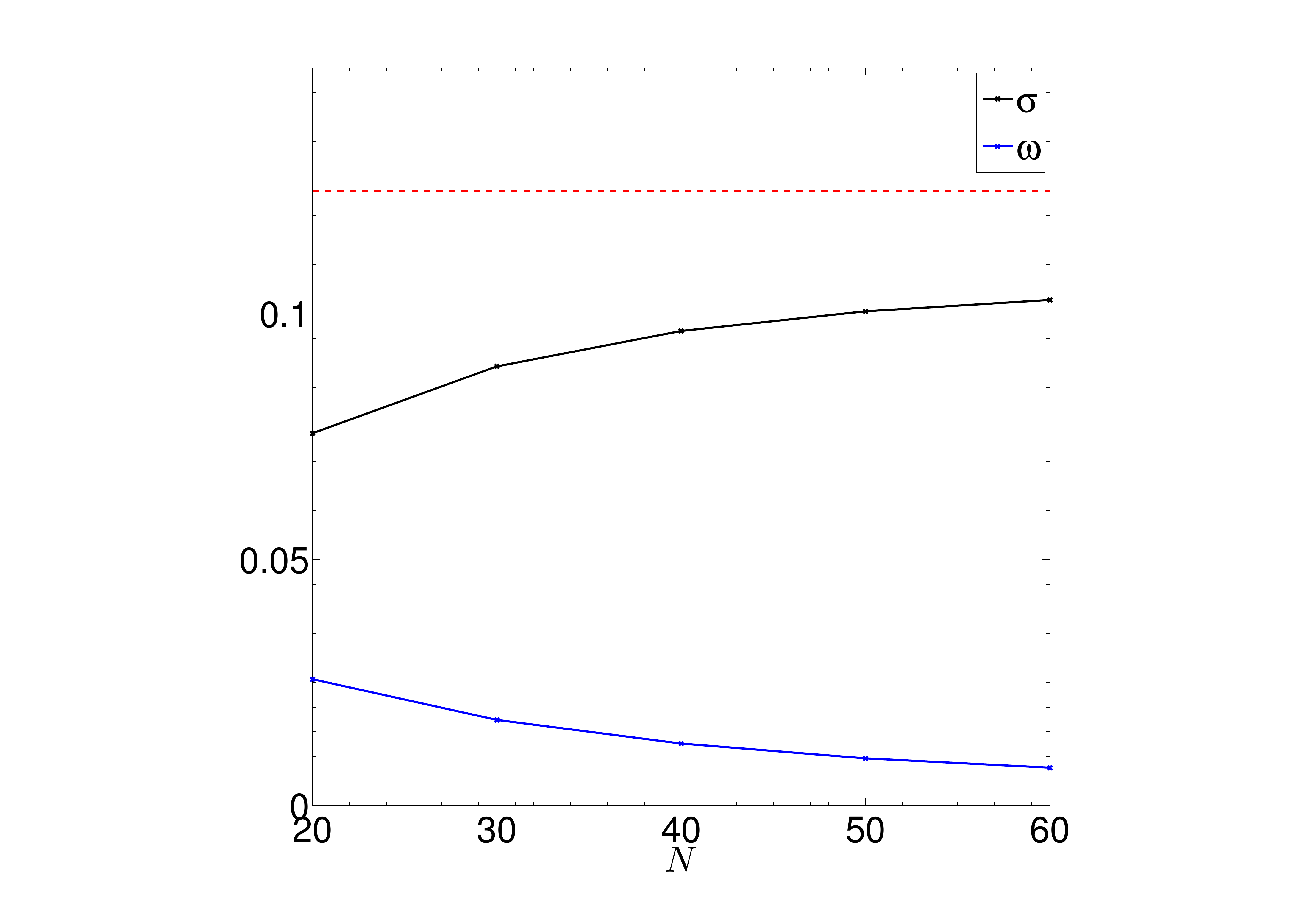} }
    \end{center}
  \caption{Growth rate and frequency of the fastest growing surface mode as a function of numerical resolution $N=N_{R}=N_{z}$ in an isothermal disc without viscosity, with $\epsilon=0.05$, $q=-1$, $z_{0}=5\epsilon $ and $R_1=1.0025$. These results were computed using an Arnoldi method. The growth rate increases with resolution $N$, and the frequency correspondingly decreases, indicating a lack of convergence. The red-dashed line is the maximum vertical shear rate.}
 \label{2Db}
\end{figure}

\begin{figure}
 \begin{center}
    \subfigure[$N_{R}=N_{z}=50,\nu=10^{-8},\omega=-0.0419+0.0681i$]{\includegraphics[trim=7cm 1cm 6cm 2cm, clip=true,width=0.33\textwidth]{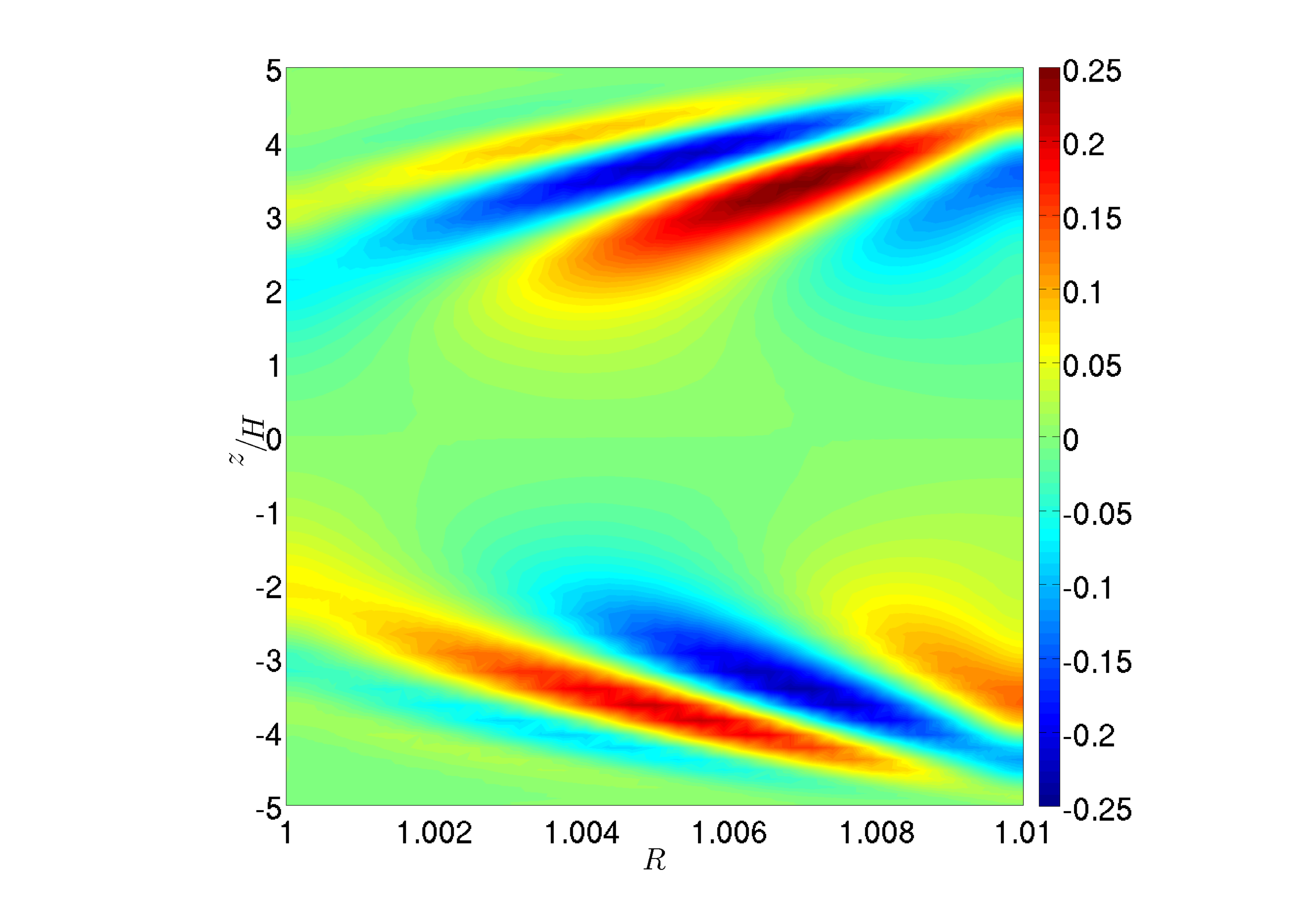}}
    \subfigure[$N_{R}=30,N_{z}=120,\nu=0,\omega=0.0092+0.0607\mathrm{i}$]{\includegraphics[trim=7cm 1cm 6cm 2cm, clip=true,width=0.33\textwidth]{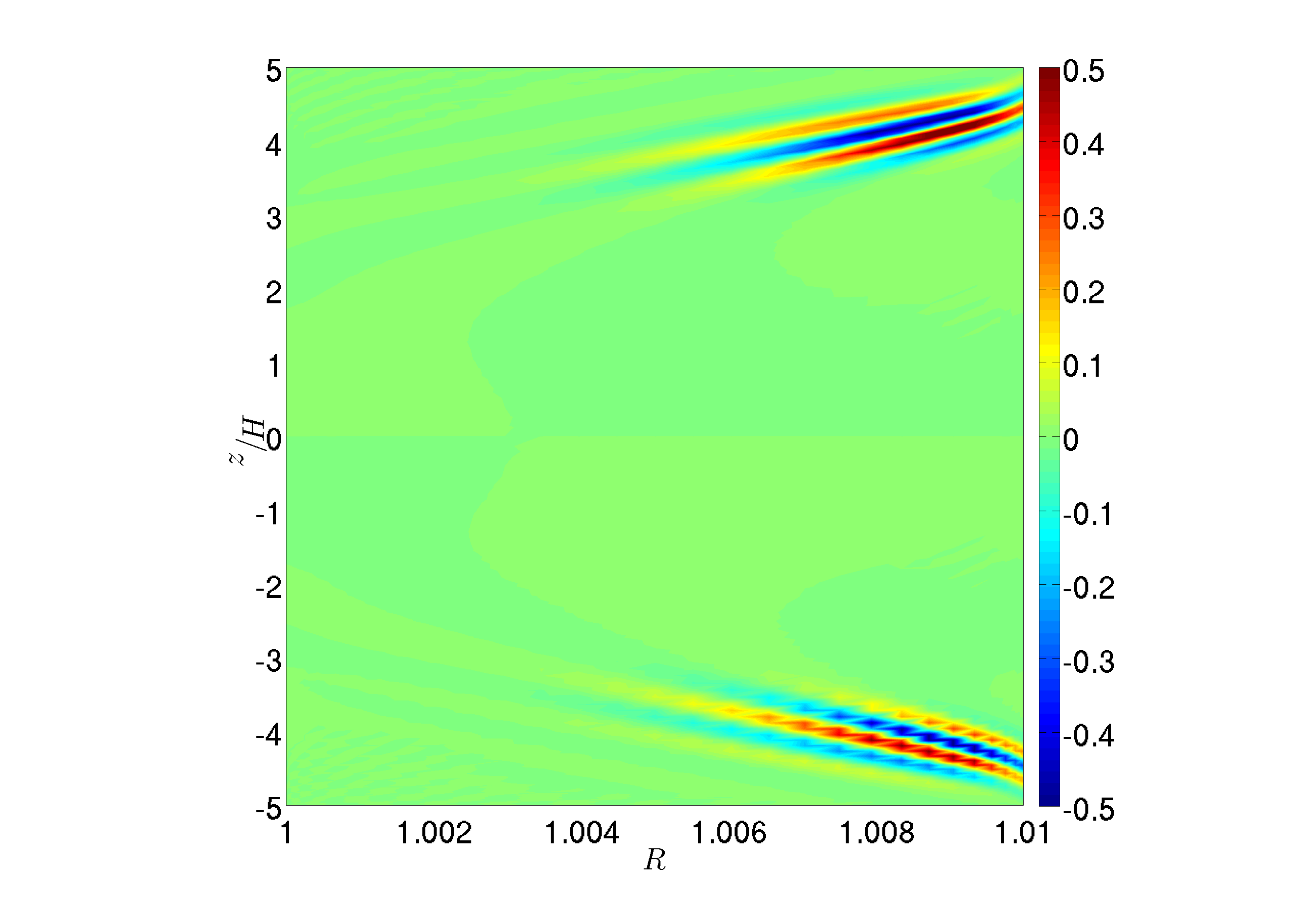}} \\
    \end{center}
  \caption{Illustration of the vertical momentum on the $(R,z)$ plane
    for a typical surface mode with and without viscosity, in an isothermal disc
    with $\epsilon=0.05$, $q=-1$, $z_0=5\epsilon $ and $R_1=1.01$. The amplitude is arbitrary.}
  \label{2Dc1}
\end{figure}

\begin{figure*}
 \begin{center}
    \subfigure[$\omega=0.0257+0.0757\mathrm{i}\;\; (N_{R}=N_{z}=20)$]{\includegraphics[trim=5cm 0cm 5cm 1cm, clip=true,width=0.32\textwidth]{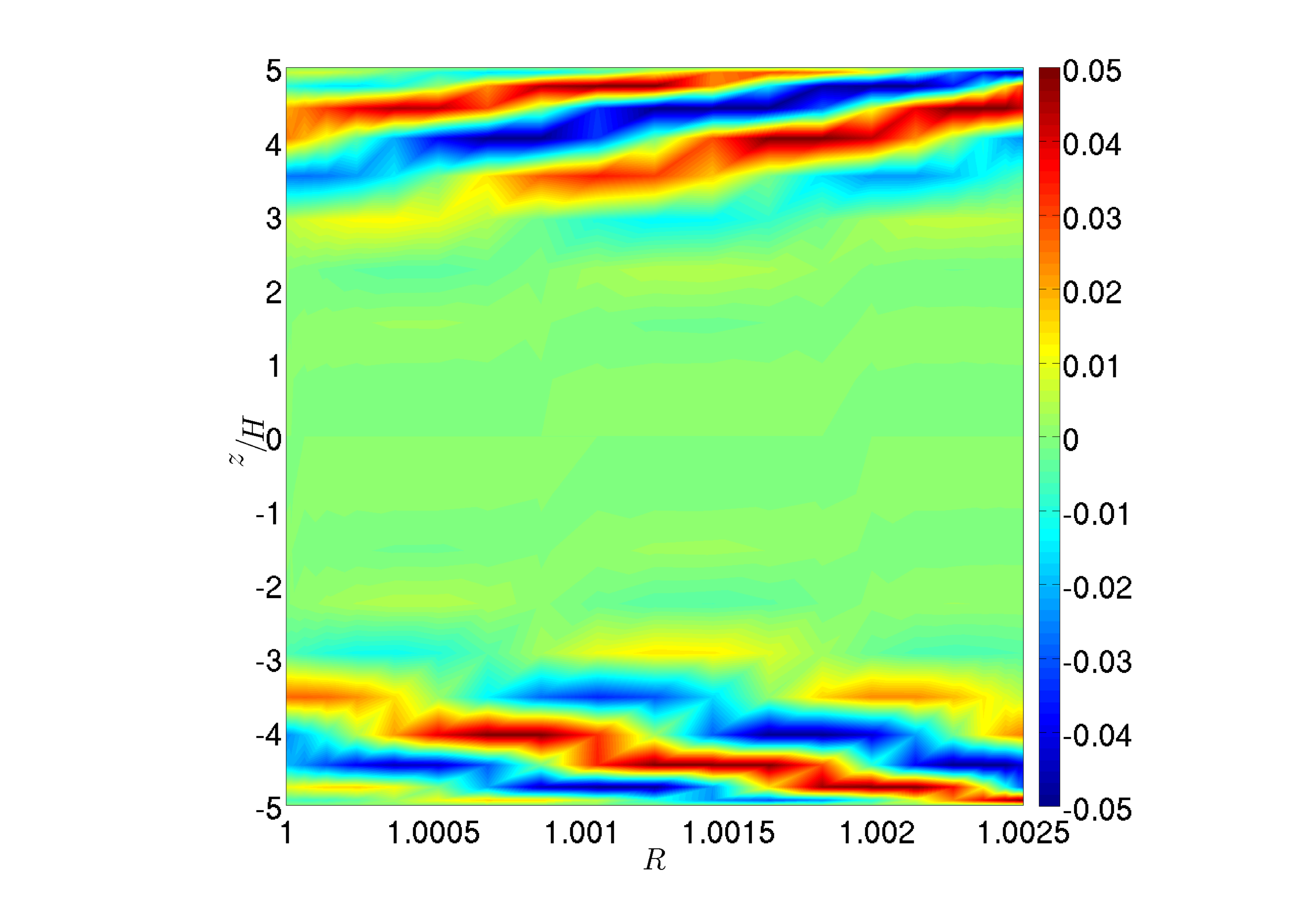} }
     \subfigure[$\omega=0.0174+0.0893\mathrm{i}\;\; (N_{R}=N_{z}=30)$]{\includegraphics[trim=5cm 0cm 5cm 1cm, clip=true,width=0.32\textwidth]{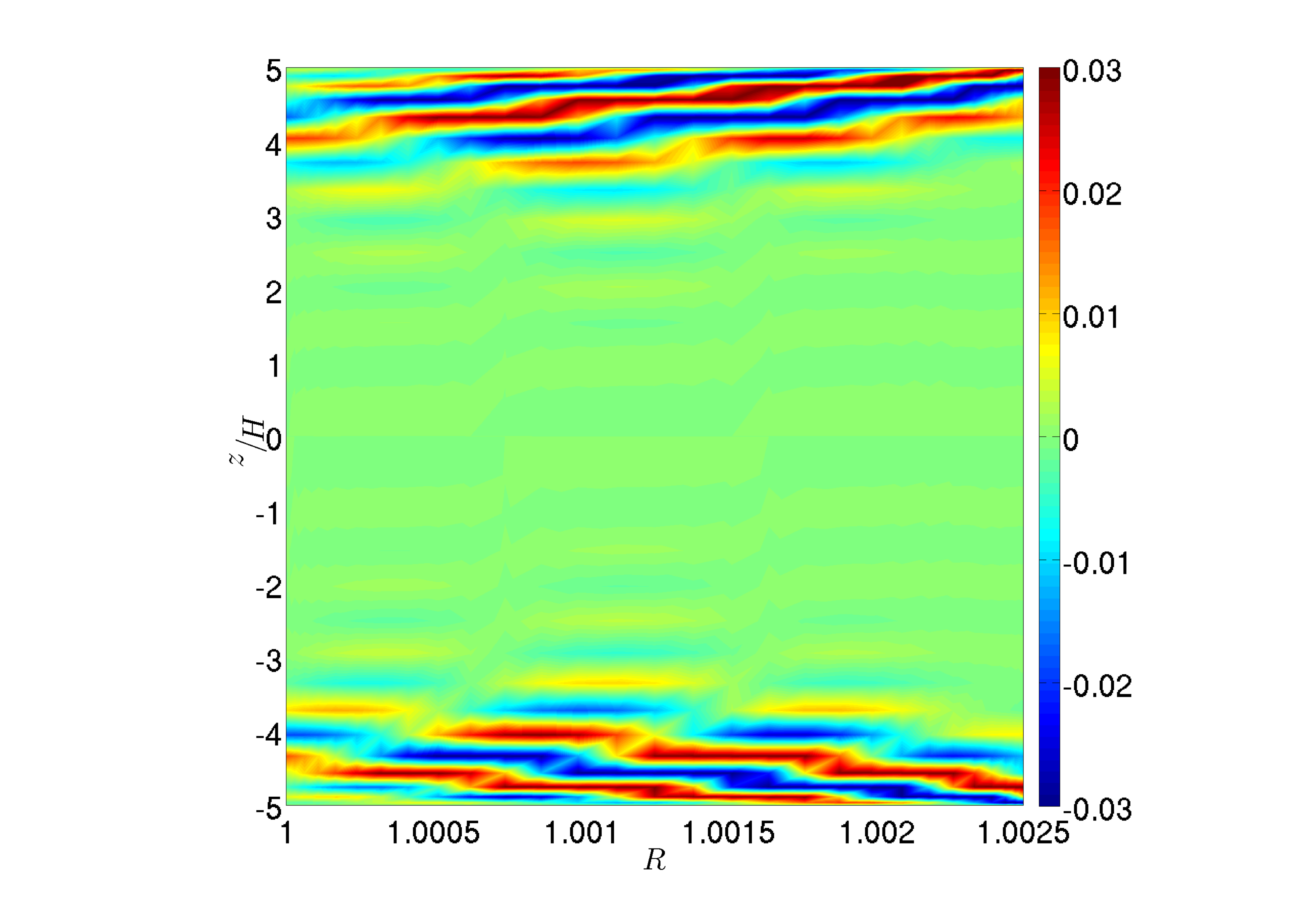} }
     \subfigure[$\omega=0.0126+0.0965\mathrm{i}\;\; (N_{R}=N_{z}=40)$]{\includegraphics[trim=5cm 0cm 5cm 1cm, clip=true,width=0.32\textwidth]{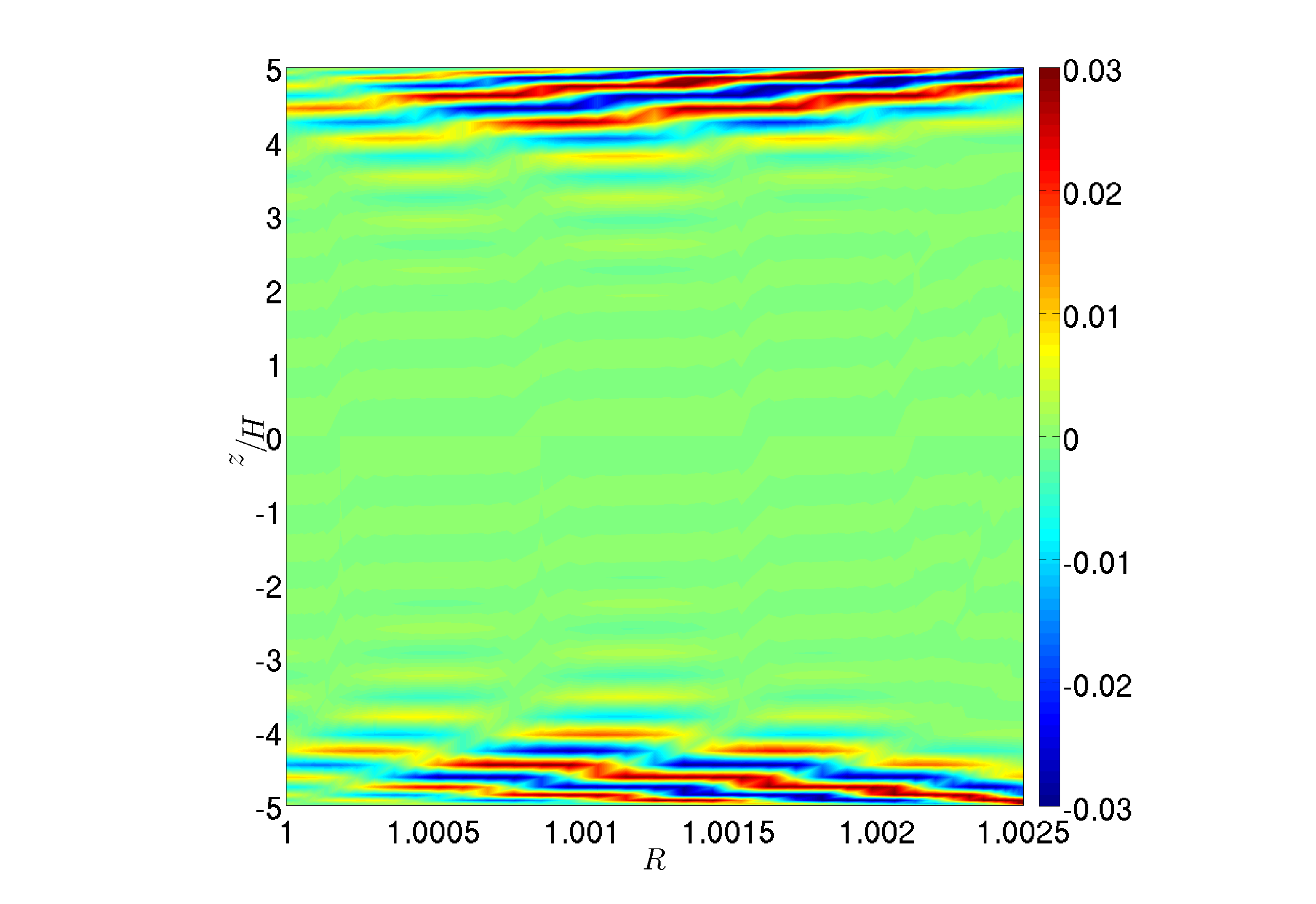} }
    \end{center}
  \caption{Illustration of the vertical momenta on the $(R,z)$ plane
    for the fastest growing surface mode without viscosity for several different
    resolutions, in a locally isothermal disc with $\epsilon=0.05$,
    $q=-1$, $z_0=5\epsilon$ and $R_1=1.0025$. These have increasing resolution from left to right panels, as indicated in the figure labels. This shows
    that the fastest growing surface modes occur on the smallest
    available lengthscales, hence this problem is ill-posed because
    modes on the numerical grid scale will always be the fastest
    growing. \textit{Note the compressed radial scale, which was chosen to capture such
    rapidly growing surface modes -- these modes are actually inclined by a small angle to the vertical, as we expect}. The amplitude in each
    panel is arbitrary.}
  \label{2Dc}
\end{figure*}

\begin{figure}
  \begin{center}
     \subfigure{\includegraphics[trim=7cm 0cm 8cm 1cm, clip=true,width=0.38\textwidth]{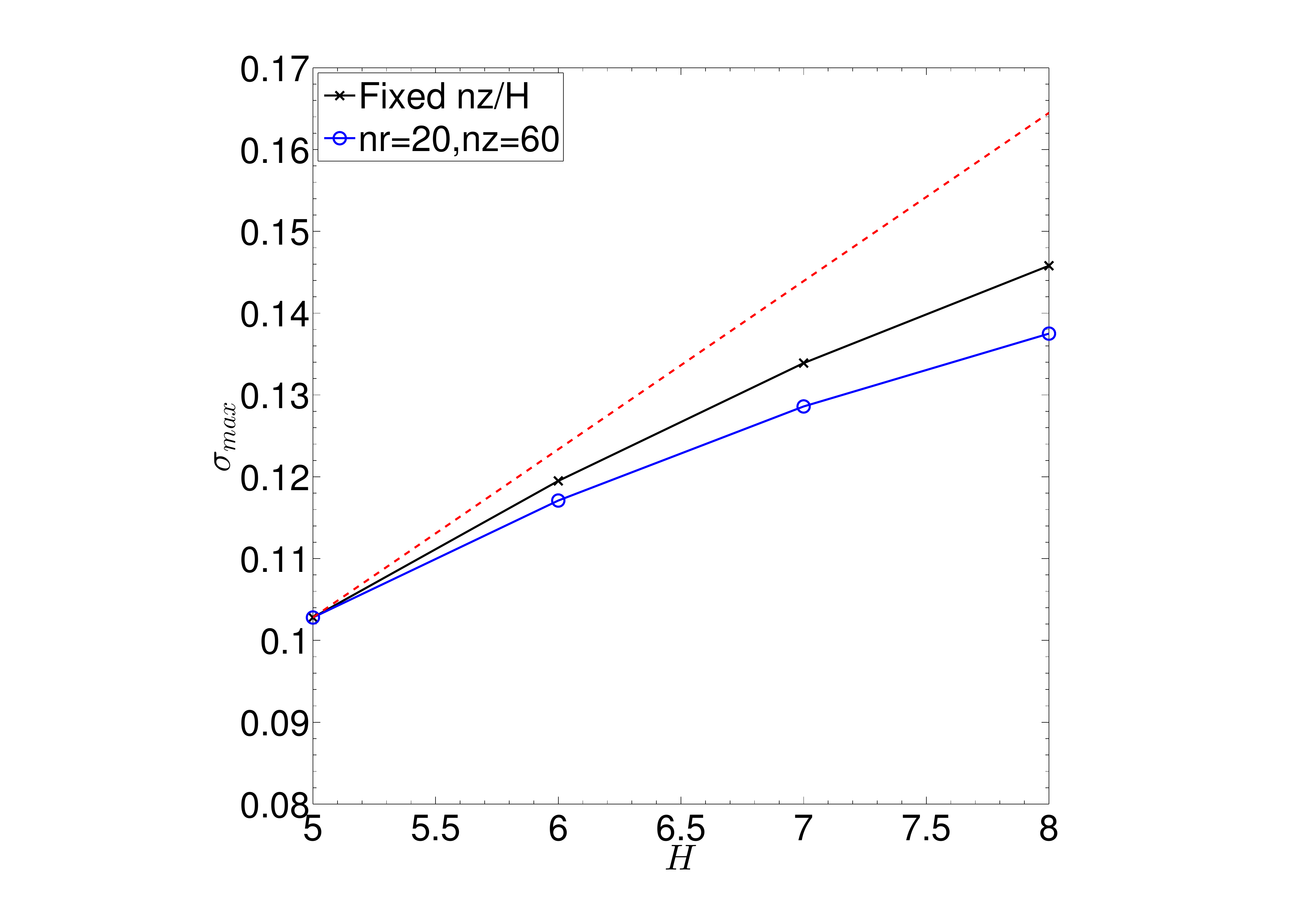} }
    \end{center}
  \caption{Growth rate of the fastest growing surface modes versus vertical domain size in a locally isothermal disc without viscosity for several values of $z_{0}=H\epsilon$, with $\epsilon=0.05$, $q=-1$ and $R_1=1.0025$. Results were computed using an Arnoldi method with fixed resolution $N_{R}=20, N_{z}=60$ (blue circles and line) and with $N_{R}=20$ and a fixed value of $N_{z}/H=12$ (black crosses and line). The growth rate of the fastest growing mode increases with $H$, indicating the lack of convergence as we increase the vertical size of our domain.}
 \label{2Da1}
\end{figure}

Just as in \S~\ref{VSIisoboundaries}, we obtain a class of rapidly
growing short-wavelength surface modes, which come in pairs with
oppositely signed frequencies. For illustration, we show an example of a typical surface mode with and without viscosity in Fig.~\ref{2Dc1}.

Without viscosity, 
the fastest growing mode always occurs on the smallest available
lengthscale. To demonstrate this, in the top panel of Fig.~\ref{2Db} 
we plot the growth rate and (real) frequencies of the fastest growing
mode versus resolution $N$ in a disc with $\epsilon=0.05$ and $q=-1$,
adopting a domain with $R_{0}=1.0025$ for several different
resolutions $N=N_{R}=N_{z}\in[20,30,40,50,60]$. Such a small radial
domain is chosen in order to better capture the rapidly growing
surface modes. We also plot the vertical momenta for the fastest
growing mode for several different resolutions in Fig.~\ref{2Dc}. As
we increase the resolution, the fastest growing mode moves to lower
frequency and exhibits increasingly shorter lengthscales. In addition,
its growth rate increases as we increase the resolution, gradually
tending towards the maximum vertical shear rate (which is $|q|H_{0}
\Omega_{0}/(2R_{0})\approx 0.125$). This is because these modes become
increasingly localised in the vicinity of the vertical boundary, where
the vertical shear is maximal. Note that Fig.~\ref{2Db} does
\textit{not} indicate convergence as $N$ is increased, because it is a
\textit{different} mode that is plotted (that is most unstable)
 for each $N$, and
this mode always tracks the grid-scale. This is demonstrated further in Fig.~\ref{2Dc}. Since these modes always occur on the grid-scale, they are necessarily the most poorly resolved modes. 

Another pathology of the isothermal model, that we first showed in \S~\ref{VSIisoboundaries}, is that the maximum growth rate depends on the vertical domain size. To further demonstrate this, we illustrate the fastest growing mode for calculations with $z_{0}= H\epsilon$, with $H\in[5, 6, 7,8]$ on Fig.~\ref{2Da1} in a radial domain with $R_{0}=1.0025$. We plot results using both a fixed resolution of $N_{R}=20$ and $N_{z}=60$ (blue circles and line), and for fixed $N_{R}=20$ and $N_{z}/H=12$ (black crosses and line). The growth rate continues to increase as we increase $H$. This is simply explained by the dependence of the vertical shear rate, which increases without bound as $O(z)$. However, the finite vertical resolution does not fully capture the fastest growing mode for each $H$ (even with fixed $N_{z}/H$), therefore the dependence on $H$ is slightly weaker than the linear extrapolation (red dashed line).

The maximum growth rate for the surface modes compares reasonably  
well with the numerical simulations of \cite{Nelson2013} for similar
parameters (they obtain a growth rate of $0.125$ whereas our maximum 
growth rate is approximately $0.1$ for $N=60$, for example -- 
the difference is due to the inevitably smaller resolution in our case).
However, growth rates are unconverged with respect to both (a) the
numerical grid resolution and (b) the vertical domain size. 
The fastest growing mode occurs on the smallest available lengthscale,
which is
 always the grid-scale. In
 addition, the fastest
 growing mode occurs at the boundaries in $z$. Since the
 boundary is artificial imposed and this choice 
 sets the maximum growth rate,
 our results are strongly dependent on an arbitrary parameter $H$,
 which far from ideal. These problems will also afflict nonlinear
 simulations of the VSI in locally isothermal models.

\subsection{Viscous calculations}

Although the dependence on the vertical boundary cannot be removed,
convergence with respect to resolution can be addressed via the
inclusion of viscosity. This is necessary for the problem to be
well-posed. 
As we will demonstrate here, even when we include viscosity,
 the dominant modes prefer the smallest available lengthscales. 

\begin{figure*}
  \begin{center}
  \subfigure[$R_{o}=1.5, N_{R}=N_{z}=60$]{\includegraphics[trim=6cm 0cm 7cm 1cm, clip=true,width=0.38\textwidth]{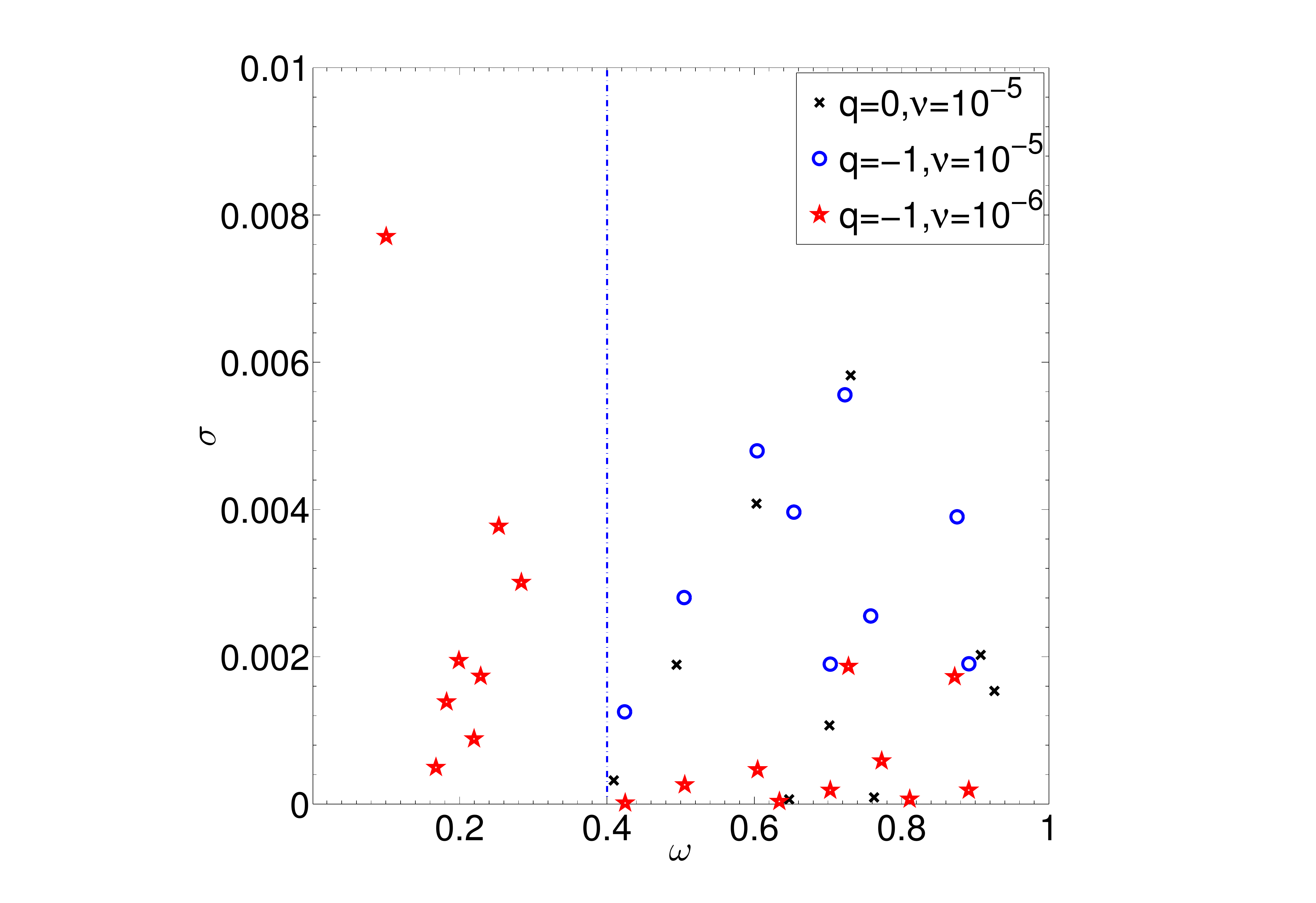} }  \hspace{1.5cm}
  \subfigure[$R_{o}=1.05, N_{R}=30, N_{z}=150$]{\includegraphics[trim=6cm 0cm 7cm 1cm, clip=true,width=0.38\textwidth]{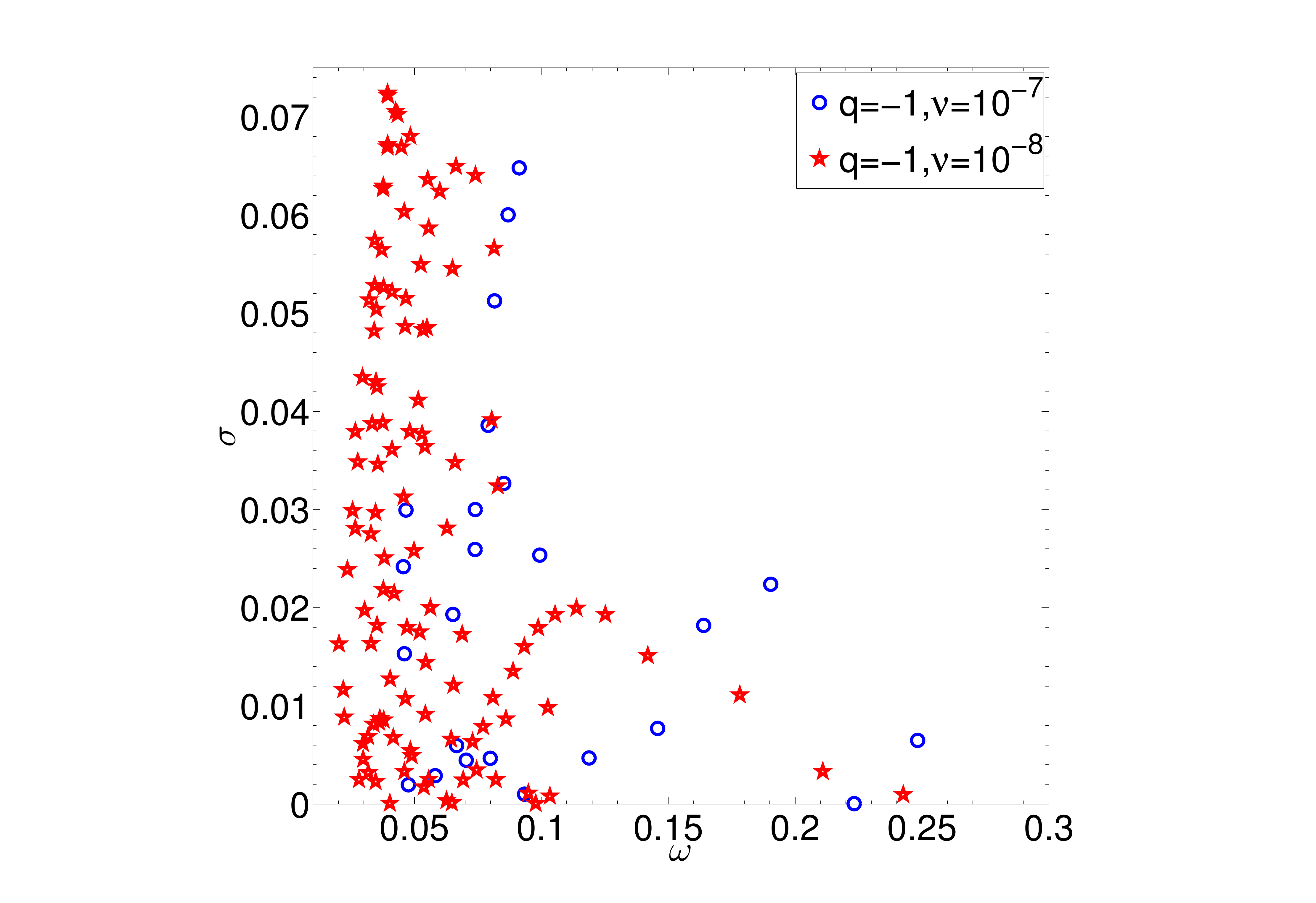} }
      \end{center}
  \caption{Growth rate of unstable modes versus their (real)
    frequencies in a locally isothermal disc with $\epsilon=0.05$, $z_{0}=5\epsilon$, $R_{1}=1.5$ and $1.05$ for several viscosities (resolution is listed in the figure labels). The dot-dashed line in the left panel demarcates the low-frequency VSI modes from the viscously overstable modes. The VSI appears only when $\nu\lesssim 10^{-6}$. As the viscosity is decreased, there are more unstable modes, which occur on shorter lengthscales, and their maximum growth rates increase. (We caution that some of the eigenvalues in the right panel for $\nu=10^{-8}$ in particular do not appear to be converged with the adopted resolution -- however, we do not expect any of our conclusions to be adversely affected by this.)}
 \label{2Dd}
\end{figure*}

In Fig.~\ref{2Dd}, we plot the unstable modes on the complex frequency
plane for converged modes in a domain with
$z_{0}=5\epsilon$ in various radial domains with several values of $\nu$. Note that
 these figures are busier than say Fig.~\ref{5} because they include modes
 of all resolvable vertical \textit{and} radial quantum numbers; in Fig.~\ref{5}, the
 radial wavenumber $k$ is restricted to take only one value.
 
The left panel shows unstable modes when $R_{0}=1.5$. Two classes
of instability are obtained in this case: the VSI with $\omega\lesssim
0.4$ and the viscous overstability with $\omega\gtrsim 0.4$, which
have been separated visually by the blue dot-dashed vertical line. The
viscous overstability
\citep{Kato1978,KleyPapLin1993,LatterOgilvie2006} preferentially
excites modes with little vertical structure and long radial
wavelengths $k^{-1} \gtrsim O(H_{0})$, with frequencies comparable
with the local rotational frequency and growth rates $O(\nu
k^2)$. This can be distinguished from the VSI, which occurs only when
$q\ne 0$, preferentially excites waves with short radial wavelengths,
$\omega\ll 1$, and has growth rates bounded above by the maximum
vertical shear rate. When $q=0$, only the viscous overstability
persists -- this is modified by vertical shear when $q=-1$, but is
found to remain in the same region of the complex plane. 
Given that our focus here is on the VSI, we will not discuss these modes any further.

The VSI is absent if the viscosity is too large ($\nu\gtrsim 10^{-6}$) since it preferentially excites (in this case body) modes with short radial wavelengths which are damped by a weak viscosity. We note that a similar order of magnitude for $\nu$ is required to obtain the VSI in numerical simulations \citep{Nelson2013}. The fastest growing mode when $\nu=10^{-6}$ is a body mode that has $n=1$ (i.e., the fundamental ``corrugation mode") with a radial wavelength of approximately 0.05 (similar to the top left panel of Fig.~\ref{2Da}).

The right panel in Fig.~\ref{2Dd} plots the unstable modes in a
smaller radial domain of $R_{0}=1.05$, for several viscosities. The
smaller domain is chosen so as to permit a greater radial resolution.
This demonstrates that as we decrease the viscosity, there are (a)
more unstable modes and (b) the fastest growing mode has a larger
growth rate, in tune with our expectations from the reduced model.
The increase in the number of unstable modes is because the smaller $\nu$ the shorter the
viscous cutoff; consequently, there is a wider range of lengthscales
that are potentially unstable (cf.~Fig.~\ref{7}). When $\nu=10^{-6}$ there is only one unstable mode in this case, which we omit for clarity.

We also observe that the fastest growing modes occur on the shortest 
available lengthscales, since the mode frequency decreases with
decreasing $\nu$.
 However, we note that there are many unstable modes, 
with the fastest growth rate being somewhat smaller than 
the maximum magnitude of the vertical shear ($0.125$)
 as a result of viscous damping.  Surface modes appear, and become dominant when $\nu\lesssim 10^{-7}$.

\subsection{Summary}

Our results in this section are in accord with the
reduced model introduced by \cite{Nelson2013} and revisited in
\S~\ref{VSIisoreduced} but with imposed vertical boundaries.
We obtain the same classes of modes
and they have similar growth rates in both models (and compared with
numerical simulations) in a finite vertical domain. We have
highlighted that the VSI preferentially
excites ultra short-scale disturbances which occur on the smallest
available lengthscales, be they numerical or viscous. 
 Adoption of explicit
viscosity is therefore required to obtain results that are converged with
respect to resolution. 

True isothermal discs do not exhibit surface modes,   
as we have explained in \S~\ref{VSIisoreduced}. Only the presence of a  
rigid boundary permits those modes to exist, but such a model is  
poorly defined given the freedom regarding our choice of $z_0$. Given
the notable lack of convergence as the vertical domain size is
varied, this indicates  that the vertically isothermal model is not
well suited for studying the linear properties of the VSI. 

\section{Discussion and conclusions}
\label{Conclusions}

We have analysed the linear stability of astrophysical discs with vertical shear, as a result of their radial variations in temperature or entropy. Such variations are expected to be present in real discs, as indicated by both observations and theory (e.g.~\citealt{AndrewsWilliams2005}; \citealt{ChiangGold1997}), and generally lead to vertical shear, thereby rendering the disc unstable to a hydrodynamic instability. Recent nonlinear simulations of the resulting vertical-shear instability have highlighted its potential to drive hydrodynamic activity in MRI-stable regions of protoplanetary discs \citep{ArltUrpin2004,Nelson2013,StollKley2014}. The aim of this work was to better understand the nature of the linear instability in two simple disc models: the locally isothermal disc with a radial power law in temperature (building on previous work by \citealt{Nelson2013}) as well as the locally polytropic disc with a radial power law in entropy.

In both models, there are two classes of unstable modes: modestly growing
(vertically) global body modes, and rapidly growing ultra short-scale
surface modes\footnote{The surface modes are somewhat analogous to the modes that appear in fingering convection, another type of double-diffusive instability (e.g.~\citealt{Brown2013}).}.
The latter only appear in discs with a vertical
surface and, though this is not the case in strictly isothermal
models, realistic discs should exhibit a density feature/transition upon
which such modes can affix themselves. 
Ironically, artificially
imposed boundaries present qualitatively correct behaviour
even if this behaviour is a numerical artefact!
The value of such models beyond the qualitative is unclear, however.

A separate issue is that surface modes preferentially occur
 on the smallest available length scales. This necessitates the
 inclusion of
 viscosity to obtain convergence in any numerical simulation, 
otherwise the results will inevitably depend on numerical resolution.
It may be that the VSI saturates on length scales much smaller
than can be reached by nonlinear global simulations, which would
consequently pump power to artificially 
larger scales and hence misrepresent the ensuing
turbulent state. This however will
be very difficult to test numerically.

We have restricted our study to the locally isothermal and locally
polytropic models, since studying the linear VSI in a more realistic
model would require the two-dimensional disc structure to be computed numerically
after accounting for the various sources of heating and cooling,
which are uncertain. In any case, it is unlikely that there is much
to be gained from doing this, owing to the similarity in its
properties in both the locally isothermal disc (with an artificial boundary) and the locally polytropic disc.

As shown by recent work, the nonlinear evolution of the VSI leads to
wave activity or turbulence, which  transports angular momentum
vertically in order to  eliminate the vertical shear
\citep{ArltUrpin2004,Nelson2013,StollKley2014}.  It may also transport
angular momentum radially to enable the disc to accrete at modest
levels. 
The amount of such hydrodynamical activity in reality depends on the
battle between the external heating of the disc, and the efficiency
of the VSI in eliminating the ensuing vertical shear. The disc must
be externally heated sufficiently strongly
or the
VSI will eventually win out, leading to a very low level or no 
turbulence (as observed in \citealt{StollKley2014}). 
That this process can occur on timescales
that are 
not very long may preclude the use of local computational models to
determine 
the transport properties of the VSI. This is unfortunate, because
global simulations of the VSI
 that are able to capture the fastest growing modes (even in the
 presence of viscosity)
 are computationally very challenging. Nevertheless, it would be
 worthwhile to 
determine the nonlinear outcome of the VSI in more realistic disc
models
(continuing from \citealt{StollKley2014}) to determine its longevity and transport properties.

An interesting byproduct of the VSI's saturation is its
radial transportation of angular momentum. If the VSI is sufficiently
active in protoplanetary dead zones, it may provide the stresses
necessary for these regions to accrete. 
The saturation of
the VSI could be controlled by 
the smallest-scale surface modes, which may be ineffectual in this
regard because of their small scales. On the other hand,
longer wavelength body modes carry greater quantities of angular
momentum, but only if the saturated state endows them with sufficient power.
If this is the case a crude upper limit on the $\alpha$ associated
  with this transport is of order $\epsilon^3\sim 10^{-4}$ (which is roughly consistent with previous simulations).
On the other hand, \cite{StollKley2014} show that
the measured $\alpha$ decreases with resolution, and is moreover
unconverged, a result that suggests that it is in fact the smallest
scales that are
controlling the transport, not the largest, and that in real discs
the value of $\alpha$ could be negligible. Obviously much more
work is required to further test these ideas.

\section*{Acknowledgments}
This research was partially supported by STFC through grants
ST/J001570/1 and ST/L000636/1. AJB is supported by the Leverhulme
Trust and Isaac Newton Trust through the award of an Early Career
Fellowship. We would like to thank John Papaloizou for his helpful comments
and for generously reading through an earlier version of the
manuscript. We also thank the referee for promptly reviewing the manuscript.

\appendix

\section{Derivation of a reduced model for the vertical-shear instability in a locally polytropic disc}
\label{derivation}

We look for linear axisymmetric perturbations
($u_{R},u_{\phi},u_{z},P^{\prime},\rho^{\prime},S^{\prime}$) 
to Eqs.~\ref{fulleqns1}--\ref{fulleqns2}, to which we also include thermal diffusion in the entropy equation of the form
$\frac{1}{\rho T} \nabla \cdot \left[\chi\nabla T\right],$
where $\chi=\chi(\rho,T,\kappa)=\frac{16 \sigma T^3}{3\kappa \rho}$ is the thermal conductivity. We will end up neglecting thermal diffusion when constructing the basic state, but will include it when considering the small-scale perturbations. The motivation is that we are considering slow perturbations with very short radial lengthscales, on which thermal diffusion is very rapid, so the perturbations evolve approximately isothermally. 

The system of linearised equations is
\begin{eqnarray}
\label{linsystem1}
\partial_{t}u_{R}  &=& 2 R\Omega u_{\phi}-\frac{1}{\rho}\partial_{R}P^{\prime} +\frac{\rho^{\prime}}{\rho^2}\partial_{R}P, \\
\partial_{t}u_{\phi} &=&-u_{R}\frac{1}{R}\partial_{R} (R^2\Omega) - u_{z}\partial_{Z}(R\Omega), \\
\partial_{t}u_{z} &=& -\frac{1}{\rho}\partial_{Z}P^{\prime}+\frac{\rho^{\prime}}{\rho^2}\partial_{Z}P, \\
\partial_{t}\rho^{\prime} &=& -u_{R}\partial_{R}\rho-u_{z}\partial_{Z}\rho-\frac{\rho}{R}\partial_{R}(Ru_{R}) - \rho\partial_{Z}u_{z},\\
\partial_{t}S^{\prime} &=&-u_{R}\partial_{R}S
-u_{z}\partial_{Z}S + \frac{1}{\rho T}\nabla \cdot\left(\chi \nabla T^{\prime}\right).
\label{linsystem2}
\end{eqnarray}

We adopt a non-dimensionalisation such that our fiducial radius has $R_{0}=1$ and $\Omega_{0}=1$, and take $\epsilon = H_{0}/R_{0}=H_{0}$ as our small parameter. To obtain a reduced model, we consider slow dynamics on vertical scales comparable with the disc thickness and radial scales that are much smaller. We also assume that the resulting velocities are comparable with the sound speed, with the exception of the radial velocity, which is assumed to be much slower. In particular, we define a slow timescale $\tau =\epsilon t$ such that $\partial_{t}=\epsilon \partial_{\tau}$, along with new radial coordinate $x$ and vertical coordinate $z$, such that
\begin{eqnarray}
&& R-1=\epsilon^{2}x, \;\;\;\;\; \partial_{R}=\epsilon^{-2}\partial_{x}, \\
&& Z=\epsilon z, \;\;\;\;\;  \partial_{Z}=\epsilon^{-1}\partial_{z}.
\end{eqnarray}
This allows us to neglect curvature and to consider the geometry as locally Cartesian, similar to the classical shearing box \citep{GoldreichLyndenBell1965,UmurhanRegev2004}. We also define scaled velocity components $(u,v,w)$ such that
\begin{eqnarray}
u_{R}=\epsilon^{2} u, \;\;\;\; u_{\phi}=\epsilon v, \;\;\;\; u_{z}=\epsilon w,
\end{eqnarray}
and appropriately scale the perturbed pressure, density and entropy for the approximation to be consistent as follows:
\begin{eqnarray}
P^{\prime}=\epsilon^3 \tilde{P}, \;\;\;\;\; \rho^{\prime}=\epsilon \tilde{\rho}, \;\;\;\;\; S^{\prime}=\epsilon \tilde{S}.
\end{eqnarray}
The background pressure is also rescaled as $P=\epsilon^2 \tilde{P}_{0}$.
We also note that $\partial_{R}$ acting on background quantities (e.g. $\rho$) is $O(1)$ and $\partial_{z}$ acting on background quantities is $O(\epsilon^{-1})$. Under the above scaling assumptions (and neglecting thermal diffusion), the basic state in \S~\ref{polybasic} can be written
\begin{eqnarray}
\rho&=&\rho_{0}\left[1-\frac{z^2}{H_{0}^2}\right]^{m}, \\
P &=&P_{0}\left[1-\frac{z^2}{H_{0}^2}\right]^{m+1}, \\
R\Omega&=&R_{0}\Omega_{0}\left[1 + \frac{q_s z^2}{4\gamma}\epsilon^2 \right]^{\frac{1}{2}},
\end{eqnarray}
where the disc thickness is
\begin{eqnarray}
H_{0}=\sqrt{2(1+m)\frac{P_{0}}{\rho_{0}}}.
\end{eqnarray}

Finally, we take $\chi=\tilde{\chi}\epsilon^{\beta}$, where $\beta$ is an ordering parameter. We require $\beta>2$ so that thermal diffusion can be neglected when constructing the basic state, since 
\begin{eqnarray}
\nonumber
-\rho T (\boldsymbol{u}\cdot\nabla S)&=&\nabla\cdot \left(\chi \nabla T\right)\\ &=&\epsilon^{-2+\beta}\left[\tilde{\chi}\partial_{z}^{2}T+\partial_{z}\tilde{\chi}\partial_{z}T\right] \nonumber \\ &&
+\epsilon^\beta \left[\tilde{\chi}\partial_{R}^{2}T+\partial_{R}\tilde{\chi}\partial_{R}T\right],
\end{eqnarray}
where the left hand side is $O(1)$.
For the perturbations, we have
\begin{eqnarray} 
\nonumber
\nabla \cdot\left(\chi \nabla T^{\prime}\right)&=&\epsilon^{-3+\beta}\tilde{\chi}\partial_{x}^2\tilde{T} + \epsilon^{-1+\beta}\tilde{\chi}\partial^{2}_{z}\tilde{T} \\ &&+ \epsilon^{-1+\beta}\partial_{R}\tilde{\chi}\partial_{x}\tilde{T} + \epsilon^{-1+\beta} \partial_{z}\tilde{\chi}\partial_{z}\tilde{T}.
\end{eqnarray}
The dominant term here is clearly the first one, and this dominates over all other terms in Eq.~\ref{linsystem2} as long as $\beta <3$, since the leading order term on the left hand side of Eq.~\ref{linsystem2} is $w \partial_{z}S$, which is $O(1)$. If $\beta<3$, the influence of a stabilising entropy gradient on the perturbations is eliminated by rapid thermal diffusion. We will therefore consider $\beta \in (2,3)$. While this choice may seem contrived, what this means physically is that we are neglecting thermal diffusion for the basic state, but we are considering it to dominate the thermal evolution of the perturbations, which evolve isothermally.

Applying these scaling assumptions to Eq.~\ref{linsystem1}--\ref{linsystem2} leads to the reduced model
\begin{eqnarray}
\label{reducedpoly1}
0&=&2v-\partial_{x} h \\
\partial_{\tau} v &=& -\frac{u}{2} - w \frac{q_s z}{2\gamma} \\
\partial_{\tau} w &=& -\partial_{z}h \\
0&=&\rho\left[\partial_{x}u + \partial_{z}w \right] + w\partial_{z}\rho
\label{reducedpoly2}
\end{eqnarray}
at leading order, where the psuedo-enthalpy perturbation is $h=\tilde{P}/\rho$. The perturbations are in radial geostrophic balance and are anelastic. 

\section{WKBJ analysis of the locally polytropic reduced model}

In this section we obtain approximate analytic solutions to
Eq.~\ref{Polyredeqn} in the limit of large $k$. To ease the
calculation, we first rescale
$\omega$ and $q_s$, introducing the quantities $\hat{\omega}=k\omega$
and $\hat{q}= kq_s/(2\gamma)$. The limit of large $k$ then becomes
the limit of large $\hat{\omega}$ and $\hat{q}$.
Next Eq.~\ref{Polyredeqn} is transformed into Schr\"{o}dinger form
\begin{align}
&\frac{d^2 H}{d z^2} + \left[\hat{\omega}^2 +
  m\frac{1-(m-1)z^2}{(1-z^2)^2} \right. \notag \\
&\left. \hskip2cm  +\text{i}\hat{q}\frac{1-(1+2m)z^2}{1-z^2}
  +\hat{q}^2z^2 \right]H=0,  \label{apB1}
\end{align}
for the new dependent variable $H= (1-z^2)^{m/2}
\text{e}^{-\text{i}\hat{q}z^2/2} h(z)$. Note that asymptotically 
Eq.~\ref{apB1} is
 the harmonic oscillator equation for all $z$ except
for small regions near the boundaries at $z=\pm 1$.  

Near the upper boundary
$z=1$, Eq.~\ref{apB1} may be approximated by
\begin{equation}
\frac{d^2 H}{d z^2} + \left[\hat{\omega}^2 + \hat{q}^2+ \frac{m(2-m)}{4(1-z)^2}-\frac{\text{i}m\hat{q}}{1-z} \right]H=0,
\end{equation}
which admits a solution in terms of generalised Laguerre functions and
the Tricomi hypergeometric function. Only the former, however is
regular at the boundary. The appropriate solution is
hence
\begin{equation}
H\approx (1-z^2)^{m/2}\text{e}^{-\varpi(1-z)}L_{-\mu}^{m-1}[2\varpi (1-z)],
\end{equation}
where $L_{-\mu}^{m-1}$ is the generalised Laguerre function, 
$\varpi=\sqrt{-\hat{\omega}^2-\hat{q}^2}$, and
\begin{equation}
\mu= \frac{m}{2}\left(1+ \frac{ \text{i}\,\hat{q}}{\varpi}\right).
\end{equation}

In the limit of large $\varpi$, the solution
takes the following asymptotic form far from the boundary,
\begin{equation}
H \sim \Gamma(m-\mu)\,\text{exp}[\phi(z)]+ \Gamma(\mu)\,\text{exp}[-\phi(z)],
\end{equation}
where $\Gamma$ is the gamma function, and the phase function is
\begin{equation}
\phi(z)= \varpi(1-z)+ \frac{\text{i}\,m\,\hat{q}}{2\varpi}\ln[2\varpi(1-z)] + \frac{\text{i}\pi\mu}{2}.
\end{equation}
This solution, in fact, should hold throughout the rest of the domain
and so we impose boundary conditions (evenness or oddness) at the mid-plane.
This yields the eigenvalue equation
\begin{equation}\label{Eigv}
\varpi + \frac{\text{i}m \hat{q}}{2\varpi}\log (2\varpi) -
\tfrac{1}{2}\log \left(\frac{\Gamma (\mu)}{\Gamma(m-\mu)}\right) + \tfrac{1}{2}\text{i}\pi(n+\mu)=0,
\end{equation}
where $n$ is an integer.

Equation \eqref{Eigv} is transcendental in $\varpi$. However, in the
non-vertically shearing case it can be solved easily and we obtain
the very simple dispersion relation
\begin{equation}\label{nsomega}
\omega = \frac{\pi}{4k}\left(2n + m\right),
\end{equation}
recalling that $n$ is an integer, and $m$ is the polytropic index. 

For general $\hat{q}$, Eq.~\ref{Eigv} must be solved via a
root-finding algorithm, usually an easier task than tackling the ODE
itself. However, when $\hat{q}\sim 1$ (meaning $k q_s \ll 1$), we find that the
wave frequencies of the unstable body modes are given by Eq.~\ref{nsomega} to
leading order, while the growth rate is
\begin{equation}
\sigma \approx \frac{m q_s}{\pi\gamma(2n+m)}\log\left[\tfrac{1}{2}\pi(2n+m)\right].
\end{equation}
The growth rate is linear in the shear, but the larger $n$, the
smaller $\sigma$, in contrast to the isothermal case. This therefore predicts the $n=1$ body mode to grow fastest at small $k$, as we have observed in Fig.~\ref{8} (though the growth rate is only correct to within an $O(1)$ factor, as might be expected for such small $k$).

Estimates for the surface mode frequencies can be obtained by equating
the first and second terms in Eq.~\ref{Eigv}, assuming that $n$ is not
too large. To leading order in large $k$, this balance yields
\begin{equation}
\omega \approx \frac{\text{i}q_s}{2\gamma} + \frac{m\ln k}{4 k}.
\end{equation}
Thus the growth rates of the surface modes are proportional to $q_s$,
and their wave frequencies are typically a factor $\ln k/k$ smaller.

\bibliography{disc}
\bibliographystyle{mn2e}
\label{lastpage}
\end{document}